\documentclass[aip,pop,sd,reprint,10pt,amsmath,amssymb,graphicx]{revtex4-1}

\usepackage{mathtools}
\usepackage{amsfonts}
\usepackage{amsmath}
\usepackage{amssymb}
\usepackage{graphicx}
\usepackage{bm}
\usepackage{dcolumn}
\usepackage{color}
\usepackage{comment} 
\def\cala{\mathcal{A}}

\def\calj{\mathcal{J}}

\def\calm{\mathcal{M}}

\def\bq{\begin{equation}}
\def\eq{\end{equation}}
\def\bqy{\begin{eqnarray}}
\def\eqy{\end{eqnarray}}

\def\al{\alpha}

\def\de{\delta}
\def\De{\Delta}

\def\ga{\gamma}
\def\Ga{\Gamma}

\def\la{\lambda}

\def\na{\nabla}
\def\om{\omega}

\def\Up{\Upsilon}

\def\bfb{\mathbf{b}}
\def\bfB{\mathbf{B}}

\def\bfF{\mathbf{F}}
\def\bfA{\mathbf{A}}
\def\bfa{\mathbf{a}}
\def\bfj{\mathbf{J}}

\def\bfQ{\mathbf{Q}}

\def\bfV{\mathbf{V}}
\def\bfx{\mathbf{x}}
\def\bfv{\mathbf{v}}

\def\bfq{\mathbf{q}}

\def\bfet{\boldsymbol{\eta}}

\def\bfPi{\boldsymbol{\Pi}}
\def\bfpi{\boldsymbol{\pi}}

\def\mfA{\mathfrak{A}}
\def\mfB{\mathfrak{B}}
\def\mfJ{\mathfrak{J}}

\def\p{\partial}

\def\na{\nabla}

\def\inta{\int\!\! d^{3}a \,}
\def\intb{\int\! \!d^{3}b \,}
\def\intx{\int\!\! d^{3}x \,}

\begin{document}


\title{Hamiltonian Magnetohydrodynamics:  Lagrangian, Eulerian,  and Dynamically
Accessible Stability - Examples with Translation Symmetry}
\author{T. Andreussi}
\affiliation{SITAEL S.p.A., Pisa, 56121, Italy}
\author{P. J. Morrison}
\affiliation{Institute for Fusion Studies and Department of Physics,\\ The University of
Texas at Austin, Austin, TX 78712-1060, USA}
\author{F. Pegoraro}
\affiliation{Dipartimento di Fisica E.~Fermi, Pisa, 56127, Italy}
\date{\today}

\begin{abstract}

Because different constraints are imposed, stability conditions for dissipationless fluids and magnetofluids may take different forms when derived within the  Lagrangian, Eulerian (energy-Casimir), or dynamical accessible frameworks.  This is in particular the case when flows are present. These differences are explored explicitly  by working out  in detail two magnetohydrodynamic examples:  convection against gravity in a stratified fluid and translationally invariant perturbations of a rotating magnetized plasma pinch. In this second example we show in explicit form how to perform the time-dependent relabeling introduced in Andreussi {\it et al.}\ [Phys.\ Plasmas {\bf20}, 092104 (2013)]  that makes it possible  to reformulate Eulerian equilibria with flows  as Lagrangian equilibria in the relabeled variables. The procedures detailed in the present article provide a paradigm that can be applied to more general plasma configurations and  in addition extended to more general plasma descriptions where dissipation is absent.

\end{abstract}
\pacs{52.30.Cv, 02.30.Xx, 47.10.Df, 52.25.Xz}
\keywords{magnetohydrodynamics, stability, Hamiltonian, Poisson bracket}
\maketitle
 


\section{Introduction}

The early plasma literature on magnetohydrodynamics (MHD) is specked with traces of a general  underlying structure: the self-adjointness of the MHD force operator in terms of the displacement $\boldsymbol{\xi}$ of the original energy principle,  the Woltjer invariants of helicity and cross helicity and their use in obtaining Beltrami states, and the representation of the magnetic and velocity fields in terms of `Clebsch' potentials  being examples.  All of these are symptoms of the fact that MHD is a Hamiltonian field theory, whether expressed in Lagrangian variables as shown by Newcomb\cite{newcomb1962} or in terms of Eulerian variables as shown by Morrison and Greene.\cite{morrison80}  General ramifications of the Hamiltonian nature of MHD were  elucidated  in our series of publications,\cite{amp0,amp1,amp2a,ampE} while in  the present work we examine explicitly the stability of stratified plasma  and of rotating  pinch equilibria within each of the three  Lagrangian, Eulerian, and dynamically accessible  descriptions. 

These particular two examples were chosen because they are at once tractable and significant.  They display difficulties one faces in  ascertaining stability within the  three approaches  and provide a means to compare and contrast stability results.   The paper is designed to serve as a `how-to' guide for application of the three approaches, providing a framework for what one might expect,  and delineating the sometimes  subtle differences between the approaches.   Here  and in our  previous papers the scope was limited to MHD, but the same Hamiltonian structure exists for all important dissipation free plasma models,  kinetic as well as fluid,  and the story we tell for  MHD applies to them as well.    (See e.g.\ Ref.~\onlinecite{morrison2005} for review.)  Recently there has been great progress in understanding the Hamiltonian structure of extended MHD,\cite{kimura,keramidas14,LMT15,HKY15,LMM15,LMM16}  the effect of gyroviscosity,\cite{MLA14}  and relativistic magnetofluid models.\cite{DMP15,KMMA15}  In addition, recent work on hybrid kinetic-fluid 
models\cite{tronci10,pjmTTC14} and gyrokinetics\cite{BBMQ15,BMBGV65} now also lie within the purview.

There are many concepts of stability of importance in plasma physics (see Sec.VI of Ref.~\onlinecite{morrison98} for a general discussion) --   here we will only be concerned with what could be referred to as formal Lyapunov stability, where at least a  sufficient condition for stability is implied  by the positive-definiteness of a quadratic form obtained from the second variation of an energy-like quantity.  This kind of stability is stronger than spectral or eigenvalue stability:   for finite-dimensional systems it  implies nonlinear stability, i.e.,  stability to infinitesimal perturbations under the nonlinear evolution of the system.  Note, nonlinear stability should not be confused with finite-amplitude stability that explores the extent of the basin of stability, a confusion that oft appears  in the plasma literature.  For infinite-dimensional systems like MHD there are technical issues that need to be addressed in order to rigorously claim that formal Lyapunov stability implies nonlinear stability (see e.g.\ Ref.~\onlinecite{rein} for an example of a rigorous nonlinear stability analysis), but the formal Lyapunov stability of our interest is a most important ingredient and it does imply linear stability.

A common practice in the plasma literature, employed e.g.\  by Chandresekhar,\cite{chandresek}  is to manipulate  the linear equations of motion in order to obtain a conserved quadratic form that implies stability.  Although this procedure shows linear stability, it cannot be used  to obtain nonlinear stability and may give a misleading answer.  This is evidenced by the Hamiltonian system, which when linearized has both of the two Hamiltonians for two linear oscillators, 
\bq
H_{\pm}=\om_1(p_1^2 +q_1^2)/2 \pm \om_2(p_2^2 +q_2^2)/2\,.
\label{sho}
\eq
Both signs of \eqref{sho} are conserved by the linear system, yet  only one arises from the expansion of the nonlinear  Hamiltonian of the system.   Nonlinear Hamiltonians that give rise to linear Hamiltonians of the form of  $H_-$ can in fact be unstable (see Ref.~\onlinecite{MP90} for an example), and are prototypes for systems with negative energy modes. 
This example shows why  the formal Lyapunov stability, our subject,  is stronger than spectral or eigenvalue stability.
To reiterate, throughout by stability we will mean  formal  Lyapunov stability 

 The remainder of the paper is organized as follows:  in Sec.~\ref{sec:basics} we review basic ideas of the three approaches, giving essential formulas so as to make the paper self-contained.  Of note is the new material of Sec.~\ref{ssec:compform} that summarizes various comparisons between the approaches.  This is followed by our convection example of Sec.~\ref{sec:convection} and our pinch example of Sec.~\ref{sec:pinch}.  These sections are  organized  in parallel with Lagrangian, Eulerian (or so-called energy-Casimr), and dynamically accessible stability treated in order, followed by a subsection on comparison of the results.  Finally,  we conclude in Sec.~\ref{sec:conclusions}.


\section{Basics}
\label{sec:basics}

In what follows we will consider the stability of MHD equilibria  that are solutions to the following equations:
\bqy
&&\rho_e \bfv_e\cdot\nabla \bfv_e= -\nabla p_e + \bfj_e\times\bfB_e + \rho_e \nabla \Phi_e\,,
\label{Emom}\\
&&\nabla\times(\bfv_e \times\bfB_e)=0\,,
\label{Emag}\\
&& \nabla\cdot(\rho_e \bfv_e)=0\,,
\label{Eden}\\
&& \bfv_e\cdot \nabla s_e =0\,, 
\label{Eent}
\eqy
for the equilibrium velocity field $\bfv_e(\bfx)$, magnetic field $\bfB_e(\bfx)$,  current density $4\pi \bfj_e = \nabla\times \bfB_e$, density field $\rho_e(\bfx)$, and entropy/mass field $s_e(\bfx)$.  Here $\Phi(\bfx,t)$ represents an  external gravitational potential.  The pressure field is assumed to be determined by an internal energy function $U(\rho,s)$,  where $p=\rho^2\p U/\p\rho$ and the temperature is given by $T=\p U/\p s$.  For the ideal gas $p=c\rho^{\ga}\exp(\la s)$, with $c,\la$ constants and $\rho U= p/(\ga-1)$.   MHD has four thermodynamical variables $\rho, s, p$, and $T$. The assumption of local thermodynamic equilibrium implies that knowledge of two of these variables at all points $\bfx$ is  sufficient to determine the other two, once the $U$ appropriate to the fluid under consideration is specified.

For   static equilibria with  $\bfv_e\equiv 0$,    the only equation to solve is 
\bq
\nabla p_e= \bfj_e\times\bfB_e +  \rho_e \nabla \Phi_e\,. 
\label{Estat}
\eq
Equation \eqref{Estat} is one equation for several unknown quantities; consequently, there is freedom to choose profiles such as those for the current and pressure as we will see in our examples.

If we neglect the gravity force by removing $\nabla\Phi_e$,   Eq.~\eqref{Estat} leads as usual to the Grad-Shafranov equation, e.g., by noting that $\bfB_e\cdot \nabla p=0$ implies pressure is a flux function. However, unlike the barotropic case where $p$ only depends  on $\rho$,  in general this does not imply that $\rho$ and $s$ are flux functions, since their  combination in $p(\rho,s)$  could cancel out their variation on a flux surface. Thus, as far as  static ideal MHD is concerned,   because only $p$ occurs in the equilibrium equation,  density and temperature on a flux function can vary while pressure is constant.  The MHD static equilibrium equations give no information/constraints on this variation.  

When gravity is included,  Eq.~\eqref{Estat} still is only one constraining equation for several unknown quantities. 
In Sec.~\ref{sec:convection} we consider  stratified equilibria both with and without a magnetic field and we will investigate there  the role played by entropy. 

For  stationary equilibria the full set of Eqs.~\eqref{Emom}--\eqref{Eent} must be solved.  Because in general  there are many possibilities, we will restrict our analysis to the rotating  pinch example of Sec.~\ref{sec:pinch}, where we describe the equilibrium in detail.  

\subsection{Lagrangian formulae}
\label{ssec:lagB}

The Hamiltonian  for MHD in Lagrangian variables is  
\bqy
H[\mathbf{q},\boldsymbol{\pi}]&=&\inta
\bigg[
\frac{ \pi_{i} \pi^{i}}{2\rho_{0}}
+ \rho_{0}U\left(  s_{0},\rho_{0}/\calj\right)\nonumber\\
 &{\ }& \hspace{.5 in} + %
\frac{\partial q_{i}}{\partial a^{k}}  \frac{\partial q^{i}}{\partial a^{\ell}}
\frac{B_{0}^kB_{0}^{\ell}}{8\pi \calj} + \rho_0 \Phi(q,t)
\bigg]\,, 
\label{H_Lagr}%
\eqy
where $(\mathbf{q},\boldsymbol{\pi})$ are the conjugate fields with $\boldsymbol{q}(\mathbf{a},t)=(q^1,q^2,q^3)$ denoting  the position of a fluid element at time $t$ labeled by $\mathbf{a}=(a^1,a^2,a^3)$ and $\boldsymbol{\pi}(\mathbf{a},t)$ being its momentum density.  In \eqref{H_Lagr} the quantities $s_0$, $\rho_0$, and $\mathbf{B}_0$ are fluid element attributes that only depend on the label $\mathbf{a}$, and $\calj:=\det (\p q^i/\p a^j)$.  Also, $A^i_j\,  \p q^j/\p a^k=\calj \de^i_k$, where $A^i_j$ denotes elements of the cofactor matrix of $\p q/\p a$.    In a general coordinate system $\pi^i=  g^{ij}(\bfq) \, {\pi}_i$  where $g^{ij}$ is the metric tensor.   This Hamiltonian together with the canonical Poisson bracket
\begin{equation}
\left\{ F,G\right\}  =\inta  \left(  \frac{\delta F}{\delta q^i}%
\frac{\delta G}{\delta \pi_i}-\frac{\delta G}{\delta q^i}%
\frac{\delta F}{\delta \pi_i}\right)\,, 
\label{cbkt}%
\end{equation}
renders the equations of motion in the form
\begin{equation}
\dot{\pi_{i}}=\left\{  \pi_{i},H\right\} =-\frac{\de H}{\de q^i} \quad {\rm and} \quad
\dot{q}^i=\left\{q^i,H\right\} =\frac{\de H}{\de \pi_i} \,, 
\label{PBeqs}%
\end{equation}
where `\,$\cdot$\,' denotes time differentiation at constant label $\bfa$ and ${\de H}/{\de q^i}$ is the usual functional derivative.  The results of these calculations can be found in Appendix \ref{app:LanEOM}  and further details can be found in  Refs.~\onlinecite{morrison98,amp2a}.

In Ref.~\onlinecite{amp2a} we introduced the general time-dependent relabeling transformation $
\bfa=\mathfrak{A}(\bfb,t)$,  with  the inverse $\quad \bfb=\mathfrak{B}(\bfa,t)$, which gave 
 rise to the new dynamical variables
\bq
\bfPi(\bfb,t)={\mfJ}\,  \bfpi(\bfa,t)\,,\qquad\bfQ(\bfb,t)=\bfq(\bfa,t)\,,
\label{genlab}
\eq
and  the new Hamiltonian  
 \bqy
\tilde H[\bfQ,\bfPi]
&=& H - \intb\,  \bfPi\cdot (\bfV \cdot \nabla_b \bfQ)\,,
\label{add}\\
&=&\intb \Big[
\frac{ \Pi_i \, \Pi^i}{2\tilde\rho_0}
-  \Pi_i  V^j \frac{\p Q^i}{\p b^j} \nonumber\\
&+& \tilde\rho_0\,U\left(\tilde s_0,\tilde\rho_0/\tilde\calj\right)
 +
 \frac{\partial Q_{i}}{\partial b^{k}}  \frac{\partial Q^{i}}{\partial b^{\ell}}
\frac{\tilde{B}_{0}^k \tilde{B}_{0}^{\ell}}{8\pi \tilde{\calj}}
\Big]\,,
\nonumber
\\
&=& K + H_f + W\,,
\label{tildeH_MHD}
\eqy
where $K$ is the kinetic energy, $H_f$ is the fictitious term  due to the relabeling, and $W$ represents the sum of the internal and magnetic field energies.  In the first equality  of \eqref{tildeH_MHD},  
 \bq
\bfV(\bfb,t):= \dot\mfB\circ\mfB^{-1}=\dot\mfB(\mfA(\bfb,t),t)\,,
\label{VR}
\eq
which is  the label velocity,  $\nabla_b:=\p /\p \bfb$, and   $H$ is to be written in terms of the new variables.  In the second equality we used  $d^3a=\mfJ\,   d^3b $, with $\mfJ:= \det(\p a^i/\p b^j)$,    $\tilde \rho_0 =\mfJ\,  \rho_0$,  $\tilde\calj:=\det(\p Q^i/\p b^j) = \calj \mfJ$,   and $\tilde\rho_0/\tilde\calj= \rho_0/ \calj$, which follows from  mass conservation  $\rho_0d^3a= \tilde \rho_0 d^3b$.  The relabeled  entropy is $\tilde{s}_0(\bfb,t)=s_0(\mathfrak{A}(\bfb,t))$. 

From \eqref{PBeqs} it is clear that extremization of Hamiltonians give equilibrium equations.  For the Hamiltonian $H[\bfq,\bfpi]$ of \eqref{H_Lagr} this gives static equilibria, while for $\tilde{H}[\bfQ,\bfPi]$  of \eqref{tildeH_MHD} one obtains  stationary equilibria.   This was the point of introducing the relabeling:  it allows us to express stationary equilibria in terms of Lagrangian variables, which would ordinarily be time dependent, as  time-independent orbits with the moving labels.   

The equilibrium equations are 
\bqy
0=\p_t\bfQ_e&=&\frac{\bfPi_e}{\tilde \rho_0} -\bfV_e \cdot\nabla_b \bfQ_e\,,
\nonumber\\
0=\p_t\bfPi_e&=&- \nabla_b\cdot\left( \bfV_e \otimes \bfPi_e\right) +  \, \bfF_e\,,
\label{reftraj}
\eqy
where $\bfF_e$ comes from the $W$ part of the Hamiltonian. From (\ref{reftraj}) the equilibrium equation follows, 
\bq
\nabla_b\cdot(\tilde{\rho}_0\,  \bfV_e \bfV_e\cdot\nabla_b \bfQ_e)=   \bfF_e\,.
\label{equil}
\eq
Using $\bfb=\bfQ_e(\bfb)= \bfq_e(\mfA_e(\bfb,t),t)=\mfB_e(\bfa,t)$  and   the definition of $\bfV$ of \eqref{VR}, 
$\bfV(\bfb,t)=\dot\mfB_e(\mfA_e(\bfb,t),t)=\bfv_e(\bfb)$,
where $\bfv_e(\bfb)$  denotes an Eulerian equilibrium state, we obtain upon setting $\bfb=\bfx$  the usual stationary equilibrium equation,
\bq
\nabla\cdot(\rho_e \bfv_e \bfv_e)=  \bfF_e\,,
\eq
where $\rho_e(\bfx)$ is  the usual  equilibrium density.  It can be shown that  $\bfv_e\cdot \na s_e=0$, $ \na\cdot (\rho_e \bfv_e)=0$, and $\bfv_e\cdot \nabla \bfB_e - \bfB_e\cdot \nabla \bfv_e  +  \bfB_e \nabla\cdot \bfv_e=0$, follow from the Lagrange to Euler map.  Further details of this relabeling transformation are given in Ref.~\onlinecite{amp2a}, while application to our rotating pinch example of Sec.~\ref{sec:pinch} is worked out in Appendix \ref{app:atrt}.  

For stability,   we expand as follows:
\bq
\bfQ= \bfQ_e(\bfb,t) + \bfet(\bfb,t)\,,
\quad
\bfPi= \bfPi_e(\bfb,t) + \bfpi_{\eta}(\bfb,t)\,,
\label{linLag}
\eq
and calculate the second variation of the
Hamiltonian  in terms of the relabeled canonically conjugate variables $(\boldsymbol{\eta},\bfpi_{\eta})$  giving 
\bqy
\delta^{2}H_{\mathrm{la}}\left[Z_e;\boldsymbol{\eta},\boldsymbol{\pi}_{\eta}\right]  &=&
\frac{1}{2}\intx \bigg[
  \frac{1}{\rho_e}\big|  \boldsymbol{\pi}_{\eta}%
-\rho_e\mathbf{v}_e\cdot\mathbf{\nabla}\boldsymbol{\eta}\big|^{2}%
\nonumber\\
&{\ }& \hspace{.75 in} + \boldsymbol{\eta}\cdot\mathfrak{V}_e\cdot\boldsymbol{\eta}
\bigg]\,, 
\label{d2H_Lagr}%
\eqy
which depends on  the time independent equilibrium quantities $Z_e=(\rho_e, s_e,\mathbf{v}_e,\bfB_e)$,  i.e.,  the operator $\mathfrak{V}_e$\ has  no explicit time dependence.  (Again, see in Refs.~\onlinecite{amp2a,morrison98} for details.)   The functional
\bqy
\delta^{2}W_{\mathrm{la}}\left[ Z_e; \boldsymbol{\eta}\right]  &:=&\frac{1}{2}%
\intx   \boldsymbol{\eta}\cdot\mathfrak{V}_e\cdot
\boldsymbol{\eta}
\nonumber\\
&=&\frac{1}{2}\intx \Big[  \rho_e\left(
\mathbf{v}_e\cdot\nabla\mathbf{v}_e\right)  \cdot\left(  \boldsymbol{\eta}\cdot \nabla
\boldsymbol{\eta}\right)
\nonumber\\
&{\ }&
\hspace{.2 in} -\rho_e\left|  \mathbf{v}_e%
\cdot\nabla\boldsymbol{\eta}\right|^{2}\Big]
+\delta^{2}W\left[
\boldsymbol{\eta}\right]\,,
\label{FLdW}
\eqy
 is identical to that obtained  by Frieman and Rotenberg\cite{Frieman1960}, although obtained here in an alternative and more general manner.

The energy $\delta^{2}W_{\mathrm{la}}$ can be transformed in the more familiar expression of 
Ref.~\onlinecite{Bernstein1958b}, %
\bqy
\delta^{2}W_{\mathrm{la}}\left[ Z_e; \boldsymbol{\eta}\right]  &=&\frac{1}{2}\!\intx\!
 \bigg[
\rho_{e}\frac{\partial p_{e}}{\partial\rho_{e}}\left(  \mathbf{\nabla}%
\cdot\boldsymbol{\eta}\right)^{2}
+ \left(  \mathbf{\nabla}\cdot
\boldsymbol{\eta }\right)  \left(  \mathbf{\nabla}p_{e}\cdot
\boldsymbol{\eta}\right)
\nonumber\\
&& \hspace{-1.2 cm} + \frac{\left\vert \de \bfB \right\vert^{2} }{4\pi}
+ \mathbf{J}_{e}\times\boldsymbol{\eta}\cdot\de \bfB
- \nabla\cdot(\rho_e\boldsymbol{\eta}) (\boldsymbol{\eta}\cdot\nabla\Phi_e) 
\bigg] \,,
\label{eq:dW_Ber}%
\eqy
where  $4\pi \mathbf{J}_{e}= \mathbf{\nabla}\times\mathbf{B}_{e}$ is the equilibrium
current and $\de \bfB :=\mathbf{\nabla}\times\left(\boldsymbol{\eta}\times \mathbf{B}_{e}\right)$.

For completeness  we record the first order Eulerian perturbations that are induced by the 
Lagrangian variation  written in terms of the  displacement  $\boldsymbol{\eta}$: 
\begin{align}
\delta\rho_{\mathrm{la}} &  =- {\nabla}\cdot\left( \rho_{e}\boldsymbol{\eta}\right)
\label{eq:laDrho} \\
\delta\mathbf{v}_{\mathrm{la}} &  =\boldsymbol{\pi}_{\eta}/\rho_e - 
\boldsymbol{\eta}\cdot\mathbf{\nabla v}_e
\nonumber \\
&= {\partial\boldsymbol{\eta}}/{\partial t}+\mathbf{v}_e%
\cdot \nabla\boldsymbol{\eta}-\boldsymbol{\eta}\cdot\nabla \mathbf{v}_e
\label{eq:laDM}
\\%
\delta {s}_{\mathrm{la}}&=-  \boldsymbol{\eta}\cdot{\nabla}s_e 
\label{eq:laDs}%
\\
\delta\mathbf{B}_{\mathrm{la}}&= - {\nabla}\times
\left(  \mathbf{B}_{e}   \times   \boldsymbol{\eta}
\right)
\label{eq:laDB}%
\end{align}
where $\de s_{\mathrm{la}}$ can be replaced by the  pressure perturbation, 
$\delta p_{\mathrm{la}} =-\gamma p_e {\nabla}\cdot\boldsymbol{\eta} 
-\boldsymbol{\eta}\cdot\mathbf{\nabla}p_e$, that is often used.


\subsection{Eulerian formulae}
\label{ssec:eulB}

The Hamiltonian  for MHD in Eulerian variables is  
\bqy
H[Z]&=&\intx
\bigg[
\frac{\rho }{2}|\bfv|^2
+ \rho \, U(s,\rho) \nonumber\\
&&\hspace{1 cm}  + \frac{ |\bfB|^2}{8\pi} +\rho \Phi
\bigg]\,. 
\label{eq:H_Euler}%
\eqy
where $Z=(\rho,s,\bfv,\bfB)$.  When \eqref{eq:H_Euler} is substituted into the noncanoncal Poisson bracket $\{F, 
G\}_{\mathrm{nc}}$ of  Ref.~\onlinecite{morrison80} one obtains the Eulerian equations of motion in the form 
$\p Z/\p t=\{Z, H\}_{\mathrm{nc}}$.    Because the noncanonical Poisson bracket  $\{F, G\}_{\mathrm{nc}}$ is degenerate, i.e.\ three exist a functional $C$ such that $\{F, C\}_{\mathrm{nc}}=0$ for all functionals $F$,  Casimir invariants $C$ exist and equilibria are given by extremization of the energy-Casimir functional $\mathfrak{F}= H + C$.  For MHD with no symmetry the Casimirs are
\bq
C_s= \intx \rho\,  {\cal S} (s)\,,
\label{entCas}
\eq
and the magnetic and cross helicities, 
\bq
C_B=\intx \bfA\cdot\bfB,\quad\mathrm{and}\quad  C_v=\intx \bfv\cdot\bfB\,, 
\label{hels}
\eq
respectively.  By manipulation of the MHD equations, the  helicities were shown by Woltjer\cite{W1,W2,W3,W4} to be invariants ($C_v$ requiring the barotropic equation of state) and used by him to predict plasma states.  Woltjer's ideas pertaining to   magnetic helicity  were adapted by Taylor\cite{T1,T2}  to describe reversed field configurations.  The invariant of \eqref{entCas} and Woltjer's helicities were shown to be Casimir invariants in Ref.~\onlinecite{morrison82}.  (See Refs.~\onlinecite{padhye96,padhye96fluid} for further discussion.)

 An important point to note is that knowledge of the Casimirs determines this additional physics, but this knowledge must come from physics outside of the ideal model.

Special attention has been given to the equilibrium states obtained by extremizing the energy subject to the Woltjer invariants, perhaps because these are the states for which Casimirs are at hand.  
(See Refs.~\onlinecite{YM14a,YM14b} for discussion of the Casimir deficit problem.)  However, 
 we will see in Sec.~\ref{ssec:Daform} that {\it all} MHD equilibria are obtainable  from the variational principle with directly constrained variations,   the  dynamically accessible variations, rather than using Lagrange multipliers and helicities etc.  

In the case were translational symmetry is assumed,  all variables are assumed to be independent of a coordinate $z$  with 
\bqy
\mathbf{B}  &=&B_{z}\mathbf{\hat{z}}+\mathbf{\nabla}\psi\times\mathbf{\hat
{z}}
\label{Bpsi},\\
\mathbf{M}  &=&M_{z}\mathbf{\hat{z}}+\mathbf{\nabla}\chi\times\mathbf{\hat
{z}}+\mathbf{\nabla}\Upsilon,
\label{Mchi}
\eqy
where   $\, \chi,\Up$ and $\psi$ are ``potentials'', $\mathbf{M}=\rho \mathbf{v}$,  $M_z=\rho v_z$ and $\mathbf{\hat{z}}$ is the unit vector in the symmetry direction.  The Hamiltonian  then becomes
\begin{eqnarray}
 H_{TS}[Z_s]   &  =&\intx \bigg[  \frac{M_{z}^{2}}{2\rho}
+\frac{\left\vert\nabla\chi\right\vert ^{2}}{2\rho}
+\frac{\left\vert \nabla\Upsilon\right\vert^{2}}{2\rho}
\nonumber\\
&+& \frac{\left[  \Upsilon,\chi\right] }{\rho}
+\frac{\left\vert\mathbf{\nabla}\psi\right\vert ^{2}}{8\pi}
+\frac{B_{z}^{2}}{8\pi}+\rho U +\rho\Phi \bigg]  \,, 
\label{Hts}
\end{eqnarray}
where $Z_s=(\rho,s,M_z,\chi,\Up,\psi,B_z)$.  With this symmetry assumption, the set of Casimir is expanded and is sufficient to obtain a variational principle for the equilibria considered here.   However,  because of this symmetry assumption it is only possible to obtain stability results restricted to perturbations consistent with  this assumption.

In Refs.~\onlinecite{amp0,amp1} the translationally  symmetric noncanonical Poisson brackets were
obtained for both  neutral fluid and MHD dynamics. For the case of a neutral fluid, which we consider in Sec.~\ref{ssec:Econv} for convection, the Poisson bracket for translationally symmetric  flows was given in Ref.~\onlinecite{amp0}.  This bracket with the Hamiltonian of \eqref{Hts}, where the magnetic energy terms involving $B_z$ and $\psi$ are removed, gives the compressible Euler's equations for fluid motion.   The translationally symmetric fluid Poisson bracket has the following Casimir invariants:
\begin{eqnarray}
C_{1}&=&\intx\, \rho\,  {\cal S} \left(s,v_{z}, \left[s,v_{z} \right]/\rho,\ldots\right),\label{eq:C1}
\\
C_{2}&=&\intx\,  \big( \mathbf{\nabla}{\cal A}(s)\cdot\mathbf{\nabla}\chi+ \left[\Upsilon,{\cal A}(s)\right] \big)/\rho
\nonumber\\
&=&\intx \,{\cal A}(s)\, \mathbf{\hat{z}\cdot\nabla}\times\mathbf{v} \,,
\label{eq:C2}
\end{eqnarray}
where $[f,g]=\mathbf{\hat{z}}\cdot \nabla  f\times \nabla g$.  The second Casimir applies if $v_z$ depends only on $s$, which will suit our purpose, i.e., the energy-Casimir variational principle $\de \mathfrak{F}=0$  will give our  desired equilibria.

For  the case of MHD it was shown  in Refs.~\onlinecite{amp0,amp1} that the following are the  Casimir invariants with translational symmetry:
\begin{eqnarray}
C_{s}&=&\intx \rho\, \mathcal{J}\Big( s,\psi, \left[s,\psi\right]/\rho  , \Big[   
\left[s,\psi\right] /\rho ,\psi\Big]/\rho , 
\nonumber\\
&&\hspace{2 cm}  \Big[ s, \left[ s,\psi\right]/\rho  \Big]/\rho  ,...\Big)  \,,
\label{eq:casimir1_z}
\\
C_{B_z}  &  =&\intx B_{z}\mathcal{H}\left(  \psi\right)\,,
\label{eq:casimir2_z}\\
C_{v_z}  &  =&\intx \rho v_{z}\mathcal{G}\left(  \psi\right)\,,
 \label{eq:casimir3_z}%
\end{eqnarray}
and, if the  entropy is assumed to be a flux function, i.e., $\left[\psi,s\right]  =0$,  then \eqref{eq:casimir1_z} collapses to 
\bq
C_{s}=\intx \rho\, \mathcal{J}(\psi)\,,
\eq
and there is the additional cross helicity Casimir, 
\bqy
C_{v}&=&\intx\left(
{v_{z}B_{z}}\,  {\cal F}^{^{\prime}}(\psi)+\frac{1}{\rho}\mathbf{\nabla
}{\cal F}(\psi)\cdot\mathbf{\nabla}\chi+\frac{\left[  \Upsilon,{\cal F}(\psi)\right]}{\rho}
\right) 
\nonumber\\
&=&\intx   \mathbf{v\cdot B} \, {\cal F}^{^{\prime}}(\psi)\,.
\label{eq:casimir4_z}%
\eqy
where  ${\cal S},\, {\cal A},\, {\cal J},\, {\cal H},\, {\cal G}$,  and ${\cal F} $ are arbitrary functions of their arguments with prime denoting  differentiation with respect to argument.

For both the neutral fluid and MHD  equilibria 
that satisfy $\delta\mathfrak{F}=0$ a sufficient condition for stability follows if the second variation   $\delta^{2}\mathfrak{F}$ can be shown to be positive definite.  For MHD it was shown in  Refs.~\onlinecite{amp2a,ampE}  that $\delta^{2}\mathfrak{F}$ could be put into the following diagonal form: 
\bqy
\delta^{2}\mathfrak{F}[Z_e;\de Z_s]&=&\intx
\Big[
a_{1}\left|\delta\mathbf{S}\right|^{2} 
+a_{2}\left(\delta Q\right)^{2} + a_3 (\de R_z)^2
\nonumber\\
&{\ }&\hspace{1 cm} +a_{4}\left|\delta\mathbf{R}_{\perp}\right|^{2}  
+ a_{5}\left(\delta\psi\right)^{2}
\Big]\,, 
\label{d2F_diag}%
\eqy
where the variations $(\de \mathbf{S},\de \mathbf{R}, \de Q, \de \psi)$ are linear combinations of $(\de \mathbf{v},  \de \mathbf{B},\de \rho,\de \psi)$.  The coefficients $a_i$ for $i=1-5$ depend on space through the equilibrium and were given first explicitly in Ref.~\onlinecite{amp2a} (and corrected in Ref.~\onlinecite{ampE}).  Note, for these calculations the external potential $\Phi$ was omitted.

Upon extremizing over all variables except $\de\psi$ and then back substituting the resulting algebraic relations,   \eqref{d2F_diag} becomes 
\bqy
\delta^{2}\mathfrak{F}[Z_e;\de\psi]&=&\intx \bigg[b_{1}\left\vert \mathbf{\nabla}\delta\psi\right\vert ^{2}+b_{2}\left(\delta\psi\right)^{2}
\nonumber\\
&&\hspace{1 cm}
+b_{3}\left|  { \mathbf e}_\psi \times \mathbf{\nabla}\delta\psi \right|^{2}\Big]\,,
\label{eq:d2Fgen}
\eqy
where $ {\mathbf e}_\psi = {\mathbf{\nabla}\psi}/{\left|\mathbf{\nabla}\psi\right|}$ and 
\bqy
b_{1} &=&\frac{1-\calm^{2}}{4\pi}\frac{c_{s}^{2}-\calm^{2}\left(c_{s}^{2}+c_{a}^{2}\right)}{c_{s}^{2}-\calm^{2}\left(c_{s}^{2}+c_{a}^{2}\right) 
+ \frac{\calm^{4}}{4\pi\rho}|\nabla \psi|^{2}}\,,
 \label{eq:b1}\\
b_{2}&=&\nabla\cdot\left[\frac{\partial}{\partial\psi}\left(\frac{\calm^{2}}{4\pi}\right)\mathbf{\nabla}\psi\right]
\nonumber\\
&& \hspace{1.3 cm} -\frac{\partial^2}{\partial\psi^{2}}\left(p+\frac{B_{z}^{2}}{8\pi}+\frac{\calm^{2}}{4\pi}|\nabla \psi|^{2}\right)\,, 
\label{eq:b2}\\
b_{3} & =&
\, \frac{1-\calm^{2}}{4\pi}-b_{1}\,.
\label{eq:b3}
\eqy
where the Alfv\'en-Mach number   $\calm^{2}:=4\pi\mathcal{F}^{2}/\rho<1$ has been assumed.  Here 
\bq
c_{a}^{2}=B^{2}/\left(4\pi\rho\right) \qquad \mathrm{and}\qquad c_{s}^{2}=\partial p/\partial\rho
\label{velos}
\eq
are  the Alfven  and  the sound speed, respectively.    

Thus, stability in this MHD context rests on whether or not \eqref{eq:d2Fgen} is  definite, and for the neutral fluid equilibria we treat here, which include a gravity force, the same is true for  the corresponding functional.


\subsection{Dynamically accessible formulae}
\label{ssec:Daform}

Extremizing the  Hamiltonian of  \eqref{eq:H_Euler} without constraints gives trivial equilibria. With energy-Casimir the constraints are  incorporated  essentially by using Lagrange multipliers.   Dynamically accessible variations, as introduced in Ref.~\onlinecite{MP89}, restrict the variations to be those generated by the noncanonical Poisson bracket and in this way assures that all kinematical constraints are satisfied.  The  first order dynamically accessible variations,  obtained directly from the noncanonical Poisson bracket of Ref.~\onlinecite{morrison80},   are the following:
\begin{align}
\delta\rho_{\mathrm{da}}  &  =   \mathbf{\nabla}\cdot\left( \rho\mathbf{g}_{1}\right)\, ,
\label{eq:var_0_1}\\
\delta\mathbf{v}_{\mathrm{da}}&=\mathbf{\nabla}   g_{3}  + s \nabla g_{2} 
+  \left( \nabla
\times\mathbf{v}\right)\times \mathbf{g}_{1}
\nonumber\\
& \hspace{1.1 cm} 
+  \mathbf{B}\times\left(   {\nabla}\times\mathbf{g}_{4}\right)/\rho
\label{eq:var_0_2}\\
\delta s_{\mathrm{da}}  &  =  \mathbf{g}_{1}  \cdot{\nabla}  s \, ,
\label{eq:var_0_3}\\
\delta\mathbf{B}_{\mathrm{da}}  &  =  \mathbf{\nabla}\times\left(
\mathbf{B}\times\mathbf{g}_{1}\right)  \,, 
\label{eq:var_0_4}%
\end{align}
where the freedom of the variations is embodied in the arbitrariness of $\mathbf{g}_1$, $g_2$, $g_3$,  and $ \mathbf{g}_4$.   Using these in the variation of the Eulerian Hamiltonian gives  
\bqy
\de H_{\mathrm{da}}&=& \intx\Big[\left(v^2/2 +(\rho U)_\rho + \Phi\right)\de \rho_{\mathrm{da}} +\rho \bfv\cdot\delta\mathbf{v}_{\mathrm{da}}
\nonumber\\
&&\hspace{1.5 cm} + \rho U_s\, \delta s_{\mathrm{da}}  +  \bfB\cdot\delta\mathbf{B}_{\mathrm{da}} /4\pi \Big]\,, 
\nonumber\\
&=& \intx\Big[\mathbf{g}_{1}\cdot\big(\rho \bfv\times( \nabla\times\bfv) - \rho \nabla v^2/2 
\nonumber\\
&&  \hspace{.2  cm}  - \rho \nabla h + \rho T \nabla s + \bfj\times \bfB  \big)  -  g_2\nabla\cdot(\rho s\bfv)
\nonumber\\
&&  -  g_3\nabla\cdot(\rho \bfv) + \mathbf{g}_4\cdot \nabla\times (\bfv\times\bfB)
\Big]= 0\,,
\label{daH}
\eqy
whence it is seen that the vanishing of the terms multiplying the independent   quantities $\mathbf{g}_1$, $g_2$, $g_3$,  and $ \mathbf{g}_4$
gives precisely the Eulerian equilibrium equations  \eqref{Emom}--\eqref{Eent}.

Next, stability is assessed by expanding the Hamiltonian to second order using the dynamically accessible constraints to this order (see Refs.~\onlinecite{morrison98,amp2a} for details), yielding the following expression:
\bqy
\delta^{2}H_{\mathrm{da}}\left[ Z_e; \mathbf{g}
\right]
&=&\intx
\rho\big|
 \delta\mathbf{v}_{\mathrm{da}} -\mathbf{g}_{1}\cdot\mathbf{\nabla v} 
 +\mathbf{v}\cdot \nabla\mathbf{g}_{1}\big| ^{2}
\nonumber\\
&{\ }& \qquad + \delta^2 W_{\mathrm{la}}\left[ \mathbf{g}_{1}\right] \,.
\label{d2H_da}%
\eqy
If in \eqref{d2H_da} $ \delta\mathbf{v}_{\mathrm{da}}$ were independent and arbitrary we could use it to nullify the first term and then upon setting  $\mathbf{g}_1=-\boldsymbol{\eta}$, we would see that dynamically accessible stability is identical to Lagrangian stability.  However, as we will see in Sec.~\ref{ssec:compform},  this is not always possible.


\subsection{Comparison formulae}
\label{ssec:compform}

In our calculations of stability we obtained the quadratic energy expressions of \eqref{d2H_Lagr}, \eqref{d2F_diag},  and \eqref{d2H_da}, which can be  written in terms of various Eulerian perturbation variables 
\bq
\mathfrak{P}:=\{\delta \rho, \delta \mathbf{v}, \delta s, \delta\mathbf{B}\}\,.
\eq 
In the case of the Lagrangian energy of \eqref{d2H_Lagr}, the set of perturbations $\mathfrak{P}_{\rm la}$ as  given by Eqs.~\eqref{eq:laDrho}--\eqref{eq:laDB} are constrained, while for the energy-Casimir expression of \eqref{d2F_diag} the perturbations $\mathfrak{P}_{\rm ec}$ are entirely unconstrained provided they satisfy the translation symmetry we have assumed.  Similarly the perturbations  for the energy expression  \eqref{d2H_da}, $\mathfrak{P}_{\rm da}$ of \eqref{eq:var_0_1}--\eqref{eq:var_0_4},  are constrained.    In our previous work of Ref.~\onlinecite{amp2a} we established the inclusions 
\[
\mathfrak{P}_{\mathrm{da}}\subset\mathfrak{P}_{\mathrm{la}}
\subset\mathfrak{P}_{\mathrm{ec}}\,, 
\]
which led to the  conclusions   
 \[
\mathfrak{stab}_{ec}\Rightarrow\mathfrak{stab}_{\mathrm{la}}\Rightarrow\mathfrak{stab}_{\mathrm{da}}\,,
\]
viz.,  dynamically accessible stability is the most limited  because its perturbations are the most constrained,  while energy-Casimir stability is the most general, when it exists, for its perturbations are not constrained at all.  We 
wish to explore further  the differences between these kinds of stability by exploring, in particular, the differences between Lagrangian and dynamically accessible perturbations.

From \eqref{d2H_da} it is clear that if $\de\bfv_{\mathrm{da}}$ is arbitrary, independently of  $\mathbf{g}_1$, then the first term of this expression can be made to vanish.  This would reduce $\delta^{2}H_{\mathrm{da}}$ to   the energy expression obtained for Lagrangian stability, making the two kinds of stability equivalent.   Given that  there are  five components of $g_2, g_3$ and $\mathbf{g}_4$,  in addition to $\mathbf{g}_1$, one might think that this is always possible.   However, as pointed out in  Ref.~\onlinecite{amp2a} this is not always possible and whether or not it is  depends on the state or equilibrium under consideration.  We continue this discussion here. 

Consider first a static equilibrium state that has entropy as a flux function and no equilibrium flow.  Thus,  for this case, the cross helicity $C_v$ of \eqref{hels} vanishes.  For a dynamically accessible  perturbation  
\bqy
\de C_v&=&\intx \de\bfv_{\mathrm{da}}\cdot \bfB_e=\intx ( \mathbf{\nabla}   g_{3}  + s_e \nabla g_{2})\cdot \bfB_e
\nonumber\\
&=&-  \intx    g_{2}\, \bfB_e \cdot  \nabla s_e=0\,,
\eqy
 where the last equality assumes $g_{3}$ is single-valued and the vanishing of surface terms, as well as $s_e$ being a flux function. The fact that $\de C_v=0$ for this case is not a  surprise since it is a Casimir, but we do see clearly that if $s$ were not a flux function, then a perturbation $\de\bfv_{\mathrm{da}}$ could indeed create cross helicity.   Because of the term $ {\partial\boldsymbol{\eta}}/{\partial t}$  of \eqref{eq:laDM},  which can be chosen arbitrarily, it is clear that $\de\bfv_{\mathrm{la}}$ can create cross helicity for any equilibrium state, supplying clear evidence  that  $\de\bfv_{\mathrm{da}}$ is not completely general.  
 
 Although $\de\bfv_{\mathrm{da}}$  is not completely general, it was noted in Ref.~\onlinecite{morrison98} that for static equilibria the  first term of \eqref{d2H_da} becomes
 \bq
\intx
\rho\big|
 \delta\mathbf{v}_{\mathrm{da}}\big| ^{2}
 \label{static}
\eq
and this can be made to vanish independent of $\mathbf{g}_1$ by choosing $g_2=g_3=0$ and $\mathbf{g}_4=0$.   Thus, for static equilibria the Lagrangian and dynamically accessible approaches must give the same necessary and sufficient conditions for stability, i.e. 
\[
 \mathfrak{stab}_{\mathrm{la}}\Leftrightarrow\mathfrak{stab}_{\mathrm{da}}\,,
\]

As another example consider the variation of the  circulation integral  $\Ga=\oint_c \bfv\cdot\mathbf{dx}$ on  a fixed closed contour $c$  for an equilibrium with $\bfv_e\equiv 0$ and $\bfB_e\neq 0$. Clearly $\de\bfv_{\mathrm{la}}$ can generate any amount of circulation.  However, for a  dynamically accessible variation 
\bqy
\de \Ga&=&\oint_c \de\bfv_{\mathrm{da}} \cdot\mathbf{dx}
\\
&=& \oint_c s_e \nabla g_{2}\cdot\mathbf{dx}
+  \oint_c  (\nabla \times \mathbf{g}_{4})\cdot(\mathbf{dx} \times \bfB_e)/\rho_e
\nonumber
\eqy
and we can draw two conclusions:   In the case where $c$ is a closed magnetic field line 
$\mathbf{dx} \parallel \bfB$ and  $\de \Ga$ becomes
\bqy
\de\Ga_B&=&\oint_c s_e \nabla g_{2}\cdot\mathbf{dx}
= \oint_c ( \nabla (s_e g_{2}) - g_2\nabla s_e)\cdot\mathbf{dx}
\nonumber
\\
&=&-   \oint_c g_2\nabla s_e\cdot\mathbf{dx}\,,
\eqy
whence we see clearly that if $\nabla s_e$ is everywhere parallel to $\bfB_e$, then $\de\Ga_B=0$ and otherwise this is not  generally true.   Alternatively, suppose the contour $c$ lies within a level set of $s_e$, for which it need not be true that $\bfB_e\cdot\nabla s_e=0$ along $c$.   For this case
\bq
\de\Ga_s =\oint_c  (\nabla \times \mathbf{g}_{4})\cdot(\mathbf{dx} \times \bfB_e)/\rho_e
\eq
which in general does not vanish. If a magnetic field line were to lie within a surface of constant $s_e$, then in the general case, $\bfB_e\cdot \nabla s_e=0$ otherwise surfaces of constant $s_e$ would be highly irregular, i.e., if $\bfB_e\cdot \nabla s_e\neq 0$, then $\bfB_e$ cannot lie within a level set of $s_e$. 

We point out that similar arguments can be supplied for cases where $\bfv_e\neq 0$, e.g., variation of the fluid helicity $\de C_{\om}=2\intx \boldsymbol{\omega}\cdot \de\bfv_{\mathrm{da}}$ 
for an equilibrium with $\bfB_e\equiv 0$ becomes
\bqy
\de C_{\om}&=&2\intx \boldsymbol{\omega}_e\cdot  \left( s_e \nabla g_{2}  +  \boldsymbol{\omega}_e \times \mathbf{g}_{1}\right)
\\
&=&2\intx \boldsymbol{\omega}_e\cdot   \nabla g_{2} \, s_e
= -2\intx g_{2} \,  \boldsymbol{\omega}_e\cdot   \nabla s_e
\nonumber
\eqy
which vanishes if $\boldsymbol{\omega}_e$ is perpendicular to  $\nabla s_e$ or if the entropy is everywhere constant.

In summary, the  general conclusion is that $\de\bfv_{\mathrm{da}}$, unlike $\de\bfv_{\mathrm{la}}$, is not completely arbitrary and the degree of arbitrariness depends on the equilibrium.   We also point out that although we are here interested in perturbations  away from equilibrium states, for the purpose of assessing stability, the conditions we have described  apply to perturbations away from any state, equilibrium or not. 

Now we turn to our examples.  For the remainder of this paper we drop the subscript `$e$' on equilibrium quantities, so as to avoid clutter.


\section{Convection}
\label{sec:convection}

For this first example we consider thermal convection in static equilibria,  both with and without a magnetic field.  This example has been well studied by various approaches, e.g., heuristic arguments that mix Lagrangian and Eulerian ideas were given  in Ref.~\onlinecite{landau1975} for the neutral fluid.  Here our analysis will be done separately  in purely Lagrangian and purely Eulerian terms, and it will  illustrate   the role played by entropy in determining stability.

 We suppose the equilibrium has stratification in the $\hat{\mathbf{y}}$-direction  due to gravity,  i.e.\ $\Phi= g y$, with $\rho$ and $s$ dependent only on $y$.   Thus the only equation to be solved  for the neutral fluid is 
\bq
 \frac{dp}{dy} =   -  \rho  \frac{d\Phi}{dy}  = -  \rho g  \,. 
\label{Estat0}
\eq
 If a magnetic field of the form $\bfB=B(y)\hat{\mathbf{x}}$ is supposed, then the equilbirum  equation is the following:
\bq
 \frac{dp}{dy} =  -  \frac{dB}{dy}\frac{B}{4\pi}  -  \rho  \frac{d\Phi}{dy}  =JB -  \rho g  \,. 
\label{Estat2}
\eq 
 
For barotropic fluids, $s$ is constant everywhere and is eliminated from the theory, i.e., $U(\rho)$ alone.  Thus, \eqref{Estat0} (together with $U(\rho)$) determines completely the thermodynamics at all points $y$ by integrating
\bq
\frac{p_{\rho}}{\rho} \frac{d\rho }{dy} =-g 
\eq
giving $\rho(y)$ and consequently $p(y)$.  For this special case,  no further information is required.    However, in the general case where $p(\rho,s)$, \eqref{Estat0} is not sufficient and one needs to know more about the fluid, since now we have 
\bq
\frac{p_{\rho}}{\rho} \frac{d\rho }{dy}   + \frac{p_{s}}{\rho} \frac{d s}{dy}=-  {g} \,, 
\eq
which is insufficient because we have only one equation for the two unknown quantities $\rho$ and $s$.    Thus,   knowledge of additional physics is required, which could come from boundary or initial conditions, solution of some heat or transport equation with constitutive   relations, etc.  
  
Next consider the case of MHD where 
\bq
\frac{dp}{dy} =p_{\rho} \frac{d \rho }{dy} + p_s  \frac{d s}{dy}= JB - \rho g \,.
\label{conequil}
\eq
If gravity is absent MHD differs from that of the stratified fluid because only the pressure enters and the thermodynamics of $\rho$ and $s$ do not explicitly enter the equilibrium equation.  We will consider the case where gravity is present. 

Thus, in general, equilibria depend on two kinds of conditions:  force balance, as given in our cases of interest by \eqref{Estat0} or \eqref{Estat2} and thermodynamics.   For latter convenience we record here several thermodynamic relations: 
\bqy
p&=&\rho^2 U_{\rho}\quad\mathrm{and}\quad  T=U_s
\label{EOS}\\
c_s^2&=& \left.\frac{\p p}{\p \rho}\right|_s=(\rho^2U_{\rho})_{\rho}=\rho(\rho U)_{\rho\rho}
\label{sound}\\
\left.\frac{\p p}{\p s}\right|_{\rho}&=&- \left.\frac{\p \rho}{\p s}\right|_p c_s^2=\rho^2U_{\rho s}
\label{dpds}
\eqy
where, without confusion, we use  subscripts on $U$ to denote partial differentiation with the other thermodynamic variable held constant and the subscript of $c_s$ denotes `sound'.   

The  coefficient of thermal expansion,  $\alpha$,  is given by 
\bq
\alpha =  -\frac1{\rho}\left.\frac{\p \rho}{\p T}\right|_p
\eq
and for typical fluids 
\bqy
\left.\frac{\p p}{\p s}\right|_{\rho}&=&\frac{\al}{\rho}>0 \quad\mathrm{and} \quad
\left.\frac{\p \rho}{\p s}\right|_p<0\,.
\label{thermoIneq}
\eqy
If the pressure is given by  $p=c\rho^{\ga}\exp(\la s)$, then
$c_s^2= {\ga p}/{\rho}$, as it is often written.


\subsection{Lagrangian convection}
\label{ssub:Lconvection}

\subsubsection{Lagrangian convection equilibria}
\label{sssec:LconvectionEquilbria}

From \eqref{PBeqs} Lagrangian equilibria must satisfy
\begin{equation}
\dot{\pi_{i}}=-\frac{\de H}{\de q^i}=0 \quad {\rm and} \quad
\dot{q}^i  =\frac{\de H}{\de \pi_i}=0 \,, 
\end{equation}
whence if follows from \eqref{H_Lagr} that $\pi_i\equiv 0$ and 
\bqy
0=\dot\pi_i&=& 
-A_i^j\frac{\p}{\p a^j}\left(\frac{\rho_0^2}{\calj^2}U_{\rho} 
 + \frac1{2\calj^2}\frac{\p q^k}{\p a^l}   \frac{\p q_k}{\p a^m}B^l_0 B^m_0
\right)
\nonumber\\
&& + B_0^j\frac{\p}{\p a^j}\left(\frac1{\calj} \frac{\p q_i}{\p a^l}   B_0^l
\right)
-  \rho_0 \frac{\p \Phi}{\p q^i}\,,
\label{LagEq}
\eqy
which  is the Lagrangian variable form of  the static Eulerian equilibrium equation \eqref{Estat}.  (See e.g.\  Refs.~\onlinecite{newcomb1962, padhye96,morrison98} for further details.) 

Because we are investigating  equilibria that only depend on the variable $y$ and have magnetic fields of the form $\bfB=B(y)\hat{\mathbf{x}}$,   we only consider   the $\hat{\mathbf{y}}$-component of \eqref{LagEq}, which  is the  Lagrangian variable form of  the static Eulerian equilibrium equation  \eqref{Estat2}.

\subsubsection{Lagrangian convection stability}
\label{sssec:LconvectionStability}

The second variation of the energy about this equilibrium is the usual expression  given in Ref.~\onlinecite{Bernstein1958b}.  For static equilibria this is obtained by setting  $\boldsymbol{\eta}\equiv\boldsymbol{\xi}$ in  \eqref{eq:dW_Ber}, and we know
that the stability of such  configurations is determined by this second variation of the potential energy.   We will  manipulate the energy expressions to facilitate comparison with results obtained in Sec.~\ref{ssec:Econv}.  Cases with and without ${B}= 0$ are considered.

\bigskip

\noindent\underline{Case ${B}= 0$}:

\medskip

 By exploiting the equilibrium equation we obtain
\begin{eqnarray}
\delta^{2}W_{\mathrm{la}} & = &\frac{1}{2} \intx\, \bigg[ \frac{1}{\rho}\left.\frac{\partial p}{\partial\rho}\right|_{s}\!\Big[\left(\rho\mathbf{\nabla}\cdot\boldsymbol{\xi}\right)^{2}+2\left(\mathbf{\nabla}\cdot\boldsymbol{\xi}\right)\left(\mathbf{\nabla}\rho\cdot\boldsymbol{\xi}\right) \nonumber  \\
 &  &  \hspace{-.8cm} +\left(\mathbf{\nabla}\rho\cdot\boldsymbol{\xi}\right)^{2}\Big]+\frac{1}{\rho}\left.\frac{\partial p}{\partial s}\right|_{\rho}\! \!\left[\rho\mathbf{\nabla}\cdot\boldsymbol{\xi}+\mathbf{\nabla}\cdot\left(\rho\boldsymbol{\xi}\right)\right]\left(\mathbf{\nabla}s\cdot\boldsymbol{\xi}\right) \nonumber
  \bigg]\,.
 \nonumber
\end{eqnarray}
In conventional `$\de W$' stability analyses one would consider conditions for positivity of the above as a quadratic expression in terms of  $\boldsymbol{\xi}$.  However, for our present purposes we rewrite it in terms of 
\begin{equation}
\delta\rho_{\mathrm{la}}=-\mathbf{\nabla}\cdot\left(\rho\boldsymbol{\xi}\right)\qquad\delta s_{\mathrm{la}}=-\mathbf{\nabla}s\cdot\boldsymbol{\xi},\label{eq:d_la}
\end{equation}
which with 
\[
\frac{1}{\rho}\left.\frac{\partial p}{\partial s}\right|_{\rho}\left(\mathbf{\nabla}\rho\cdot\boldsymbol{\xi}\right)\left(\mathbf{\nabla}s\cdot\boldsymbol{\xi}\right)=\frac{1}{\rho}\left.\frac{\partial p}{\partial s}\right|_{\rho}\frac{\left(\mathbf{\nabla}\rho\cdot\boldsymbol{\xi}\right)}{\left(\mathbf{\nabla}s\cdot\boldsymbol{\xi}\right)}\left(\delta s_{\mathrm{la}}\right)^{2},
\]
yields 
\begin{eqnarray}
\delta^{2}W_{\mathrm{la}}&=&\frac{1}{2} \intx \bigg[\frac{1}{\rho}\left.\frac{\partial p}{\partial\rho}\right|_{s}\left(\delta\rho_{\mathrm{la}}\right)^{2}+\frac{2}{\rho}\left.\frac{\partial p}{\partial s}\right|_{\rho}\delta\rho_{\mathrm{la}}\delta s_{\mathrm{la}}  
\nonumber\\
&&\hspace{1.5cm}  -\frac{1}{\rho}\left.\frac{\partial p}{\partial s}\right|_{\rho}\frac{\left(\mathbf{\nabla}\rho\cdot\boldsymbol{\xi}\right)}{\left(\mathbf{\nabla}s\cdot\boldsymbol{\xi}\right)}\left(\delta s_{\mathrm{la}}\right)^{2}\bigg].\label{eq:d2W}
\end{eqnarray}
Now, using \eqref{dpds} we can rearrange this equation as   
\begin{eqnarray}
\delta^{2}W_{\mathrm{la}}&=&\frac{1}{2}\intx \frac{c_{s}^{2}}{\rho}\Bigg[\left(\delta\rho_{\mathrm{la}} - 
\left.\frac{\partial\rho}{\partial s}\right\vert _{p}\delta s_{\mathrm{la}}\right)^{2} 
\nonumber\\
&& + \left.\frac{\partial\rho}{\partial s}\right\vert _{p}\left(\frac{\left(\mathbf{\nabla}\rho\cdot\boldsymbol{\xi}\right)}{\left(\mathbf{\nabla}s\cdot\boldsymbol{\xi}\right)}-\left.\frac{\partial\rho}{\partial s}\right\vert _{p}\right)\left(\delta s_{\mathrm{la}}\right)^{2}\Bigg]\label{eq:d2Wbis}\,.
\end{eqnarray}

We will see that \eqref{eq:d2Wbis}  is of the same form as that  of (\ref{eq:d2F}) of  Sec.~\ref{ssec:Econv},  obtained via the energy-Casimir functional,  yet here the perturbations $\delta\rho$ and $\delta s$ are both constrained to depend on $\boldsymbol{\xi}$ according to \eqref{eq:d_la}.

Examination of \eqref{eq:d2Wbis} reveals that  positivity of the second term is sufficient for positivity of $\delta^{2}W_{\mathrm{la}}$, viz.
\begin{equation}
\frac{\nabla\rho\cdot\boldsymbol{\xi}}{\nabla s \cdot\boldsymbol{\xi}}   <  \left.\frac{\partial\rho}{\partial s}\right\vert _{p} 
\label{eq:cond_la}
\end{equation}
Given that the equilibrium only depends on the  variable $y$, in which the systems is stratified, \eqref{eq:cond_la} gives the following sufficient condition for stability
\bq
\frac{d\rho/dy}{d s/dy}  < \left.\frac{\partial\rho}{\partial s}\right\vert _{p}<0 \,.
\label{eq:cond_s}
\eq
If the equilibrium is stably stratified, i.e.,  $d\rho/dy<0$,  then ${ds }/{dy}$ must be positive and we would have a threshold involving the density and entropy scale lengths.

However let us proceed further.  Define
\bq
\De=\left.\frac{\partial\rho}{\partial s}\right\vert _{p} - \frac{d\rho/dy}{d s/dy} =\left.\frac{\partial\rho}{\partial s}\right\vert _{p} - \frac{d\rho}{ds}
\label{DE}
\eq
where in the second term of the second equality we have replaced the coordinate $y$ by $s$, which is possible if $ds/dy$ does not vanish.  Observe in the definition of $\De$ of  \eqref{DE} this second term depends on the equilibrium profiles, while the first term is of a thermodynamic nature. 
So far, the sufficient condition for stability $\De>0$ does not account for the fact that $d\rho/dy$ and $ds/dy$ are not independent but are related through the equilibrium equation \eqref{Estat0}.  To address this we first rewrite the expression for $\De$ using
\bqy
\frac{dp }{dy} &=& \left.\frac{\p p}{\p s}\right|_{\rho} \frac{d s}{dy}  +  \left.\frac{\p p}{\p \rho}\right|_{s}  \frac{d\rho }{dy}
\nonumber\\
&=&- c_s^2  \left.\frac{\p \rho}{\p s}\right|_{p}  \frac{ds }{dy} +  c_s^2 \,  \frac{d \rho}{dy} 
\label{dpdy}
\eqy
resulting in 
\bq
 \De=-\frac1{c_s^2}\frac{dp/dy}{ds/dy}=-\frac1{c_s^2}\frac{dp}{ds}\,.
\label{NewDe}
\eq
where use has been made of \eqref{sound} and \eqref{dpds}.   Now inserting  \eqref{Estat0} into \eqref{NewDe} yields
for the ${B}= 0$ case the following condition:
\bq
 \De=\frac1{c_s^2}\frac{\rho g}{ds/dy}>0\,, 
\label{Scond}
\eq
and because  $\rho g>0$ we obtain the compact sufficient  condition for stability
\bq
 \frac{d s}{dy}>0\,.
 \label{entropyB0}
\eq
We will see that an identical condition is obtained in the Eulerian energy-Casimir context (see  Eq.~(\ref{eq:d2Fbis})).

Now, given that ${d s}/{dy}>0$ we can use  \eqref{dpdy} to obtain a condition on $d\rho/dy$, 
\[
  c_s^2 \,   \frac{d \rho}{dy} + \rho g
=   c_s^2  \left.\frac{\p \rho}{\p s}\right|_{p}  \frac{ds }{dy} <0
\]
which implies
\bq
  \frac{d\rho}{dy} < -\frac{ \rho g}{c_s^2}<0\,.
  \label{denTrsh}
 \eq
 Upon defining the scale height $L^{-1}= \rho^{-1} |d\rho/dy|$, \eqref{denTrsh} is seen to be equivalent to $c_s^2> Lg$.  Thus the system is stable to convection if the free fall kinetic energy is  smaller than twice the kinetic energy at the sound speed.  Or, equivalently, if  the free fall speed through a distance $L$ is smaller than $\sqrt{2} c_s$.

The above procedure leading to \eqref{entropyB0} and \eqref{denTrsh}  was designed for comparison with Sec.~\ref{ssec:Econv}.  However, the conventional `$\de W$' stability analysis  proceeds with an extremization over $\boldsymbol{\xi}$ that  takes  account of any possible stabilization effect due to the first positive definite term of \eqref{eq:d2Wbis}.  To this end we let 
\bq
\boldsymbol{\xi}(\bfx)= \big(\xi_x(y),\xi_y(y),\xi_z(y)\big)e^{i(kz + \ell x)}/2 + c.c.
\label{Fxi}
\eq  
and rewrite   \eqref{eq:d2Wbis} as 
\bqy
\delta^{2}W_{\mathrm{la}}&=&\frac12\int_0^\infty dy\Bigg[-\left(\frac{\rho g^2}{c_s^2} + g  \frac{d\rho }{dy}\right) |\xi_y|^2 
\nonumber\\
&& + \rho\, c_s^2\left|\frac{d\xi_y}{dy} + i\ell \xi_x + i k \xi_z - \frac{\rho g}{c_s^2} \xi_y\right|^2\, 
\Bigg]
\label{dwconvect}
\eqy
Given any $\xi_y(y)$, one can choose $\xi_x$ and $\xi_z$  that make  the second term vanish.  Thus the 
smallest value of $\delta^{2}W_{\mathrm{la}}$ is given by 
\bq
\delta^{2}W_{\mathrm{la}}=-\frac12\int_0^\infty dy \left(\frac{\rho g^2}{c_s^2} + g \frac{d\rho }{dy} \right) |\xi_y|^2 
\eq
which yields \eqref{denTrsh} as a  necessary and sufficient condition for stability.  Thus \eqref{denTrsh}  is in fact a counterpart equivalent to  $ds/dy>0$.   Another equivalent condition exists in terms  of  the  temperature:
\bq
\frac{d T}{dy}> \frac{gT}{\rho c_p}\left.\frac{ \p \rho}{\p T}\right|_p
\label{tempcon}\,, 
\eq
which follows in a manner similar to  \eqref{denTrsh}.

 Lastly, for an ideal gas,    \eqref{denTrsh} and  \eqref{tempcon} become, respectively, 
\[
  \frac{d\rho}{dy}   < - \frac{\rho^2 g}{\ga p} \qquad \mathrm{and}\qquad  \frac{d T}{dy}> -\frac{g}{ c_p}\,.
\] 
 
Observe,  \eqref{eq:cond_s}  could be satisfied with $ds/dy<0$ and $d\rho/dy>0$.  But, the stability condition $ds/dy>0$, which came from \eqref{Scond},  implies $d\rho/dy<0$.  Thus it is not possible to have stability unless the fluid density is stably stratified.

\bigskip

\noindent\underline{Case ${B}\neq 0$}:

\medskip

 The case with ${B}\not= 0$ has been studied 
extensively, e.g.\  in the early works on interchange instability of  Refs.~\onlinecite{S06,thompson,Kruskal54,tserkovnikov,newcomb61,yu66,greene68}.  
For this application, Eq.~\eqref{eq:dW_Ber} can be written as follows:
\bqy
\delta^{2}W_{\mathrm{la}}   &=&\frac{1}{2}\!\intx\!
 \bigg[
\rho \, c_s^2  \left(  \mathbf{\nabla}%
\cdot\boldsymbol{\xi}\right)^{2}
+ \left(  \mathbf{\nabla}\cdot
\boldsymbol{\xi}\right)  \left(  \mathbf{\nabla}p\cdot
\boldsymbol{\xi}\right)
\nonumber\\
&& \hspace{-1.2 cm} + \frac{\left\vert \de \bfB \right\vert^{2} }{4\pi}
+ \mathbf{J} \cdot (\boldsymbol{\xi}\times\de \bfB)
- g (\boldsymbol{\eta}\cdot\hat{\mathbf{y}})  \nabla\cdot(\rho\boldsymbol{\xi}) 
\bigg] \,,
\label{eq:dW_conv}%
\eqy
where again all equilibrium quantities depend only on $y$, which we use together with  \eqref{Fxi} to rewrite this as 
\bqy
\delta^{2}W_{\mathrm{la}}&=&\frac12\!\int_0^\infty\! \!\!dy\Bigg[\frac{B^2}{4\pi}\bigg(k^2\Big(|\xi_y|^2 + |\xi_x|^2\Big) +\left|\frac{d\xi_y}{dy} +i\ell \xi_x\right|^2
\bigg)
\nonumber\\
&& 
\vspace{1cm}+\  \rho\, c_s^2 \left|\frac{d\xi_y}{dy} + i\ell \xi_x + ik \xi_z \right|^2 -  g \, \frac{d\rho }{dy} \,  |\xi_y|^2 
\nonumber\\
&&\vspace{1cm} - \ 2  \rho g \,  \xi_y \bigg(\frac{d\xi_y}{dy} + i\ell \xi_x + ik \xi_z \bigg) 
\Bigg]\,, 
\label{dwBcon}
\eqy
where, following Ref.~\onlinecite{newcomb61},  the displacements $\xi_y$, $ i\ell \xi_x$, and $ik \xi_z$ can be taken to be real-valued. 
By minimizing this functional the following necessary and sufficient condition for interchange stability of  Tserkovnikov\cite{tserkovnikov},  can be obtained:
\bq
\frac{d \rho}{dy} <-  \frac{\rho g}{c_s^2 + c_a^2}<0\,,
\label{zerk}
\eq
where recall $c_a^2=B^2/(4\pi\rho)$.

In Ref.~\onlinecite{newcomb61}  Newcomb rearranges \eqref{dwBcon} and minimizes it in the limit $k\rightarrow 0$ by choosing
$i\xi_z\rightarrow {g\xi_y}/({k c_s^2})$ for arbitrary $\xi_y$.  With this approach he obtains the  more stringent stability condition of  \eqref{denTrsh}, the condition for the case  without   $B$.  Newcomb's singular approach allows  displacements  that  interchange  
plasma elements containing long segments along magnetic field lines,  relieving  local fluid pressures.  In Ref.~\onlinecite{yu66}
it is shown that this amounts to the plasma being  least stable against these long quasi-interchange displacements because the  restoring force due to the magnetic field tension vanishes.

\subsection{Eulerian convection}
\label{ssec:Econv}

\subsubsection{Eulerian convection equilibria}

\bigskip

\noindent\underline{Case ${B}= 0$}:

\medskip

Using the Casimir invariants of \eqref{eq:C1}  and \eqref{eq:C2},  hydrodynamic equilibria with translational symmetry are obtained as extrema of the following energy-Casimir functional:
\bqy
\mathfrak{F}&=&\intx \bigg[\frac{1}{2}\rho |\bfv|^{2}+\rho U\left(\rho,s\right)+\rho\Phi+\rho {\cal S}\left(s\right)
\nonumber\\
& &\hspace{1 cm} - {\cal A}\left(s\right)\mathbf{\hat{z}\cdot\nabla}\times\mathbf{v}\bigg]\,,
\label{eq:EC}
\eqy
where  $v_{z}=v_{z}\left(s\right)$.  Variation of \eqref{eq:EC} will automatically yield equations that are cases of \eqref{Emom}--\eqref{Eent} with $B_e\equiv 0$.  Because $v_{z}\left(s\right)$,  we have   $\mathbf{v}\cdot\mathbf{\nabla}s=0$ and $\mathbf{v}\cdot\mathbf{\nabla}v_{z}=0$. Variation with respect to $\bfv$ yields 
\begin{equation}
\rho\mathbf{v}_{\perp}=\mathbf{\nabla}{\cal A}\times\mathbf{\hat{z}},\label{eq:mass_bis}
\end{equation}
while variation with respect to $\rho$ and $s$, respectively, yield
\begin{eqnarray}
\frac{1}{2}|\bfv|^{2}+\Phi+\rho U_{\rho}+U +  {\cal S}& = &0,\label{eq:bern}\\
\rho v_{z}v_{z}^{\prime}+\rho U_{s}+\rho {\cal S}^{\prime}-{\cal A}^{\prime}\mathbf{\hat{z}\cdot\nabla}\times\mathbf{v} & = & 0,\label{eq:euler}\,.
\end{eqnarray}

For  our  case of interest with $\mathbf{v}=0$, we merely set $\cala\equiv 0$, whereupon the first variation,    \begin{equation}
\delta\mathfrak{F}=\intx\big[\left(\rho U_{\rho}+U+\Phi +\mathcal{S}\right)\delta\rho+\rho\left(U_{s}+{\cal S}^{\prime}\right)\delta s\big]\,, 
\end{equation}
gives rise to 
\begin{eqnarray}
\Phi+\rho U_{\rho}+U + {\cal S} & = & 0 \\
U_{s} + {\cal S}^{\prime}& = & 0\,,
\end{eqnarray}
where recall for our analyses we choose $\Phi= g y$. 

\bigskip

\noindent\underline{Case ${B}\neq 0$}:

\medskip

For case with equilibrium magnetic field we choose the following special case for the Casimir of \eqref{eq:casimir1_z}:
\bq
C_s=\intx\rho\,  {\cal S}\left(s,\psi\right)\,,
\label{redC}
\eq
which with  the Hamiltonian 
\bq
H=\intx\Big[ \frac{1}{2}\rho |\bfv|^{2}+\rho U+\frac{|\nabla\psi|^{2}}{8\pi}+\rho gy \Big]\,,
\label{redH}
\eq
 gives upon varying $\mathfrak{F}=H+C_s$, 
\bqy
\frac{\delta \mathfrak{F}}{\delta\mathbf{v}}&=&\rho\mathbf{v}=0 
\nonumber
\\
\frac{\delta \mathfrak{F}}{\delta\psi}&=&-\Delta\psi+\rho {\cal S}_{\psi}=0
\nonumber
\\
\frac{\delta \mathfrak{F}}{\delta s}&=&\rho U_{s}+\rho {\cal S}_{s}=0
\nonumber
\\
\frac{\delta \mathfrak{F}}{\delta\rho}&=&\rho U_{\rho}+U+gy+{\cal S}=0
\nonumber
\eqy
which imply
\bq
\nabla\left(\rho U_{\rho}+U\right)+\frac{1}{\rho}\nabla^2\psi \,\nabla \psi -U_{s}\nabla s=-g\,.
\label{preeq}
\eq
 Equation \eqref{preeq} gives for our case  with  stratification in $y$  the equilibrium equation   \eqref{Estat2}.

\subsubsection{Eulerian convection stability}
\label{sssec:EconvectionStability}

Now we examine  $\delta^{2}\mathfrak{F}$ for our two cases and look for conditions that make this quantity positive definite, conditions that will be  sufficient conditions for stability. 

\bigskip

\noindent\underline{Case ${B}= 0$}:

\medskip

The second variation is 
\bqy
\delta^{2}\mathfrak{F}&=&\intx\Big[\left(\rho U_{\rho\rho} 
+ 2U_{\rho}\right)\left(\delta\rho\right)^{2}
\nonumber\\ 
&& \hspace{1 cm} + 2\left(\rho U_{\rho s}+U_{s}+{\cal S}^{\prime}\right)\delta\rho\,\delta s
\nonumber\\ 
&&  \hspace{1 cm}+\rho\left(U_{ss}+{\cal S}^{\prime\prime}\right)\left(\delta s\right)^{2}\Big]\,.
\label{2vaEc}
\eqy
 By  exploiting the equilibrium equations, \eqref{2vaEc} can be rewritten as 
\bqy
\delta^{2}\mathfrak{F}&=&\intx\frac{c_{s}^{2}}{\rho}\bigg[\left(\delta\rho\right)^{2} 
-2\left.\frac{\partial\rho}{\partial s}\right\vert _{p}\delta\rho\,\delta s
\nonumber\\
&&\hspace{2cm} +\left.\frac{\partial\rho}{\partial s}\right\vert _{p}\frac{d\rho}{dy}\frac{dy}{ds}\left(\delta s\right)^{2}\bigg]
\,,
\label{eq:d2F}
\eqy 
where we used \eqref{EOS} and \eqref{dpds}, and the derivative of the equilibrium equation $U_{s} + {\cal S}^{\prime}=0$ with respect to $y$, 
\bq
{\cal S}^{\prime\prime}+U_{ss}+U_{s\rho}\frac{d\rho}{dy}\frac{dy}{ds}=0\,.
\eq
Next, we use 
\bqy
\left(\delta\rho\right)^{2}-2\left.\frac{\partial\rho}{\partial s}\right\vert _{p}\delta\rho\delta s
&=&\left(\delta\rho-\left.\frac{\partial\rho}{\partial s}\right\vert _{p}\delta s\right)^{2}   
\\
&&\hspace{1 cm}-\left(\left.\frac{\partial\rho}{\partial s}\right\vert _{p}\right)^{2}\left(\delta s\right)^{2}
\nonumber
\eqy
obtaining 
\bqy
\delta^{2}\mathfrak{F}&=&\intx \frac{c_{s}^{2}}{\rho}\Bigg[\left(\delta\rho-\left.\frac{\partial\rho}{\partial s}\right\vert _{p}\delta s\right)^{2}
\label{eq:d2Fbis}\\
&&\hspace{1 cm} 
+\left.\frac{\partial\rho}{\partial s}\right\vert _{p}\left(\frac{d\rho/dy}{{ds}/{dy}}-\left.\frac{\partial\rho}{\partial s}\right\vert _{p}\right)\left(\delta s\right)^{2}\Bigg] \,, 
\nonumber
\eqy
an expression of the form of \eqref{eq:d2Wbis}.  Thus, as in Sec.~\ref{sssec:LconvectionStability},   stability is again determined by positivity of the quantity $\De$  of \eqref{DE} and all of the conditions of that section are reproduced as sufficient stability conditions.

In Eq.~\eqref{eq:d2Fbis}, unlike the case of \eqref{eq:d2Wbis},  $\de\rho$ and $\de s$ are independent so a sharper sufficient condition cannot be pursued  by relying on the positivity of the first term, even though in the $\de^2 W_{\mathrm{la}}$ formulation this did not materialize.  Also, the approach here gives $ds/dy>0$ as a sufficient condition for stability  (or equivalently \eqref{denTrsh}), while the $\de^2W_{\mathrm{la}}$ formulation shows that this condition is both necessary and sufficient

\bigskip

\noindent\underline{Case ${B}\neq 0$}:

\medskip

Now consider the second variation of $\mathfrak{F}= H + C_s$ with $H$ given by \eqref{redH} and  $C_s$ given by  \eqref{redC} with 
\[
\mathcal{S}\left(\psi,s\right)=\mathcal{K}\left(\psi\right)+\mathcal{L}(s)\,,
\]
which is general enough to describe the equilibria of our interest as given by \eqref{Estat2}.
This leads to 
\bqy
\delta^{2}\mathfrak{F} &=& \intx\Big[\left(\rho U\right)_{\rho\rho}\left(\delta\rho\right)^{2}+2\big[\left(\rho U\right)_{\rho s}+{\cal L}_{s}\big]\de\rho \, \delta s
\nonumber\\
&& 
+\rho\big(U_{ss}+{\cal L}_{ss}\big)\left(\delta s\right)^{2}
 +\left|\nabla\delta\psi\right|^{2}+2\mathcal{K}_{\psi} \delta\psi\, \delta\rho 
 \nonumber
 \\
&&   +\rho \mathcal{K}_{\psi\psi}\left(\delta\psi\right)^{2}\Big]\,.
\label{d2FB}
\eqy
Rewriting  \eqref{d2FB} in terms of equilibrium quantities and manipulating then gives
\begin{eqnarray}
\delta^{2}\mathfrak{F}  &=& \intx\bigg[\frac{c_{s}^{2}}{\rho}\left(\delta P\right)^{2}
+ \frac{p_{s}}{\rho} \De\left(\delta s\right)^{2}  + 2\frac{Jp_{s}}{\rho c_{s}^{2}} \, \delta s\,\delta\psi 
\nonumber\\
 &&  +\rho\left(\mathcal{K}_{\psi\psi}-\frac{J^{2}}{\rho^{2}c_{s}^{2}}\right)\left(\delta\psi\right)^{2}\bigg]\,,
\label{d2fconE}
\end{eqnarray}
where use has been made of the definition of $\De$ of \eqref{DE},  the current density $J$,   defined by
\begin{equation}
-J=\nabla^2\psi=\rho \mathcal{K}_{\psi}\,,
\label{JK}
\end{equation}
the thermodynamic expressions of  \eqref{EOS} and the following,  which is a consequence of the equilibrium equation, 
\begin{equation}
U_{ss}+{\cal K}_{ss} =-U_{s\rho}\frac{d\rho}{ds}\,,
\end{equation}
which implies
\bqy
U_{ss}+{\cal K}_{ss}&-&\frac{1}{c_{s}^{2}}\frac{p_{s}^{2}}{\rho^{2}}=-U_{s\rho}\frac{d\rho}{ds}-\frac{1}{c_{s}^{2}}\frac{p_{s}^{2}}{\rho^{2}}
\nonumber\\
&=&-\frac{p_{s}}{\rho^{2}}\left(\frac{d\rho}{ds}+\frac{p_{s}}{c_{s}^{2}}\right)=\frac{p_{s}}{\rho^{2}}\De \,. 
\eqy
In addition we have introduced the new variable $\delta P$ defined by
\begin{equation}
\delta P=\delta\rho+\frac{p_{s}}{c_{s}^{2}}\delta s-\frac{J}{c_{s}^{2}}\delta\psi\,.
\end{equation}

Next, we collect the terms with $\delta s$ to obtain 
\begin{eqnarray}
\delta^{2}\mathfrak{F}  &=& \intx\Bigg[\frac{c_{s}^{2}}{\rho}\left(\delta P\right)^{2} +  \left|\nabla\delta\psi\right|^{2} 
\label{eq:dF2_temp} \\
&+& \frac{p_{s}}{\rho} \De 
\bigg[ \delta s -\frac{J}{c_{s}^{2}\De}  \delta\psi  \bigg]^{2}
\nonumber\\
& + &
 \rho\bigg[\mathcal{K}_{\psi\psi}-\frac{J^{2}}{\rho^{2}c_{s}^{2}}
-\frac{J^{2}p_{s}}{\rho^{2}c_{s}^{4}\De}
\bigg]    
\left(\delta\psi\right)^{2}\Bigg]\,.
 \nonumber
\end{eqnarray}
If we introduce the variation
\begin{equation}
\delta Q=\delta s-\frac{J}{c_{s}^{2}\De}\, \delta\psi
\end{equation}
and we use the gradient of \eqref{JK} 
\begin{equation}
\nabla J=\frac{J}{\rho}\nabla\rho-\rho \mathcal{K}_{\psi\psi}\nabla\psi\,,
\end{equation}
which for equilibria that depend only on the $y$ coordinate can be
written as
\begin{equation}
\rho \mathcal{K}_{\psi\psi}=\frac{J}{\rho}\frac{d\rho}{d\psi}- \frac{dJ}{d\psi}
= -\rho\frac{d (J/\rho)}{d\psi}
\end{equation}
or
\begin{equation}
\rho \mathcal{K}_{\psi\psi}=\frac{J}{\rho}\frac{d\rho/ds}{d\psi/ds}- \frac{dJ/ds}{d\psi/ds}\,, 
\end{equation}
then the last term of Eq.~(\ref{eq:dF2_temp}) can be rewritten as
\bqy
&&\mathcal{K}_{\psi\psi}-\frac{J^{2}}{\rho^{2}c_{s}^{2}} 
-\frac{J^{2}p_{s}}{\rho^{2}c_{s}^{4}\De} \nonumber\\
 &&\hspace{1cm} =-\frac{d(J/\rho)}{d\psi}+\frac{J^{2}}{\rho^{2}c_{s}^{2}\De} \frac{d\rho}{ds} \,.
\eqy
Then,  finally 
\begin{eqnarray}
\delta^{2}\mathfrak{F}  &=& \intx\Bigg[\frac{c_{s}^{2}}{\rho}\left(\delta P\right)^{2} 
+ \frac{p_{s}}{\rho}\De (\de Q)^{2} + \left|\nabla\delta\psi\right|^{2}
\nonumber \\
& + &
 \rho\bigg(-\frac{d(J/\rho)}{d\psi}+\frac{J^{2}}{\rho^{2}c_{s}^{2}\De} \frac{d\rho}{ds}\bigg)   
\left(\delta\psi\right)^{2}\Bigg]\,.
\label{eq:dF2_temp2}
\end{eqnarray}
From the energy expression of  \eqref{eq:dF2_temp2} we can immediately read off the following sufficient conditions for stability:
\bqy
0&<&\De = -\left(\frac{d\rho}{ds}+\frac{p_{s}}{c_{s}^{2}}\right)
\label{DeIneq}\\
0&<&-\frac{d(J/\rho)/dy}{d\psi/dy}+\frac{J^{2}}{\rho^{2}c_{s}^{2}\De} \frac{d\rho/dy}{ds/dy}\,, 
\label{dpsiIneq}
\eqy
where recall the form of $\De$ of \eqref{DeIneq} is equivalent to that of \eqref{DE}.

 In the case with $B=0$ we had the two free functions, $\rho$ and $s$ and one stability inequality.  Thus we were able to obtain separate conditions on the equilibrium profiles of  $\rho$ and $s$ for stability.  In the present case we again have one equilibrium equation, but now with three profiles  $\rho, s, $ and $B$  and two inequalities.  Again we should expect to obtain independent conditions on the profiles $\rho, s, $ and $B$.  However,  even the condition of \eqref{denTrsh}, which has clear physical meaning, is not immediately implementable because $c_s$ depends on $y$ through both $\rho$ and $s$.   Similarly, the inequalities \eqref{DeIneq} and \eqref{dpsiIneq} require the profiles for their determination.  In practice one may construct a family of equilibria with profiles that depend on one or more parameters  and then seek thresholds in parameter space.

Inequalities \eqref{DeIneq} and \eqref{dpsiIneq} can be written in various ways.   For example, using  the equilibrium equation \eqref{conequil}, 
\begin{equation}
\frac{d p}{d y} =  c_s^2 \bigg(\frac{d \rho}{d y }   - \frac{\partial \rho}{\partial s}  \bigg|_p \frac{d s}{d y } \bigg) = -  g \rho - (B^2)'/(8\pi)\,,
\end{equation}
the inequality $\De > 0$  can be rewritten as 
\begin{equation} \Delta = -\frac1{c_s^2}\frac{dp}{ds}=
 \frac{g \rho + (B^2)'/2}{c_s^2 \,ds/dy} >0\,.
\end{equation}
Consequently, if  $d p /dy$ is negative for stability we must have $d s/dy >0$ and, conversely,  we must have $d s/dy <0$ if,  due to $B$   decreasing sufficiently fast with height,   we have $dp/dy >0$.  This is effectively the  threshold against the magnetized  Rayleigh-Taylor instability.
 Thus, as for the case with ${B}= 0$,  $dp /ds <0$  ensures stability.  
Also note, as in the ${B}= 0$ case,  a  critical  point arises  if  for some $y$ we have $d p/dy =0$ unless at the same point 
we also have $ds/dy =0$,  in which case  one then  has to look deeper into the limit.

If $dp/dy <0$ and $ds/dy >0$ we obtain from \eqref{DeIneq} an inequality for $d\rho/ dy$ analogous to  the inequality (\ref{denTrsh}), in particular,    $d\rho/ dy$ must be negative because $p_s/c^2_s>0$; however,  this inequality  is different from the ``Tserkovnikov'' inequality of \eqref{zerk}.   If $dp/dy >0$ and $ds/dy <0$ we obtain a reversed  inequality, i.e.,   $d\rho/ dy$ must be positive.

This implies that in the inequality (\ref{dpsiIneq}), if $\Delta$ is positive,  the  second term is always negative and thus for  $B >0$ we obtain the condition 
\begin{equation}
 \label{DeIneq1}
{d(J/\rho)/dy}<0, \quad {\rm or} \quad {dJ/dy}< (J /\rho) (d\rho/dy).
\end{equation}
Consider the two cases of decreasing and increasing magnetic fields: for a magnetic field decreasing with height,  $ J= - d B/dy >0$, so 
\begin{equation}
 \label{DeIneqJ1}
d \, \ln{J}/dy < d \, \ln{\rho}/dy
\end{equation}
 and if $d\rho/dy <0$ we can use the inequality obtained before for $d\rho/dy$ and obtain an inequality that involves  the second derivative of the magnetic field and the density profile.  Similarly, if $ J= - d B/dy <0$, 
\begin{equation}
 \label{DeIneqJ2}
d \, \ln{|J|}/dy > d \, \ln{\rho}/dy
\end{equation}
 and if $d\rho/dy >0$ we can use the reverse  inequality obtained before for $d\rho/dy$ and again obtain an inequality that involves  the second derivative of the magnetic field and the density profile.
These cases  above   do not exhaust all possibilities.  It is perhaps best to consider  families  of equilibria and investigate  parameter dependencies  as mentioned above.


\subsection{Dynamically accessible convection}
\label{ssub:Dconvection}
\subsubsection{Dynamically accessible convection equilibria}
\label{sssec:DconvectionEquilbria}

In Sec.~\ref{ssec:Daform} we showed how the general dynamically accessible variations of  \eqref{eq:var_0_1}--\eqref{eq:var_0_4}, when inserted into the first variation of the Hamiltonian \eqref{daH},  give rise to the general MHD equilibrium equations of  \eqref{Emom}--\eqref{Eent}.    Thus, equilibria that are solutions of \eqref{Estat2}, with or without the magnetic field, are extremal points of this kind of variation, and we can proceed to assess stability by examination of  the energy expression of \eqref{d2H_da}.
 
\subsubsection{Dynamically accessible convection stability}
\label{sssec:DconvectionStability}

For static equilibria the first term of  \eqref{d2H_da} reduces to  the form of \eqref{static}.  As noted in Sec.~\ref{ssec:compform} this term vanishes if $g_3=g_2 =\mathbf{g}_4\equiv 0$.  Thus,   choosing $\mathrm{\bf g}_1$ proportional to $\boldsymbol{\xi}$,  the condition for dynamically accessible stability in the case of static equilibria  is determined by $\de^2 W_{\rm la}$, viz.\  the Lagrangian energy expression.  In both the cases with and without a magnetic field this is the usual $`\de W'$ energy, for each case respectively, and thus dynamically accessible stability in both cases is identical to that for Lagrangian stability. 

\subsection{Convection comparisons}
\label{ssec:ConCom}

Results for the case with equilibria $B=0$ can be summarized succinctly:  the Lagrangian and dynamically accessible approaches both give the simple necessary and sufficient condition for stability,  $ds/dy>0$,  or equivalently the inequality of \eqref{denTrsh} on $d\rho/dy$,   while the Eulerian energy-Casimir approach gives this same result, but only as a sufficient condition for stability and only applicable to the case with the imposed translational symmetry. 

For  case of equilibria $B\neq 0$  the situation is more complex, although it again must be true, in light of the general discussion of Sec.~\ref{ssec:compform},  that the Lagrangian and dynamically accessible approaches must give the same necessary and sufficient condition for stability, viz.\  that of \eqref{zerk}.   However, this necessary and sufficient condition is much simpler  than the inequalities of  \eqref{DeIneq} and \eqref{dpsiIneq} obtained by the energy-Casimir method and, again,  these inequalities are only applicable to the case with the imposed translational symmetry and only give  sufficient conditions for stability. 
  Moreover, the energy-Casimir inequalities  depend on an extra derivative with respect to $y$ of at least one of the equilibrium profiles;  e.g.\   \eqref{DeIneq}  contains a derivative of the current $J$, which can be eliminated in terms of two derivatives of the pressure $p$, but cannot easily be eliminated entirely.

 If one inserts the Lagrangian variations of \eqref{eq:laDrho}--\eqref{eq:laDM},  adapted to the convection example,  into  $\de^2\mathfrak{F}$  of  \eqref{d2FB},  then $dJ/dy$ is removed.  In the context of our convection example the relevant connection is provided by $\de \psi_{\mathrm{la}} = \mathbf{\xi}\cdot\nabla\psi= \xi_y \psi'$, with prime denoting y-differentiation.  Whence,  the line-bending term  of  \eqref{d2FB} becomes 
 \bqy
 |\nabla \de \psi|^2&=&(\xi'_y \psi' +\xi_y \psi'' )^2=\xi'_y \psi'^2 
 \nonumber\\
 &&\hspace{1 cm} +\ \xi^2_y \psi''^2+ 2 \xi_y \xi_y' \psi''\psi'
 \label{LB}
 \eqy
  and one finds upon integrating the last term of \eqref{LB} by parts,  a term proportional to $J'$.  This term cancels the $J$ term  of $(\de \psi)^2$ (the same cancellation was shown to occur in the context of the magnetorotational instability in Ref.~\onlinecite{pjmTT13}).  As noted in Sec.~\ref{sec:basics}
 (cf.~Refs.~\onlinecite{morrison98,amp2a}) such a correspondence by constraining the Eulerian variations  in general connects energy-Casimir and Lagrangian stability.

\section{Rotating pinch}
\label{sec:pinch}

Now  we  investigate the stability of the  azimuthally symmetric rotating pinch,  again within  the  Lagrangian, Eulerian energy-Casimir,   and dynamically accessible frameworks.  This example is chosen to illustrate two features introduced in Sec.~\ref{sec:basics} associated with  the inclusion of an equilibrium velocity field:  the relabeling transformation that removes time dependence from a Lagrangian state associated with  a stationary  Eulerian equilibrium and the origin of the difference between Lagrangian and dynamically accessible stability.    As in Sec.~\ref{sec:convection},  we begin by discussing the  plasma equilibrium configurations of interest  by solving directly  the Eulerian MHD equations (\ref{Emom})--(\ref{Eent}) without  referring  specifically to any of the three frameworks.

 We use cylindrical coordinates $\left(r,\phi,z\right)$
 and consider    plasma equilibrium configurations  where all  equilibrium quantities (including entropy)  depend only on the radial
coordinate $r$:
\begin{eqnarray}\label{MHDsol}
\mathbf{B}& = & B_{z}(r) \hat{\mathbf{z}}+B_{\phi}(r) \hat{\boldsymbol{\phi}},\\
\mathbf{v}& = & v_{z}(r) \hat{\mathbf{z}}+v_{\phi}(r) \hat{\boldsymbol{\phi}},\\
\rho& = &\rho(r), \quad  s=  s(r), \label{MHDsols}\\
B_{\phi}&=&\hat{\boldsymbol{\phi}}\cdot \mathbf{\nabla}\psi\times\hat{\mathbf{z}}=-\frac{d\psi (r) }{dr}. 
\label{MHDsolB}
\end{eqnarray}
Equation (\ref{MHDsols}) implies that  $p= p_e(r)$. 
From Eqs.~\eqref{Emom}--\eqref{Eent} we obtain  the generalized Grad-Shafranov  equation  for the  flux function $\psi (r)$ \begin{eqnarray}
\label{GGSS}
\frac{1}{r}\frac{d}{dr}\left(\frac{1-\calm^{2}}{4\pi}rB_{\phi}\right) & & -\frac{1}{\psi_{r}}\frac{d}{dr}\left(p+\frac{B_{z}^{2}}{8\pi}\right) 
 \nonumber \\
&& + \frac{d}{dr}\left(\frac{\calm^{2}}{4\pi}B_{\phi}\right)=0,
\end{eqnarray}
where  
\[
\calm(r)  = \left[ \frac{ 4\pi\rho(r) v_\phi^2(r)}{B_\phi^2(r)}\right]^{1/2}
\] is the poloidal Alfv\`en Mach number.
   Note that $v_z(r) $ does not appear in (\ref{GGSS}) and in the following it will be  set equal to zero.

In  (\ref{GGSS}) we need to assign three free functions. We will  assign $B_z(r), B_\phi(r)$,  and  $v_\phi(r)$ and  treat  (\ref{GGSS}) as an equation for $p(r)$ that can be written as 
\begin{equation}\label{GGSS1}
\frac{1}{B_\phi}\frac{d}{dr}\left(p +\frac{B_{z}^{2}}{8\pi}\right) +\frac{d}{dr}\left(\frac{B_{\phi}}{4\pi}\right) +\frac{1 - \calm^{2}}{4\pi}\, \frac{B_{\phi}}{r} =0\,.
\end{equation} 
For the sake of simplicity we will examine the case  of an  isothermal plasma configuration as it makes the relationship between $p$ and $\rho$ linear, and also  makes  $\calm^{2}$ linear in $p$.
A further simplification is obtained by taking the current density $J_z$  to be uniform. 
By defining a dimensionless radial variable $r$ in terms of a characteristic length $r_0$, the latter assumption leads to $B_{\phi}  = B_0 r $ and (\ref{GGSS1}) becomes 
\begin{equation}\label{GGSS2}
\frac{d}{dr^2}\left({\hat p}+\frac{\hat{B}^{2}}{2}\right) = -[ 1-   {\hat p} (r) \, w(r)^2/2]\, ,
\end{equation}
where  we have  set
\bqy
p(r)&=& c_s^2 \rho(r) = {\hat p}(r)\, B_0^2/(4\pi)\,, 
\nonumber\\
\hat{B}(r)   &=&    B_{z}/ B_0\,, \quad \mathrm{and}
\nonumber\\
\calm^2(r) &=& ({4\pi p }/{B_0^2})\, [{v_\phi^2}/{(r c_s)^2}]   = {\hat p} (r) \, w^2(r)
\nonumber\,,
\eqy
with  $\hat{B}$ being the dimensionless magnetic field,  $\hat{p}$  the dimensionless pressure, $w(r)  ={ v}_\phi / (r c_s) $  the dimensionless rotation rate,  and $c_s$ the sound velocity in the isothermal case. 

For a configuration where $B_z$  is uniform and the plasma rotation  is rigid with rotation frequency $\Omega$,   Eq.~(\ref{GGSS2})  takes the elementary form 
 \begin{equation}
\frac{d  \, {\hat p} (r)}{d\, r^2}   = -[ 1-   {\hat p} (r) \, w^2/2] \, ,
\label{GGSSis}
\end{equation}
where $w = \Omega r_0/  c_s$.  While a  uniform $B_z$  field does not alter these  equilibrium configurations, it will be shown  to affect  their  stability.  Assuming  $ w^2/2<1$ we obtain 
 \begin{equation}
{\hat p} (r) = \frac{2}{w^2} \left[ 1 -\left(1 -\frac{w^2}{2}\right)\, \exp{\left(\frac{w^2 r^2}{2}\right)} \right]  \label{isoteq},
\end{equation}
where $ {\hat p} (0) = 1$,  ${\hat p} (\bar r) = 0$ for ${\bar r}^2 = - ({2}/{w^2})\ln{(1 - w^2/2)}$. Equation (\ref{isoteq}) describes a one-parameter family of equilibria.   In the absence of rotation this configuration reduces to the standard parabolic pinch with 
$\bar{r}  = 1$ and ${\hat p} (r) = 1 -r^2$, while for  $w^2 \to 2$ we have $\bar{r} = \infty$ and $\hat{p}(r) \equiv 1$.

\subsection{Lagrangian pinch}
\label{ssec:Lpinch}

\subsubsection{Lagrangian pinch equilibria}
\label{sssec:LpinchEquil}

For the rotating  pinch the appropriate Hamiltonian is that of \eqref{H_Lagr} with  $\Phi \equiv 0$ and, as before, the pinch equilibrium equations should follow from Eqs.~\eqref{PBeqs} adapted to the pinch geometry.  In particular,  with the  cylindrical coordinate system  with indices $i,j\in\left\{r,\phi,z\right\}$, $\mathbf{a}=\left({a}^r, {a}^\phi,{a}^z\right)$,    $\mathbf{q}=\left({q}^r, {q}^\phi,{q}^z\right)$, with
 $|\boldsymbol{\pi}|^{2}=g^{ij}(\mathbf{q})\,\pi_{i}\,\pi_{j}=\pi^{i}\pi_{i}$, 
and
 \bq
 \label{Metric}
g^{ij}(\mathbf{q})=\begin{bmatrix}
1&0&0\\
 0& (q^r)^{-2}& 0 \\
  0& 0&1
\end{bmatrix} \,.
\eq
From \eqref{Metric} we obtain 
\bqy
\pi^r&=&g^{rr}\pi_r=\pi_r 
\nonumber\\
\pi^\phi&=&g^{\phi\phi}\pi_\phi=  \pi_\phi/(q^r)^2\,,  
 \nonumber\\
 \pi^z &=&g^{zz}\pi_z=\pi_z\,,
 \nonumber
\eqy
and  similarly $B^r=B_r$, $B^\phi=B_\phi/(q^r)^2$,  and $B^z=B_z$.

As shown in Appendix \ref{app:LanEOM},  the  equations of motion in terms of the Lagrangian variables  $\left({q}^r, {q}^\phi,{q}^z\right)$ follow from Eqs.~\eqref{PBeqs},  and are 
\bqy
 \label{LCq}
\dot{q}^{i}&=&g^{ij}\frac{\pi_{j}}{\rho_{0}} \quad \mathrm{and} 
\\
  \label{LCpi}
\dot{\pi}_{i}&=&\delta_{i}^{r}\, \frac{\pi^{\phi} \pi_{\phi}}{q^{r}\rho_{0}}
-{\mathcal{J}}\frac{\partial}{\partial q^{i}}\left[\left(\frac{\rho_{0}}{{\mathcal{J}}}\right)^{2}U_{\rho}\right] \\
 && - \, \delta_{i}^{r}\, \frac{{\mathcal{J}}}{q^r}\frac{B^{\phi} B_{\phi}}{4\pi}
 +{\mathcal{J}}{B^{j}}\frac{\partial}{\partial q^{j}}\left(\frac{B_{i}}{4\pi}\right)-{\mathcal{J}}\frac{\partial}{\partial q^{i}}\left(\frac{B^{2}}{8\pi}\right)\,.
\nonumber 
\eqy
Transforming  (\ref{LCq}) and  (\ref{LCpi})  to Eulerian variables, we first obtain  the intermediate form
\bqy
\rho_{0}\dot{q}^{r}&=&g^{rr}\pi_{r}=\pi_{r}
\nonumber\\
\rho_{0}\dot{q}^{\phi}&=&g^{rr}\pi_{\phi}= \pi_{\phi}  /r^{2}
\nonumber\\
\rho_{0}\dot{q}^{z}&=&g^{zz}\pi_{z}=\pi_{z}
\label{eq:qdot}
\eqy
and 
\begin{align}
\dot{\pi}_{r} & =\frac{\pi_{\phi}^{2}}{r^{3}\rho_{0}}-{\mathcal{J}}\, \nabla\left(p+\frac{B^{2}}{8\pi}\right)\cdot\hat{\mathbf{r}} 
+{\mathcal{J}}\, \frac{\mathbf{B}\cdot\nabla\mathbf{B}}{4\pi}\,\cdot\hat{\mathbf{r}}\,, \label{eq:pidot1}\\
\dot{\pi}_{\phi} & =-r{\mathcal{J}}\, \nabla\left(p+\frac{B^{2}}{8\pi}\right)\cdot\hat{\boldsymbol{\phi}} +r{\mathcal{J}}\, \frac{\mathbf{B}\cdot\nabla\mathbf{B}}{4\pi}\,\cdot\hat{\boldsymbol{\phi}}\,,  \label{eq:pidot2} \\
\dot{\pi}_{z} & =-{\mathcal{J}}\, \nabla\left(p+\frac{B^{2}}{8\pi}\right)\cdot\hat{\mathbf{z}}  +{\mathcal{J}}\, \frac{\mathbf{B}\cdot\nabla\mathbf{B}}{4\pi}\,\cdot\hat{\mathbf{z}}\,,
\label{eq:pidot3}
\end{align}
from which, using 
\begin{align*}
\dot{\pi}_{r} &  =\rho_{0}\frac{D}{Dt}v_{r}\left(\mathbf{q},t\right)
,\\
\dot{\pi}_{\phi} & =\rho_{0}\frac{D}{Dt}\big(q^{r}v_{\phi}\left(\mathbf{q},t\right)\big)
,\\
\dot{\pi}_{z} &=\rho_{0}\frac{D}{Dt}v_{z}\left(\mathbf{q},t\right)
\,,
\end{align*}
with  $ D/{Dt}= {\partial_t}+ \dot{q}^{i}  \,{\partial }/{\partial q^{i}}$ and $\dot{\bfq}(\bfa,t)=\bfv(\bfx,t)$, 
we recover the  cylindrical components of the  Eulerian equation of motion, 
\begin{equation} \label{EULEUL}\rho\left(\frac{\partial\mathbf{v}}{\partial t}+\mathbf{v}\cdot\nabla\mathbf{v}\right)  =-\nabla\left(p+\frac{B^{2}}{8\pi}\right) +\frac{\mathbf{B}\cdot\nabla\mathbf{B}}{4\pi}\, .
\end{equation}
\bigskip

The rotating pinch equilibrium configuration  of this section  corresponds to  
\begin{equation}
 \dot{\pi}_{r}= \dot{\pi}_{z}    =  \dot{\pi}_{\phi}  = 0 , \quad \dot{q}^{r}= \dot{q}^{z}    = 0,\quad \dot{q}^{\phi} =  \Omega ,
\end{equation} 
with $\rho_0(b^r) = p(r)/c_s^2$ where $p(r)$ is  given by (\ref{isoteq}).  Because ${q}^{\phi} =  \Omega t + a^{\phi}$, we see explicitly that stationary Eulerian equilibria correspond to time-dependent Lagrangian trajectories.

Next, we consider  the relabeling transformation introduced in Ref.~\onlinecite{amp2a} and described in Sec.~\ref{sec:basics}, 
\[
\mathbf{a}=\mathfrak{U}\left(\mathbf{b},t\right)\;\longleftrightarrow\;\mathbf{b}=\mathfrak{B}\left(\mathbf{a},t\right)
\]
where $\mathbf{a}=\mathfrak{A}\left(\mathbf{b},t\right)$ is given by 
\bq
\label{relab}
 a^{r}=b^{r},\hspace{.2 cm} a^{\phi}=b^{\phi}-\Omega\left(b^{r}\right)t,\hspace{.2 cm}  a^{z}=b^{z}- V^{z}\left(b^{r}\right) t
 \eq
 and  $\mathbf{b}=\mathfrak{B}\left(\mathbf{a},t\right)$ is given by 
 \bq
 b^{r}=a^{r},\hspace{.2 cm}  b^{\phi}=a^{\phi}+\Omega\left(a^{r}\right)t,\hspace{.2 cm}  b^{z}=a^{z}+V^{z}\left(a^{r}\right)t \,, 
  \label{relabb} 
\eq
with 
$\mathfrak{J}:=\left|\partial a^{i}/\partial b^{j}\right| =1$,   with  
\bq
 \mathbf{V}\left(\mathbf{b},t\right):=\dot{\mathfrak{B}}\circ\mathfrak{B}^{-1}=\dot{\mathfrak{B}}\left(\mathfrak{A}\left(\mathbf{b},t\right),t\right)
 \nonumber
 \eq
given by
\bq
 \label{appar}
V^{r}=0,\quad V^{\phi}= \Omega\left(b^{r}\right),\quad V^{z}=V^{z}\left(b^{r}\right) \,.
\eq

By inserting (\ref{appar}) into the transformed Hamiltonian of (\ref{tildeH_MHD})  (see Appendix \ref{app:atrt}) we obtain the  ``time-relabeled" equations of motion corresponding to  \eqref{LCq} and \eqref{LCpi} (see \eqref{Erelab1} and \eqref{Erelab2}).
Then in the relabeled variables  by  explicitly  setting $\partial/\partial t =0 $, \,  $Q^i = b^i$ and by assigning the functions $B_0^i$ and $\rho_0 U$ as functions of $b^i$ consistently with the choices made in Sec.~\ref{sec:pinch}, 
these equations yield  the equilibrium equations in the relabeled form of  \eqref{eq:PIdot-bis-1}--\eqref{relab1}.

Thus, we have shown that the equilibrium equation of \eqref{GGSS2}  describes the  reference state 
$(\bfQ_{e}, \bf\Pi_{e})$ that follows from 
\bq
\frac{\de \tilde H}{\de \bfPi} = 0  \qquad {\rm and} \qquad\frac{\de \tilde H}{\de \bfQ} = 0\,. 
\eq
Given that our equilibrium corresponds to the vanishing of the first variation of the Hamiltonian $\tilde H$ of \eqref{tildeH_MHD},  we can expand as in \eqref{linLag} to address stability via the  energy principle described  in Sec.~\ref{sssec:LpinchStab}.

\subsubsection{Lagrangian pinch stability}
\label{sssec:LpinchStab}

Now, to address stability we expand  $\tilde H$ by inserting \eqref{linLag} (see also   Eq.~(27) of Ref.~(\onlinecite{amp2a})),    where  the  reference state  is our pinch  equilibrium of Sec.~\ref{sssec:LpinchEquil}.     This leads to  the   second variation of the Hamiltonian $\tilde H$ written   in terms of the canonically conjugate variables $(\boldsymbol{\eta},\bfpi_{\eta})$ as  given by  \eqref{d2H_Lagr} with $\delta^{2}W_{\mathrm{la}}\left[\boldsymbol{\eta}\right] $ defined by \eqref{FLdW} with \eqref{eq:dW_Ber}.  Due to the arbitrariness of $\boldsymbol{\pi}_{\eta}$ we can  make the first term of   \eqref{d2H_Lagr}  vanish,  so that a sufficient stability condition for the configuration (\ref{reftraj})  is given by $\delta^{2}W_{\mathrm{la}}\left[  \boldsymbol{\eta} \right]>0$.    We will proceed further by minimizing $\delta^{2}W_{\mathrm{la}}$ for our pinch example.

In order to be able to compare the Lagrangian stability conditions with those obtained in the Energy-Casimir framework we  restrict our analysis to perturbations $\boldsymbol{\eta }$ that do not depend on $z$.

Working  out  terms of  \eqref{FLdW} with \eqref{eq:dW_Ber} for our example, we obtain in  cylindrical curvilinear coordinates
\bqy
\label{dw1}
 &&\rho \left({\bf v}_{\phi} \cdot\nabla {\bf v}_{\phi}  \right)  \cdot\left(  \boldsymbol{\eta}\cdot \nabla
\boldsymbol{\eta}\right) =   - (\rho v_{\phi}^2 /r) \,\big[ \eta_r\partial _r  \eta_r  
\\
&& \hspace{3cm} +  \, (\eta_\phi /r) \partial _\phi \eta_r \,  - \, \eta_\phi^2/r\big]\,, 
\nonumber\\
\label{dw2}  
&&-\rho\, |{\bf v}_{\phi} \cdot\nabla \boldsymbol{\eta}|^2 = - \rho ( v_{\phi} /r)^2 \, \big[ (\partial _\phi \eta_r -  \eta_\phi)^2 \  
\\
&&  \hspace{3cm}  + \,(\partial _\phi \eta_\phi  + \eta_r)^2 \, + \,(\partial _\phi \eta_z)^2 \big]\,,
\nonumber\\
&&
\label{dw3} \rho\,{\partial p}/{\partial\rho} \left(  \mathbf{\nabla}%
\cdot\boldsymbol{\eta}\right)^{2}=  \rho\,  (c_s^2 \, /r^2) \big[ \partial _r (r \eta_r) 
\\
&&  \hspace{3cm} + \partial _\phi \eta_\phi \big]^2 \,,
\nonumber
\eqy
where in \eqref{dw6} the  isothermal equation of state $\rho\, {\partial p}/{\partial\rho} = p = \rho c_s^2$ has been used,
\bqy
\label{dw4}
&&(\eta_r \, \partial_r p)   \mathbf{\nabla}\cdot \boldsymbol{\eta }   = 
\big[(\eta_r/r)\,  \partial_r p\big] \, \big[ \partial _r (r \eta_r) \, + \partial _\phi \eta_\phi \big] \,,
\label{dw5}
\\
&& |\mathbf{\nabla}\times\left(\boldsymbol{\eta}\times \mathbf{B}\right)|^2/(4\pi) 
= (B_0^2/ 4\pi)\, \big[ (\partial _r (r \eta_r)^2  \, + ( \partial _\phi \eta_r)^2\big]
 \nonumber\\  
 && \hspace{.5 cm} +  (B_0^2/ 4\pi)\, [ \partial_\phi   \eta_z  -  ( \hat{B}/r )  \big(\partial _r (r \eta_r)  
 \, +  \partial _\phi \eta_\phi \big)]^2\,, 
\\
\label{dw6}
&&\mathbf{J}\times\boldsymbol{\eta}\cdot\delta {\bf B}  = - (B_0^2/ 2\pi)\, 
\big [ \eta_r \partial _r (r \eta_r)  - \eta_\phi \partial _\phi \eta_r \big] \,,
 \eqy
where  the restriction that $\hat{B}=B_z/ B_0$  and $J_z$ be  independent of $r$   has  been used in accordance  
with the derivation in Sec.~\ref{sec:pinch}.  In the above we used the notation $\p_r:= \p/\p r$ etc., which we use throughout the present  section.

In the following we will refer explicitly to the  rigid rotation equilibrium   given by  (\ref{isoteq}) and adopt the dimensionless variables used there.   Also, we  suppose  $\boldsymbol{\eta}\sim \exp(im\phi)$ and consider azimuthally   symmetric ($m=0$) and azimuthally   asymmetric ($m\not= 0$) perturbations separately.

\bigskip

\noindent\underline{Case $m=0$}:

\medskip

If $\partial_\phi =0$ the functional  $\delta^{2}W_{\mathrm{la}} $ depends only on the radial component  
$\eta_r$ and its radial derivative:
\bqy
 \delta^{2}W_{\mathrm{la}}  \left[\boldsymbol{\eta} \right]  
&=&\! \pi\! \! \int\! \! rdr \Big(\!\!- w^2 {\hat p}\,  \big[\eta_r \partial_r (r \eta_r )\big]  +  ({\hat p}/r^2) \big[ \partial_r (r \eta_r )\big]^2  
\nonumber\\
&&\hspace{-.65cm} +    (\eta_r/r)\, (\partial_r {\hat p}) \big[ \partial_r (r \eta_r )\big]  + \big[ \partial_r (r \eta_r )\big]^2\big[ 1 +( \hat{B}/r)^2\big]
\nonumber\\
&&\hspace{2cm}  - 2 \eta_r \big[ \partial_r (r \eta_r )\big]\Big)\,; 
\label{m=0}
\eqy
then using the  equilibrium \eqref{GGSS}  this  reduces to 
\bqy
\label{m=0bis}
\delta^{2}W_{\mathrm{la}}  \big[  \boldsymbol{\eta} \big]  
&=&  \pi  \int\! rdr\Big(\! -4 \eta_r \partial_r (r \eta_r )  
\\
&& \hspace{1cm}   +  \big[ ({\hat p} + r^2 + \hat{B}^2)/r^2\big]\,  \big[\partial_r (r \eta_r )\big]^2
\Big).
\nonumber
\eqy
The first term of \eqref{m=0bis} is a divergence and vanishes by integration with 
the proper boundary conditions, while  the second term is positive definite.   Thus we conclude our pinch equilibrium is stable to  azimuthally   symmetric  perturbations. 

\bigskip

\noindent\underline{Case $m\neq 0$}:

\medskip

In this case, besides  $\eta_r$ and  $\eta_\phi$,  the functional  $\delta^{2}W_{\mathrm{la}} $ depends  also on $\eta_z$  if  $\hat{B} \not =0$.  We use the orthogonality of the different $m$-components and consider the $m^{th}$ component.
The resulting expressions, as obtained from (\ref{dw1})--\eqref{dw6}, are given in Appendix \ref{app:pinch}.

 \bigskip

\noindent\underline{Case $B_z = 0$}:

\medskip

If $\hat{B} = 0$  the displacement  $\eta_z$ along the symmetry axis of the perturbation decouples,  
 and minimization with respect to $\eta_z$ 
gives $\eta_z = 0$, provided
 \begin{equation}
\label{eta-z}
 m^2 (1 - w^2 \hat{p} ) >0  \rightarrow  w^2 < 1\,. 
 \end{equation} 
 Combining (\ref{dw1c})--(\ref{dw6c})   and using  (\ref{GGSSis})  
we can write the integrand of the functional  $\delta^{2}W_{\mathrm{la}} $ in the following matrix form:
\begin{equation} 
\label{ww}
\begin{bmatrix}
\eta_{\phi}^{*} ,\,  \eta_{r}^{*}, \,  \partial_{r}\left(r\eta_{r}^{*}\right)
\end{bmatrix}\, \cdot\,  \mathcal{W} \cdot \,  
\begin{bmatrix}\eta_{\phi}\\
\eta_{r}\\
\partial_{r}\left(r\eta_{r}\right) 
\end{bmatrix} ,
\end{equation}
where $\mathcal{W}$ is the 3x3 matrix  given by 
\begin{equation}
\nonumber
\mathcal{W}=\begin{bmatrix} m^{2}\hat{p}\, \varsigma /{r^{2}}
& im\hat{p}w^{2} & -im \hat{p}/{r^{2}}\\
-im\hat{p}w^{2} & m^{2}\varpi & 0\\
im{\hat{p}}/{r^{2}} & 0 & 1+ {\hat{p}}/{r^{2}}
\end{bmatrix}\, .
\end{equation} 
where for convenience we have defined
 \bq
 \varpi:=1 - w^2 \hat{p}\qquad\mathrm{and}\qquad \varsigma := 1-w^{2}r^{2}\,.
 \label{defs}
 \eq
 Then, to ascertain stability we use Sylvester's criterion on the matrix $\mathcal{W}$.  This criterion states that a necessary and sufficient criterion for the  positive definiteness of a  Hermitian matrix is that the leading principal minors  be positive. The first principal minor  of ${\cal W}$ is seen to be positive if 
\begin{equation}    
 \label{23}
1-w^{2} {\bar r}^{2} >0,\quad {\rm i.e.,} \quad  w^2 < 2\big(1 -\exp{(-1/2)}\big)\,, 
\end{equation}
while the second principal minor of ${\cal W}$ is positive 
if for $m=1$ (which is the worst case)
\begin{equation}
\hat{p}\left(1-w^{2}r^{2}\right)\left(1-\hat{p}w^{2}\right)-r^{2}\hat{p}^{2}w^{4}>0\,, 
\end{equation}
which implies 
\begin{equation}
w^{2}\leq\frac{1}{r^{2}+\hat{p}}<\frac{1}{\bar{r}^{2}} \,, 
\end{equation}
 and coincides with the condition given by  (\ref{23}).
Finally, the determinant of ${\cal W}$ is positive for the worst case $m=1$ if 
\bqy
&& \left(r^{2}+\hat{p}\right)\left(1-w^{2}r^{2}\right)\left(1-\hat{p}w^{2}\right) 
\label{ineqW}\\
&&
\hspace{2 cm}
-\ \hat{p}\left(1-\hat{p}w^{2}\right)-\hat{p}r^{2}w^{4}\left(r^{2}+\hat{p}\right) >0\,,
\nonumber
\eqy
which implies 
\begin{equation} 
\label{233}
w^{2}<\frac{1}{r^{2}+2\hat{p}}\,,
\end{equation}
and yields the  stronger condition $w^2  < 1/2$.

Alternatively  we can first minimize $ \delta^{2}W_{\mathrm{la}}$ with respect to $\eta_\phi$ in order to to obtain 
a quadratic form involving $\eta_r$ and $\partial_r (r \eta_r)$ only,  from which we can derive an Euler-Lagrange 
 equation. Now observe  $\eta_\phi$ enters $\delta^2 W_{\mathrm{la}}$ through a combination of terms  that  we rewrite as 
\begin{align}
\label{reduced} 
-w^{2}\hat{p} \, |m \eta_{\phi} - i \eta_{r} | ^2 \, +  \frac{\hat{p}}{r^{2}}\, |m \eta_{\phi} - i\partial_r  (r\eta_{r})| ^2 \\+ w^{2}\hat{p}\, |\eta_{r} |  ^2 - \frac{\hat{p}}{r^{2}}\, |\partial_r  (r\eta_{r})| ^2 \, .  
\nonumber 
\end{align}
In the absence of rotation,  minimization with respect to $\eta_\phi$   would lead to the incompressibility condition. 
Assuming $w^2 \bar{r}^2 <1$ we introduce the new variable ${\tilde \eta}_\phi = \, \eta_\phi [1 -w^2 r^2]^{1/2}$
and rewrite the expression (\ref{reduced}) as
\begin{equation}
\label{reduced1} 
\frac{\hat{p}} {r^2} \, |m {\tilde\eta} _{\phi} + i \alpha \eta_{r}  - i\beta \partial_r  (r\eta_{r})| ^2 
+ {\rm R} \,, 
\end{equation}
where $\alpha = w^2r^2/(1 -w^2 r^2)^{1/2}$, $\beta =  1/(1 -w^2 r^2)^{1/2}$, 
and
\bqy
\label{resto} 
 \mathrm{R} &=&  -\frac{\hat{p}} {r^2}  \Big[ \alpha^2 |\eta_r|^2 + \beta^2 | \partial_r  (r\eta_{r})|^2
 \nonumber\\ 
  && \hspace{1 cm}- \alpha \beta \, \big(\eta^*_r \partial_r  (r\eta_{r}) + \eta_r \partial_r  (r\eta^*_{r})\big)\Big]\, .
 \nonumber 
\eqy
Then minimization with respect to ${\tilde\eta} _{\phi}$ gives the following 
reduced expression  for $\delta^2 \widetilde{W}_{\mathrm{la}}$: 
\begin{align}
&\delta^2 \widetilde{W}_{\mathrm{la}} =   \pi  \int\! rdr \left\{ \left[ m^2 \varpi 
- \frac{\hat{p} w^4 r^2}{\varsigma} \right]  |\eta_{r}|^{2}\right.
\nonumber \\
&  \hspace{2 cm} +
\left ( 1 - \frac{\hat{p}}{r^2} \frac{w^2 r^2} {\varsigma} \right) |\partial_{r}\left(r\eta_{r}\right)|^{2}    
\label{redu1}  \\ 
& \hspace{2 cm} + \left. \frac{\hat{p}} {r^2}\frac{w^2r^2}{\varsigma}\, \big[
 \eta^*_r \partial_r  (r\eta_{r}) + \eta_r \partial_r  (r\eta^*_{r})\big]\right\}\, ,  
\nonumber
\end{align}
 which we can  rewrite as
 \begin{align} 
& \delta^2 \widetilde{W}_{\mathrm{la}} =  {\pi  \int\! rdr\,} \bigg[
 \Big( 1 - \frac{\hat{p}} {r^2} \frac{w^2 r^2}{\varsigma} \Big)
 |\partial_{r}\left(r\eta_{r}\right)|^{2}  
 \label{redu3}  \\ 
& \hspace{1 cm} + 
\Big( m^2 \varpi  -  {\hat{p} w^4 r^2}/{\varsigma}   
-  r \partial_r \big({ \hat{p} w^2}/{\varsigma} \big)\Big) |\eta_{r}|^{2} \bigg] \,,
\nonumber 
\end{align}
where the contribution of the last term of  $R$ has been integrated by parts.

 It can be directly verified numerically that for $|m|= 1$ the coefficient of $|\eta_{r}|^{2}$   is positive for $w^2\lesssim  0.62$. Since in this interval also the coefficient of 
$ |\partial_{r} (r\eta_{r} )|^2$ is positive, $w^2\lesssim  0.62$ provides a less restrictive  sufficient stability 
condition that falls  between the values given by  (\ref{23})  and  (\ref{233}).   We note that  an even less restrictive  condition could be identified  by solving the Euler-Lagrange equation obtained  via  variation of $\delta^2 {\tilde W}_{\mathrm{la}}$  of (\ref{redu3}) subject to  the  constraint of $\int\! rdr\,  |r\eta_{r}|^{2}$.  Such a procedure   leads to   an eigenvalue equation that  can be searched for the  lowest eigenvalue.  

 \bigskip

\noindent\underline{Case $B_z \neq 0$}:

\medskip

For  $\hat{B}\neq0$  the  component $\eta_z$ is coupled to the other components of the displacement, and 
instead of \eqref{ww} we obtain
\begin{equation}
\label{Wmatrix3}
\begin{bmatrix}\eta_{\phi}^{*} & \eta_{r}^{*} & \partial_{r}\left(r\eta_{r}^{*}\right) & \eta_{z}^{*}\end{bmatrix} \,\cdot\,  \mathcal{W}  \,\cdot\begin{bmatrix}\eta_{\phi}\\
\eta_{r}\\
\partial_{r}\left(r\eta_{r}\right)\\
\eta_{z}
\end{bmatrix} ,
\end{equation}
where the matrix ${\cal W}$ is now  the  4x4 matrix 
\bq
\begin{bmatrix}m^{2}\left({\hat{\Pi}}/{r^{2}}-\hat{p}w^{2}\right) & im\hat{p}w^{2} & -im{\hat{\Pi}}/{r^{2}} & -m^{2}{\hat{B}}/{r}\\
-im\hat{p}w^{2} & m^{2}\varpi & 0 & 0\\
im{\hat{\Pi}}/{r^{2}} & 0 & 1+{\hat{\Pi}}/{r^{2}} & -im{\hat{B}}/{r}\\
-m^{2}{\hat{B}}/{r} & 0 & im{\hat{B}}/{r} & m^{2}\varpi 
\end{bmatrix}\,,
\nonumber
\eq
where recall $\varpi=1-\hat{p}w^2$  and $\hat{\Pi}=\hat{p}+\hat{B}^{2}$. 
Proceeding as  above using Sylvester's criterion  now leads for $m = 1$  to the four  conditions
\bqy
 \label{b1}
0&<&\hat{p}\left(1-w^{2}r^{2}\right)+\hat{B}^{2}\, ,
\\
 \label{b2}
0&<&  \frac{\hat{p}+\hat{B}^{2}}{\hat{p}\left(\hat{p}+\hat{B}^{2}+r^{2}\right)} - w^2\,,
\\
\label{b3}
0&<&\frac{\hat{p}+\hat{B}^{2}}{\hat{p}\left[2\left(\hat{p}+\hat{B}^{2}\right)+r^{2}\right]}- w^{2}\,,
\\
\label{hard} 
0&<&1-w^{2}\left(r^{2}+3\hat{p}+\hat{B}^{2}\right)
\nonumber\\
&&\hspace{2 cm} +\hat{p}w^{4}\left[r^{2}+2\left(\hat{p}+\hat{B}^{2}\right)\right]\,.
\eqy
Note  that   the first two   conditions give threshold values  that increase with $\hat{B}$  while the third gives $w^2 < 1/2$ independently of $\hat{B}$,  i.e., the effect of $B_z$ would appear to  be stabilizing or neutral if we were to neglect  the coupling to $\eta_z$ that  appears  instead in the fourth condition,  where the effect of $B_z$ is destabilizing (for $w^2 < 1/2$).

The inequality (\ref{hard}) can be better cast in the form 
\begin{equation}
\label{hard1} 
w^2 \hat{b}^{2}(1 - 2 w^2 \hat{p}) < (1 - w^2\hat{p}) [ 1 - w^2(r^2 + 2 \hat{p})]\,,
\end{equation}
which, since $1 - 2 w^2 \hat{p}$ is positive for $w^2 <1/2$ and  $r < {\bar r}$,    can be used to compute the maximum value of $\hat{B}$ that yields a sufficient stability condition when $w^2 <1/2$. This yields  $\hat{B}^2 w^2  <1$  for $w^2\to 0$ and $\hat{B}^2   <1/3$  for $w^2\to 1/2^-$.  

 Alternatively we can  perform separate minimizations with respect to $\eta_z$ and $\eta_\phi$ by defining the new variables 
 \bqy
 \nonumber
 \tilde{\tilde{\eta}}_{z} &=&  \,{\eta}_{z} [1 -w^{2}\hat{p} ]^{1/2}
 \\
 \tilde{\tilde{\eta}}_{\phi} &=&  {\eta}_{\phi}\,  [1 -w^{2}[r^2 + \hat{B}^2 /(1 - w^2 {\hat p})]  ] ^{1/2}\,.
 \nonumber
 \eqy
 Provided ${w^{2}\hat{p} <1 }$  and 
 \[
 w^2 \left[r^2 + \hat{b}^2 /(1 - w^2 {\hat p})\right] < 1\,,
 \]
 i.e., $w^2 \left[{\bar r}^2 + \hat{B}^2 \right] < 1$, minimization with respect to these variables 
 gives  after integration by parts the following reduced expression:  
\bqy
\label{redufull1}  
\delta^2 \widetilde{\widetilde{W}}_{\mathrm{la}} &=&   \pi  \int\! rdr\,  \Bigg[  
 \Bigg( 
1 +\frac{\hat{p} +\hat{B}^2} {r^2} \\
&& - \ \frac{\hat{p} \left(1- \hat{B}^{2}w^2/\varpi \right)^2}{r^2\left[1 -w^{2} \left(r^2 +\hat{B}^2/\varpi\right)\right]}
 \Bigg) |\partial_{r}\left(r\eta_{r}\right)|^{2}
 \nonumber \\ 
&&  +\   \left( m^2 \varpi - \frac{\hat{p} w^4 r^2}{1 -w^2\left[r^2 + \hat{B}^2 /\varpi\right]}   \right.
 \nonumber \\
&& \left. -\ r \partial_r\left( \frac{{\hat p}w^2\left[1- {\hat{B}^{2}w^2}/\varpi \right]}{1 -w^{2}
 \left[r^2 + \hat{B}^2 /\varpi\right]} \right) \right)  |\eta_{r}|^{2}
 \Bigg]  \,.
 \nonumber 
\eqy 
Note that the minimization with respect to ${\tilde \eta}_{z}$  can be shown to have introduced a negative, i.e., destabilizing, contribution to $\delta^2 \widetilde{\widetilde{W}}_{\mathrm{la}}$.  It can be directly verified numerically that for $|m|= 1$ the coefficient of $|\eta_{r}|^{2}$   is  no longer positive for $w^2\lesssim 0.62$ if $\hat{B}^2 >0$;  e.g., for $\hat{B}^2= 1$ 
the coefficient of $|\eta_{r}|^{2}$   is   positive for $w^2\lesssim 0.46$ (this value is essentially in agreement with the result that would be obtained from (\ref{b1})--(\ref{hard}). Since in this latter interval also the coefficient of 
$|\partial_{r} (r\eta_{r} )|^2$ is positive, $w^2\lesssim  0.46 $ provides a sufficient stability condition for $\hat{B}^2 = 1$. 
As for the $\hat{B} = 0$ case a less restrictive  condition could be identified  by solving the Euler-Lagrange equation derived  by variation with  the normalization  constraint $\int\! rdr\,  |r\eta_{r}|^{2}$.

\subsection{Eulerian pinch}
\label{ssub:Epinch}

\subsubsection{Eulerian pinch equilibria}

In Ref.~\onlinecite{amp1}, which was reviewed in Sec.~\ref{ssec:eulB}, both the equilibrium and the perturbations were assumed  to be helically symmetric.  In the present section we have assumed  the equilibrium  to be  both translationally symmetric along $z$ and azimuthally symmetric along $\phi$,  while we considered perturbations  that have only translational  symmetry along $z$. Then  the full configuration is symmetric under translations along $z$.  

Now we consider the  first variation of the energy-Casimir functional $\mathfrak{F}[Z]=H_{TS}[Z]+\sum C[Z]$ 
(see Sec.~\ref{ssec:eulB} and Eq.~(1) of Ref.~\onlinecite{ampE})  with translational and rotational symmetry, which  leads to the equilibrium  equation   
\begin{align}
&\frac{1}{4\pi r}\frac{d}{dr}\left[\left(1-\frac{4\pi\mathcal{F}^{2}}{\rho}\right)r\frac{d\psi}{dr}\right]=\rho T\mathcal{S}^{\prime}-\rho\mathcal{J}^{\prime}-B_{z}\mathcal{H}^{\prime}\label{eq:sys5}\nonumber \\
&\hspace{ 2cm} -\rho v_{z}\mathcal{G}^{\prime}-\left(v_{\phi}B_{\phi}+v_{z}B_{z}\right)\mathcal{F}^{\prime} ,
\end{align}
where  now a prime denotes differentiation with respect to the flux function $\psi$ and specific equilibrium  solutions are defined by the choice of the Casimir functions    $\mathcal{F},\, \mathcal{H},\, \mathcal{J},\, \mathcal{G}$ and $\mathcal{S}$ as functions of $\psi$.   Using the definition of these Casimirs (see  Sec.~\ref{ssec:eulB}) in terms of the plasma variables this choice allows 
us to bring (\ref{eq:sys5})  into the form  of (\ref{GGSS}) and to assign the dependence on $\psi$ of the free functions in this equation.

For the isothermal case  the internal energy is $U=c^2_s\ln(\rho/\rho_0)$ to within a constant and the relevant combination of Casimirs is
\begin{eqnarray}
\mathcal{F}B_{\phi} & = &  \rho v_{\phi} \,,
\label{eq:sys1}\\
\mathcal{F}B_{z}+\rho\mathcal{G} & = &  \rho v_{z}  \,,
\label{eq:sys2}\\
 \mathcal{H}+\mathcal{F}v_{z}  & = & \frac{B_{z}}{4\pi} \,,
 \\
 \mathcal{J}+v_{z}\mathcal{G} 
 & = &  v_{z}^{2}/2+  v_{\phi}^{2}/2   +c_{s}^{2}\ln ({\rho}/{\rho_{0}}) \,.
 \label{eq:sys4}
\end{eqnarray}

The rigid rotating pinch solution that we have chosen,  has $B_z$ constant and is invariant along $z$,   as  given by  (\ref{GGSSis}), is obtained by choosing 
\begin{align}
\mathcal{F}\left(\hat{\psi}\right) & =\frac{B_{0}}{2\pi \Omega r_0}\left[1-(1-\frac{w^{2}}{2})\,\exp(-w^{2}\hat{\psi})\right]\\
\mathcal{G}\left(\hat{\psi}\right) & =-\Omega r_0 \hat{B}\\
\mathcal{H}\left(\hat{\psi}\right) & =\frac{B_{0}}{4\pi}\hat{B}\\
\mathcal{J}\left(\hat{\psi}\right) & = -c_{s}^{2}\bigg[ w^{2}\hat{\psi}
\nonumber\\
&
- \ln\left[1-(1-{w^{2}}/{2})\,\exp\left(-w^{2}\hat{\psi}\right)\right]\bigg]\,,
 \end{align}
from which by solving the generalized Grad-Shafranov equation we obtain  $\hat{\psi} = - r^2/2$ (or $B_\phi =  B_0 r$) and where,  in  accordance with (\ref{GGSSis}), the  dimensionless variables $\hat{\psi} = \psi/(r_0 B_0)$, $\hat{B}$,  and $w$ are used 
and $r$ is the scaled  radius.

 \subsubsection{Eulerian pinch stability}
 \label{sssec:EpinchStability}

Proceeding as described  in Sec.~\ref{ssec:eulB},   a sufficient stability condition  is  obtained  by considering the second variation of $\mathfrak{F}[Z]$, viz.\ Eq.~\eqref{eq:d2Fgen}. 

Starting from   (\ref{eq:d2Fgen})--(\ref{eq:b3})  we restrict  the coefficients $b_1$, $b_2$, and $b_3$  to depend only on $r$,  because  our pinch equilibrium configuration is  both azimuthally  and translationally symmetric.  For  $b_2$  defined by (\ref{eq:b2}), we obtain 
\bqy
b_{2}&=&\frac{1}{r}\frac{d}{dr}\left[\frac{\partial}{\partial\psi}\left(\frac{\calm^{2}}{4\pi}\right)r\psi_{r}\right]
\nonumber\\
&&\hspace{1 cm} -\frac{\partial}{\partial\psi^{2}}\left(p+\frac{B_{z}^{2}}{8\pi}+\frac{\calm^{2}}{4\pi}B_{\phi}^{2}\right)
\eqy
and, using \begin{equation}
\frac{df}{dr}=\frac{\partial f}{\partial r}+\frac{\partial f}{\partial\psi}\psi_{r}+\frac{\partial f}{\partial\psi_{r}}\frac{d\psi_{r}}{dr},\label{eq:id1}
\end{equation}
and  $\rho =  \rho\left(\psi,\, \psi_r\right)$,  as implicitly given by the
Bernoulli  functional $\mathcal{J}$, $b_2$ becomes 
\bqy
b_{2}&=& \frac{\partial}{\partial\psi}\left(\frac{\calm^{2}}{4\pi} 
\left[1+\frac{1}{\rho}\frac{\partial\rho}{\partial\psi_{r}}B_{\phi}\right]\right)\frac{d\psi_{r}}{dr}
 \label{eq:ib2}\\
&&  \hspace{1cm} -\frac{B_{\phi}}{r}\frac{\partial}{\partial\psi}\left(\frac{\calm^{2}}{4\pi}\right)-\frac{\partial^{2}}{\partial\psi^{2}}\left(p+\frac{B_{z}^{2}}{8\pi}\right)\,.
 \nonumber 
 \eqy
Finally, using  the equilibrium of (\ref{GGSS1}), we obtain 
\begin{align} 
\label{use}
b_{2} & =-\frac{1}{r^{2}}\frac{1-\calm^{2}}{4\pi}+\frac{1}{rB_{\phi}}\frac{d}{dr}\left(b_{1}r\frac{dB_{\phi}}{dr}\right)\,. 
\end{align}

Before proceeding, let us  consider some special limits.   If the plasma is static, i.e.,  $v_{\phi}=0$, we obtain $b_{1}=1/4\pi$, $b_{3}=0$,  and
\begin{equation}
b_{2}=-\frac{1}{B_{\phi}}\frac{d}{dr}\left[\frac{1}{B_{\phi}}\frac{d}{dr}\left(p+\frac{B_{z}^{2}}{8\pi}\right)\right].
\end{equation}

If $B_{z}=0$,  we obtain 
\begin{align} \label{bz0}
b_{1} & =\frac{1}{4\pi}-\frac{1}{4\pi}\frac{\calm^{2}}{1-\bar{\calm}^{2}},\\
b_{3} & =\frac{1}{4\pi}\frac{\calm^{2}\bar{\calm}^{2}}{1-\bar{\calm}^{2}}\,, 
\end{align}
where $\bar{\calm}^{2}=v_{\phi}^{2}/c_{s}^{2}$ is the gas dynamic Mach number, and
\begin{align}
b_{2} & =\frac{1}{rB_{\phi}}\frac{d}{dr}\left[\frac{\calm^{2}}{4\pi}\left(B_{\phi}-\frac{1}{1-\bar{\calm}^{2}}r\frac{dB_{\phi}}{dr}\right)\right]\\
&\hspace{ 2cm} -\frac{1}{B_{\phi}}\frac{d}{dr}\left(\frac{1}{B_{\phi}}\frac{dp}{dr}\right) \nonumber\,.
\end{align}

Now we return  to our analysis of  $\de^2\mathfrak{F}$ of \eqref{eq:d2Fgen} for the pinch case at hand.   For $M^2 <1$, a sufficient stability condition is provided by  $b_1> 0,\quad b_1 + b_3 > 0$ and $b_2> 0$.
Since $4\pi( b_1 + b_3 )=  1 - M^2(r)  = 1 - w^2 {\hat p} $,\,  we find  that $b_1 + b_3>0$\,  if $w<1$ independently of $B_z$.

Using  (\ref{GGSSis})  in Eqs.~\eqref{eq:b1}, \eqref{eq:b2}, and \eqref{eq:b3}  we find
 \begin{equation}\label{b3z}
4\pi b_3  =  \frac{w^4 {\hat p}\, ( 1 -  { w^2 \hat p}) r^2}{( 1 - w^2  {\hat p})  ( 1 - w^2 r^2)- w^2   \hat{B}^2  } 
\end{equation}
and thus
  \begin{align}\label{b1z}
& 4\pi b_1  =   - 4\pi b_3 +1 - \calm^2  =
\\
& ( 1 - w^2  {\hat p})  \left[ 1 - \frac{w^4 {\hat p}\,  r^2}{( 1 - w^2  {\hat p})\, ( 1 - w^2 r^2)- w^2   \hat{B}^2  }\right]\,.
 \nonumber 
\end{align}
Note that  $\partial b_1/\partial \hat{B}^2 <0$  and  $b_1> 0$ so $1 -w^2( {\hat p} + r^2  + \hat{B}^2) > 0$, 
which reduces  (in agreement with the conditions listed above (\ref{redufull1}))  to
\[
w^2( {\bar  r}^2  + \hat{B}^2) < 1\,.
\]
From $4\pi r^2 b_2 =  - 4 \pi \, b_3 + 4 \pi \,  r \, d b_1/dr $, 
we obtain  
\bqy
 \label{b2z}
4\pi b_2  &=&   - \frac{w^4 {\hat p}\, ( 1 - w^2  {\hat p}) }{( 1 - w^2  {\hat p})  ( 1 - w^2 r^2)- w^2   \hat{B}^2 }  
\\
&&- 2 \frac{d}{d \, r^2}  \left[  w^2  {\hat p} + \frac{w^4 {\hat p}\,( 1 - w^2  {\hat p})  r^2}{( 1 - w^2  {\hat p})( 1 - w^2 r^2)- w^2   \hat{B}^2  }\right] \,.
\nonumber
\eqy
Note that the value of $b_2$ decreases with increasing $\hat{B}^2$ and  that $b_2 >  0$ implies  \begin{equation}\label{b2zth}
w^2  <   \frac{3 + \hat{B}^2 - (1 + 4 \hat{B}^2 + \hat{B}^4)^{1/2}}{4 + \hat{B}^2},
\end{equation}
i.e., $w^2 < 1/2 - (3/8)\, \hat{B}^2$\,  for small $\hat{B}^2$,  and $w^2 < 1/ \hat{B}^2$\,  for large $\hat{B}^2$.  To obtain  (\ref{b2zth})  we  have exploited the fact  that $b_2$ starts to become negative at $r^2 = 0$.

For $\hat{B}^2 = 1$ we find  $w^2\lesssim  0.31$,  which is more restrictive than the condition $w^2\lesssim  0.46$ found in the Lagrangian framework  below (\ref{redufull1}). This result  is consistent  with the expectation (see Ref.~\onlinecite{amp2a})   that 
 energy-Casimir  stability conditions  are  more restrictive than the  Lagrangian stability  conditions.

The Euler-Lagrange equation associated with the extrema of (\ref{eq:d2Fgen}) subject to  the normalization  constraint  of constant $\intx \left(\delta\psi\right)^{2}$   is
\begin{equation}
\mathbf{\nabla}\cdot[b_{1}\, I\,+b_{3}\,(I-{\mathbf{e}}_{\psi}{\mathbf{e}}_{\psi})]\cdot\mathbf{\nabla}\delta\psi\,-\, ( b_{2} -\lambda) \delta\psi=0\,,
\label{Newcomb}
\end{equation}
where $\la$ is the Lagrange multiplier,  $I$ is the identity tensor,  and $(I-{\mathbf{e}}_{\psi}{\mathbf{e}}_{\psi})$
is the projector on the tangent plane to the $\psi$-surfaces. 
Writing $\delta \psi $ as 
\begin{equation} 
\delta \psi  = 
\delta {\hat \psi} (r) \, \exp{(i m\phi)}, \label{Newcomb1}  
\end{equation}
with $m$ the azimuthal wave number, (\ref{Newcomb}) becomes 
\bqy
&&\frac{1}{r} \frac{d}{d r}  \left[ r \,  b_{1}  \frac{d \, \delta {\hat \psi} (r)}{d r}\right]
\label{Newcomb2} \\
&&\hspace{1cm}  - \left[ \frac{m^2}{r^2}\, ({b_1+b_3})+\,( b_{2} - \lambda)\, \right] \delta {\hat \psi} (r) \, = \, 0\,.
\nonumber
\eqy
Note that $b_3$ becomes irrelevant for stability in the case of azimuthally symmetric perturbations.

In terms of $w$,  ${\hat p}(r)$ and $\hat{B}$,  and our shorthand $\varpi=1-w^2\hat{p}$,   (\ref{Newcomb2})  takes the form  
\bqy
&& \frac{1}{r} \frac{d}{d r} 
\left[ 
r \,  \varpi  \left(1 - \frac{w^4 {\hat p}\,  r^2}{\varpi \, ( 1 - w^2 r^2)- w^2   \hat{B}^2  }\right)
 \frac{d \, \delta  \hat{\psi}}{d r}
 \right] 
 \nonumber\\
&&
 - \Bigg[
 \varpi\, \frac{m^2}{r^2}  - \frac{\lambda}{4\pi}   + 
 \frac{w^4 \hat{p}\, \varpi}{\varpi ( 1 - w^2 r^2)- w^2   \hat{B}^2 }
 \label{NC-2}\\
  &&\hspace{1 cm} + 2 \frac{d}{d \, r^2}  \bigg(  w^2  {\hat p} 
 +    \frac{w^4 {\hat p}\,\varpi  r^2}{\varpi( 1 - w^2 r^2)- w^2   \hat{B}^2  }\bigg)  \Bigg] \, \delta {\hat \psi}  \, = \, 0\, . 
\nonumber
\eqy
Searching for the lowest eigenvalue of the Lagrange multiplier $\lambda$  as a function of $w$  in the range 
 \begin{equation}
 \label{range}
  \frac{3 + \hat{B}^2 - (1 + 4 \hat{B}^2 + \hat{B}^4)^{1/2}}{4 + \hat{B}^2} < w^2 < \frac{1}{ {\bar r}^2  + \hat{B}^2}
\end{equation}
would yield  a more  accurate  sufficient stability condition that could be  compared with the one obtained by solving the constrained Euler-Lagrange equation derived from the functional (\ref{redufull1}).  We leave it here and continue on to discuss dynamically accessible stability. 


\subsection{Dynamically accessible pinch}
\label{ssub:DApinch}
\subsubsection{Dynamically accessible pinch equilibria}
\label{sssec:DApinchEquilbria}

As discussed in Sec.~\ref{ssec:Daform}, with the dynamically accessible approach one considers the constrained variations of Eqs.~\eqref{eq:var_0_1}--\eqref{eq:var_0_4}.  Upon evaluating these expressions on the pinch equilibrium of this section, expressed by \eqref{MHDsol}--\eqref{MHDsolB},  it is straightforward  to show that $\de H_{\mathrm{da}}$  of \eqref{daH} vanishes.  For example, vanishing of the coefficients of $g_2$ and  $g_3$  give immediately that $\rho(r)s(r)v_\phi(r)$ and  $r \rho(r)v_\phi(r)$ are constant. Evaluation of the coefficients of $\mathbf{g}_1$ and $\mathbf{g}_4$ are more tedious, but must vanish since we have shown in general that \eqref{daH} gives all equilibria.

\subsubsection{Dynamically accessible pinch stability}
\label{sssec:DApinchStability}

Given that $\de H_{\mathrm{da}}=0$ we can proceed to examine  $\delta^{2}H_{\mathrm{da}}$ of \eqref{d2H_da} with the variations of \eqref{eq:var_0_1}--\eqref{eq:var_0_4} evaluated on our rotating  pinch equilibrium.  Rather than starting from scratch we will appeal to our results already obtained in Ref.~\onlinecite{amp2a}. 

For a translationally symmetric equilibrium  along the $z$-direction, the stability condition  derived from  dynamically  accessible variations may or may not coincide with that obtained in terms of the Lagrangian variations\cite{hameiri03,amp2a}.    Starting from  Eq.~(103) of Ref.~\onlinecite{amp2a}  with $\mathbf {h} = \mathbf{e}_z$ , $k=1$,  the crucial  quantity for translationally symmetric equilibria  is 
\begin{equation}
\label{crucial}
\boldsymbol{\Gamma}=\begin{bmatrix}\left\langle 2\mathbf{B}\cdot\left(\mathbf{v}\cdot\nabla\mathbf{g}_{1}\right)\right\rangle \\ 
\left\langle {\rho}\mathbf{v}_{\bot}\cdot\nabla g_{1z}
\right
\rangle 
\end{bmatrix}\,,
\end{equation}
where $\left\langle ~\right\rangle =\int_{\psi} {d^{2}x}  / |\nabla\psi| $ denotes surface integral  over a flux surface.
If  the expression of \eqref{crucial} vanishes, the two kinds of stability coincide.  

The first stabilizing term in  $\delta^{2}H_{\mathrm{da}}$ of \ref{d2H_da}, which can be eliminated in $\delta^{2}H_{\mathrm{la}}$ by minimizing over Lagrangian variations, here becomes 
\begin{equation}
\Delta=\int d^{3}x\,\rho\left|\mathbf{X}\right|^{2},
\end{equation}
where
\begin{align}
&\mathbf{X}:=\nabla g_{3}+\frac{\sigma}{\rho}\nabla g_{2}+\mathbf{v}\times\left(\nabla\times\mathbf{g}_{1}\right)\\ &+2\left(\mathbf{v}\cdot\nabla\right)\mathbf{g}_{1}+\frac{1}{\rho}\mathbf{B}\times\left(\nabla\times\mathbf{g}_{4}\right), \nonumber
\end{align}
and this term is  minimum for
\begin{equation}
\mathbf{X}_{\min}=({\Xi_{1}}/{\rho})\, \mathbf{B}+{\Xi_{2}}\, \mathbf{e}_z,
\end{equation}
where $\boldsymbol{\Xi}=\mathbb{A}^{-1}\boldsymbol{\Gamma}$, i.e.
\begin{equation}
\begin{bmatrix}\Xi_{1}\\
\Xi_{2}
\end{bmatrix}=\begin{bmatrix}\left\langle {\left|\mathbf{B}\right|^{2}}/{\rho}\right\rangle  & \left\langle {B_{z}}\right\rangle \\
\left\langle  {B_{z}} \right\rangle  & 1 
\end{bmatrix}^{-1}\begin{bmatrix}\Gamma_{1}\\
\Gamma_{2}
\end{bmatrix}.
\end{equation}

For our rotating pinch example  we obtain
\begin{equation}
\mathbb{A}=
4\pi h\begin{bmatrix} \left(B_{0}\left(r^{2}+b^{2}\right)\right)/{\rho} & \, \,  b\\
b & \, \, {1}/B_{0}
\end{bmatrix},
\end{equation}
where $\pm h$ is the height of the plasma column in the $\pm z$-directions;  ideally $h\rightarrow\infty$ but it cancels and does not appear in the result.  Finally 
\begin{equation}
\boldsymbol{\Gamma}=\begin{bmatrix}\left\langle 2\mathbf{B}\cdot\left(\mathbf{v}\cdot\nabla\mathbf{g}_{1}\right)\right\rangle \\
\left\langle \rho\mathbf{v}_{\bot}\cdot\nabla g_{1z}\right\rangle 
\end{bmatrix}=\begin{bmatrix}rVB_{0}\left\langle g_{1r}\right\rangle \\
0
\end{bmatrix}.
\end{equation}

It can be noted on general grounds that $\left\langle g_{1r}\right\rangle$ vanishes identically  for 
perturbations that average to zero after integration over the azimuthal angle (i.e.,   that do not contain an $m = 0$ component).  Since  in Sec.~\ref{sssec:LpinchStab} we have shown that  for  our  rotating pinch  example   azimuthally symmetric perturbations of our rotating pinch equilibrium are stable to Lagrangian perturbations, thus  the  restriction to dynamically accessible perturbations does not modify the stability condition.  However, for general equilibria  this is not true.


\subsection{Pinch comparisons}
\label{ssec:PinCom}

Let us now summarize and compare our three stability approaches for the rotating pinch  equilibria.   In order  to compare  the Lagrangian  and the dynamically accessible stability conditions with those obtained in the energy-Casimir framework,   it is necessary to  restrict our analysis to perturbations $\boldsymbol{\eta }$ that do not depend on $z$. This excludes  ``sausage'' or kink type instabilities.  The  results of the stability analysis for such perturbations can be  expressed as  stability  bounds on the normalized rotation frequency $w$. These bounds
are modified by  the presence of an equilibrium  magnetic field  along the symmetry  direction,  $B_z$, that    couples the  component  $\eta_z$ to the other components of the displacement leading in general to stricter bounds.

For the equilibrium  under examination,  the Lagrangian and the dynamically accessible approaches lead to  equivalent   conditions.  Although  the constraints  obeyed  by the dynamically accessible perturbations in the presence of  flows lead  to an additional  stabilizing term  that cannot be made to vanish for azimuthally symmetric perturbations,  this term does not modify the stability analysis since  azimuthally symmetric perturbations are found to be  stable even within the Lagrangian framework.  For more general equilibria than the ones considered her, this need not be the case. 

The minimization of $\delta^2 W_{\mathrm{la}}$ of  \eqref{FLdW} for our pinch case reduced to the study  of the  $3\times 3$ matrix of \eqref{ww} (the  $4\times 4$ matrix  for $B_z\not= 0$ of \eqref{Wmatrix3})  for  $|m|=1$ perturbations. Two different methods can be used: a  necessary and sufficient condition for the positivity of this matrix is provided by the Sylvester criterion  which  yields 
$w^2 < 1/2$  for $B_z= 0$  and  $w^2 B_z^2 < 1$ for  $B_z\not= 0$ and  $w^2 \to 0$. A partial minimization  procedure with respect to  $\eta_\phi$ (to $\eta_z$ and $\eta_\phi$ for $B_z\not= 0$) leads to less restrictive  conditions:  $w^2\lesssim 0.62$  for $B_z= 0$   and  $w^2\lesssim 0.46$  choosing,   e.g.,  $B_z^2= 1$.

Extremization of the  energy-Casimir  functional over all variables except  $\delta \psi$ leads  to sufficient stability bounds on $w^2$   that, similarly to the Lagrangian case,  become stricter  as $B_z^2$ increases.
As predicted in Ref.~\onlinecite{amp2a} and recalled in Sec.~\ref{sec:basics},   these bounds   are  in general  more restrictive  than  those found within the Lagrangian framework, as shown, e.g.,  by considering again $B_z^2= 1$,   in which case we find  $w^2\lesssim 0.31$. Sharper stability  conditions could  be obtained by solving the Euler-Lagrange equation associated  with this reduced  energy-Casimir functional  subject to a normalization  constraint  on   $\delta \psi$.

\section{Conclusions}
\label{sec:conclusions}
 
 To summarize, we have investigated MHD stability in the Lagrangian, Eulerian, and dynamically accessible approaches.  In Sec.~\ref{sec:basics} we reviewed general properties,  in particular, the time-dependent relabeling idea introduced in Ref.~\onlinecite{amp2a} that gives Eulerian stationary equilibria as a static state in terms of a relabeled Lagrangian variable.  New details on the general comparison of the three approaches was given in Sec.~\ref{ssec:compform}.  Then we proceeded to our two examples, the convection problem of Sec.~\ref{sec:convection} and the rotating pinch of Sec.~\ref{sec:pinch}, with comparison of the stability results for  the three methods given in Secs.~\ref{ssec:PinCom} and \ref{ssec:ConCom}, respectively.  Of note,  is the explicit incorporation  of the time-dependent relabeling for the rotating pinch, which to our knowledge is  the first time this has been done.

 As noted previously,  the methods described here for the three approaches are of general utility -- they  apply to all important plasma models, kinetic as well as fluid, when dissipation is neglected.   In fact, some time ago in Refs.~\onlinecite{MP89,MP90} the approaches were compared for the Vlasov and guiding-center kinetic equations (see also Refs.~\onlinecite{throum94,throum96,pfirsch04a,pfirsch04b}), including a dynamically accessibly calculation in this kinetic context akin to the one done here and  in Refs.~\onlinecite{hameiri03,amp2a} for MHD.  Given the large amount of recent progress on extended magnetofluid models,\cite{kimura,keramidas14,LMT15,HKY15,LMM15,LMM16,MLA14,DMP15,KMMA15} hybrid kinetic-fluid models,\cite{tronci10,pjmTTC14} and gyrokinetics\cite{BBMQ15,BMBGV65} a great many  stability calculations like the ones of this paper are now possible.

 \appendix


\section{Lagrangian Equations of Motion and Rotating  Pinch Equilibria}
\label{app:LanEOM}

 In order to obtain the MHD equations of motion from the Hamiltonian of (\ref{H_Lagr}), as described in Sec.~\ref{ssec:lagB},   we split $H$  into two terms $H=H_F+ H_B$ where $H_F$ is sum of the fluid kinetic and internal energies and  $H_B$ is the magnetic field energy given by 
\bq
H_B=\int d^{3}a\,  \frac{\partial q_{i}}{\partial a^{j}}\frac{\partial q^{i}}{\partial a^{k}}\frac{B_{0}^{j}B_{0}^{k}}{8\pi {\mathcal{J}}}\,.
 \label{eq:HB}
\eq
The functional derivative of $H_F$ is given by (see Ref.~\onlinecite{morrison98} for details)
\bq
\frac{\delta H_F}{\delta q^{i}}=
\frac{\pi_{n}\,\pi_{m}}{2\rho_{0}} \frac{\partial g^{nm}}{\partial q^{i}}  
+ \frac{\partial}{\partial a^{m}} \bigg[
 \left(\frac{\rho_{0}}{{\mathcal{J}}}\right)^{\!2}\! U_{\rho} \, 
\frac{\partial {\mathcal{J}}}{\partial q_{,m}^{j}} \, 
\bigg]
 \label{first}\,.
\eq
Using
\begin{equation}
\frac{\partial {\mathcal{J}}}{\partial q_{,m}^{i}}= 
A_{i}^{m}= \epsilon_{ijk}\epsilon^{mnl} \frac{1}{2}\frac{\partial q^{j}}{\partial a^{n}}\frac{\partial q^{k}}{\partial a^{l}}
\label{eq:d{{J}}_dq_m}
\end{equation}
and 
\begin{equation}
\frac{\partial A_{i}^{m}}{\partial a^{m}} =   \frac{\partial}{\partial a^{m}}\epsilon_{ijk}\epsilon^{mnl} \frac{1}{2}\frac{\partial q^{j}}{\partial a^{n}}\frac{\partial q^{k}}{\partial a^{l}} = 0\,,
 \label{eq:ident1}
\end{equation}
we can rewrite Eq. (\ref{first}) as
\bq
\frac{\delta H_F}{\delta q^{i}}=\frac{\pi_{n}\,\pi_{m}}{2\rho_{0}} \frac{\partial g^{nm}}{\partial q^{i}}  
+
A_{i}^{~m}\frac{\partial}{\partial a^{m}}\left[ 
\left(\frac{\rho_{0}}{{\mathcal{J}}}\right)^{\!2}\!U_{\rho}\right] \,.
\label{HFq}
\eq
Similarly for \eqref{eq:HB} we obtain
\bqy
\frac{\delta H_B}{\delta q^{i}}&=& 
\frac{\partial g_{lm}}{\partial q^{i}}\frac{\partial q^{l}}{\partial a^{j}}\frac{\partial q^{m}}{\partial a^{k}}\frac{B_{0}^{j}B_{0}^{k}}{8\pi {\mathcal{J}}}
 -\frac{\partial}{\partial a^{j}}\left( 
g_{im} \frac{\partial q^{m}}{\partial a^{k}}\frac{B_{0}^{j}B_{0}^{k}}{4\pi {\mathcal{J}}}\right) 
\nonumber\\
&&  +\frac{\partial}{\partial a^{t}}\left(
g_{lm}
\frac{\partial q^{l}}{\partial a^{j}}\frac{\partial q^{m}}{\partial a^{k}}\frac{B_{0}^{j}B_{0}^{k}}{8\pi {\mathcal{J}}^{2}}\frac{\partial {\mathcal{J}}}{\partial q_{,t}^{i}}\right)\,,  \label{eq:dmag}
\eqy
and the Lagrangian equations of motion are given by 
\bq
\dot{\pi_i}=-\frac{\delta H}{\delta q^{i}}= -\frac{\delta H_F}{\delta q^{i}}-\frac{\delta H_B}{\delta q^{i}}\,, 
\label{dotpiA}
\eq
with \eqref{first} and \eqref{eq:dmag}, and 
\bq
\dot{q^i}=-\frac{\delta H}{\delta \pi_{i}}=  \frac{\pi^i}{\rho_0} = g^{ij}\frac{\pi_j}{\rho_0}\,.
\label{dotqA}
\eq
Note that the first terms of  \eqref{first} and \eqref{eq:dmag} give the effect of non-cartesian coordinates. 

To obtain from \eqref{dotpiA} and \eqref{dotqA} the Eulerian form of the equations of motion it is convenient to recall that 
 the cofactor matrix $A_{k}^{i}$ satisfies  the identity 
\[
\delta_{j}^{i}{\mathcal{J}}= 
\frac{\partial q^{k}}{\partial a^{j}}A_{k}^{i}
\]
and consequently
 \[
\frac{\partial}{\partial q^{k}}=\frac{\partial a^{i}}{\partial q^{k}}\frac{\partial}{\partial a^{i}}= 
\frac{A_{k}^{~i}}{{\mathcal{J}}}\frac{\partial}{\partial a^{i}}.
\]
where ${\partial}/{\partial q^{k}}$ becomes $\nabla$ in the Eulerian description. Using $p=\rho^2 U_{\rho}$,   
the second term of \eqref{HFq} becomes the pressure force, and using
the flux conservation expression,
 \begin{equation}
B^{i}=\frac{\partial q^{i}}{\partial a^{k}}\frac{B_{0}^{k}}{{\mathcal{J}}} ,
\end{equation}
the last two terms of \eqref{eq:dmag} become 
\bq
\label{last}
-{\mathcal{J}}B^{j} \frac{\partial}{\partial q^{j}}\left(\frac{B_{i}}{4\pi}\right) 
+{\mathcal{J}}\frac{\partial}{\partial q^{i}}\left(\frac{B^{2}}{8\pi}\right) \,,
\eq
where we used the divergence equation ${\partial B_{0}^{j} }/{\partial a^{j}} =0$.

To facilitate our calculation of the rotating pinch equilibrium  (cf.\ Appendix \ref{app:atrt}) consider  the  cylindrical pinch geometry where 
the metric is given by \eqref{Metric}. Evidently,
\begin{equation}
\frac{\partial g^{nm}}{\partial q^{i}}=-\delta^{n}_{\phi}\, \delta^{m}_{\phi}\, \delta_{i}^{r}\frac{2}{(q^{r})^3}.
\end{equation}
and consequently
\bq
\frac{\pi_{n}\,\pi_{m}}{2\rho_{0}} \frac{\partial g^{nm}}{\partial q^{i}}  = 
-\delta_{i}^{r}\, \frac{\pi^{\phi} \pi_{\phi}}{q^{r}\rho_{0}}
\label{metriK}
\eq
and the first  term of Eq.~(\ref{eq:dmag}) is 
\begin{equation}
\delta_{i}^{r}\frac{g_{\phi\phi}}{q^{r}}\frac{\partial q^{\phi}}{\partial a^{j}}\frac{\partial q^{\phi}}{\partial a^{k}}\frac{B_{0}^{j}B_{0}^{k}}{4\pi {\mathcal{J}}} = \delta_{i}^{r}\frac{{\mathcal{J}}}{q^r}\frac{B^{\phi} B_{\phi}}{4\pi}\,. 
\label{metriB}
\end{equation}
Expressions \eqref{metriK}  and \eqref{metriB}  are of use for our equilibrium calculation.


\section{Relabeling transformation for the Pinch}
\label{app:atrt}

The canonical transformation induced by the time-dependent relabeling is generated by the functional 
\[
F \left[\mathbf{q},\boldsymbol{\Pi},t\right]=\int d^{3}a\int d^{3}b\,\,\mathbf{q}\cdot\boldsymbol{\Pi}\,\delta\left(\mathbf{a}-\mathfrak{A}\left(\mathbf{b},t\right)\right) ,
\]
and yields (see Eq.(9) of Ref.\onlinecite{amp2a})  the new Hamiltonian of \eqref{add} according to 
\[
\tilde{H}[\mathbf{Q},\boldsymbol{\Pi}]= H  +\frac{\partial F }{\partial t}\, ,
\]
with 
$\mathbf{V}\left(\mathbf{b},t\right)\rightarrow V^{\phi}(b,t)=b^{r}\Omega(b^{r})
$ 
for the  relabeling defined by Eq.(\ref{relab}). 

With an integration by parts involving the time derivatives of the delta functions  we obtain 
\begin{align}
&\frac{\partial F}{\partial t}\!
=\hspace{-1 mm}\int \hspace{-1 mm} d^{3}b\!\! \int  \hspace{-1 mm} da^{r}da^{\phi}da^{z} \delta(a^{r}-\mathfrak{A}^{r})\delta(a^{\phi}-\mathfrak{A}^{\phi})\delta(a^{z}-\mathfrak{A}^{z})
 \nonumber  \\
& 
\!\times\!\left[
\partial_{t}\mathfrak{A}^{r}\! \frac{\partial}{\partial a^{r}} \! \left(\mathbf{q}\cdot\boldsymbol{\Pi}\right) 
+\partial_{t}\mathfrak{A}^{\phi}\! \frac{\partial}{\partial a^{\phi}} \! \left(\mathbf{q}\cdot\boldsymbol{\Pi}\right) 
+ \partial_{t}\mathfrak{A}^{z}\! \frac{\partial}{\partial a^{z}} \!\left(\mathbf{q}\cdot\boldsymbol{\Pi}\right)\right]
\nonumber  
\end{align}
where $\partial_{t} $ denotes time derivative at constant label $b$. 
Using $\mathbf{Q}\left(\mathfrak{B}\left(\mathbf{a},t\right),t\right)=\mathbf{q}\left(\mathbf{a},t\right)$,  the first term in the bracket $\times[\quad]$  above becomes 
\begin{align}
&\partial_{t}\mathfrak{A}^{r}\frac{\partial}{\partial a^{r}}\left(\mathbf{q}\cdot\boldsymbol{\Pi}\right)  
=\Pi_{r}\partial_{t}\mathfrak{A}^{r}\frac{\partial{Q^r}}{\partial b^{i}}\frac{\partial\mathfrak{B}^{i}}{\partial a^{r}}\\ 
&+\Pi_{\phi}\partial_{t}\mathfrak{A}^{r}\frac{\partial{Q^\phi}}{\partial b^{i}}\frac{\partial\mathfrak{B}^{i}}{\partial a^{r}}+\Pi_{z}\partial_{t}\mathfrak{A}^{r}\frac{\partial{Q^z}}{\partial b^{i}}\frac{\partial\mathfrak{B}^{i}}{\partial a^{r}}\,.
\nonumber
\end{align}
Similar expressions follow for the other two terms. 
Collecting all the terms proportional to  $\Pi_{r}$,  we obtain
\begin{align}
&\left[\partial_{t}\mathfrak{A}^{r}\frac{\partial\mathfrak{B}^{i}}{\partial a^{r}}+
\partial_{t}\mathfrak{A}^{\phi} \frac{\partial\mathfrak{B}^{i}}{\partial a^{\phi}}  + 
\partial_{t}\mathfrak{A}^{z}\frac{\partial {\mathfrak{B}^{i}}}{\partial a^{z}}\right]\Pi_{r}  \frac{\partial }{\partial b^{i}}{Q^r}
 \nonumber\\ 
& \hspace{1cm} = - \Pi_{r} \cdot \dot{\mathfrak{B}}^{i}   \frac{\partial}{\partial b^{i}} {Q^r}\,,
\end{align}
where we used the identity 
\[
\dot{\mathfrak{B}}^{i}+\frac{\partial\mathfrak{B}^{i}}{\partial a^{r}}\partial_{t}\mathfrak{A}^{r}+\frac{\partial\mathfrak{B}^{i}}{\partial a^{\phi}}\partial_{t}\mathfrak{A}^{\phi}+\frac{\partial\mathfrak{B}^{i}}{\partial a^{z}}\partial_{t}\mathfrak{A}^{z}=0\,. 
\]
Finally,   employing  \eqref{VR}, 
\[
V^{r}=\dot{\mathfrak{B}}^{r},\quad V^{\phi}= \dot{\mathfrak{B}}^{\phi},\quad V^{z}=\dot{\mathfrak{B}}^{z}, 
\]
we obtain
$
{\partial F}/{\partial t}= - \int d^{3}b\,\,\left[\left(\mathbf{V}\cdot\nabla_{b}\mathbf{Q}\right)_{r}\Pi_{r} \right]\,.
$

With this additional term  in the Hamiltonian (\ref{tildeH_MHD}),  (\ref{LCq})  and  (\ref{LCpi}) become
\begin{align}
 \label{Erelab1}
 & \partial_{t}{Q}^{i}=\frac{\delta \tilde{H}}{\delta\Pi_{i}}=g^{ij}\frac{\Pi_{j}}{{\tilde\rho}_{0}}-V^{k}\frac{\partial Q^{i}}{\partial b^{k}}\,,
 \\
 &\text{and}\nonumber\\
&\partial_{t}\Pi_{i} = -  \frac{\delta \tilde{H}}{\delta Q^{i}} = \delta_{i}^{r}\frac{\Pi_{\phi} \Pi^{\phi}}{Q^{r}{\tilde \rho}_{0}}  
- {\tilde J}\frac{\partial}{\partial Q^{i}} 
\left[\left(\frac{{\tilde \rho}_{0}}{{\tilde J}}\right)^{2}\!\!U_\rho\right]
 \nonumber\\
 &
 \hspace{1cm} - \delta_{i}^{r}\frac{{\tilde J}}{Q^{r}}\frac{{\tilde B}_{\phi}{\tilde B}^{\phi}}{4\pi}
 +{\tilde J {\tilde B}^{j}}\frac{\partial}{\partial Q^{j}}\! \! \left(\frac{{\tilde B}_{i}}{4\pi}\right) 
 \nonumber\\
 & 
 \hspace{1cm} - {\tilde J}\frac{\partial}{\partial Q^{i}} \!\!  \left(\frac{{\tilde B}^{2}}{8\pi}\right)-\frac{\partial}{\partial b^{k}}\left(V^{k}\Pi_{i}\right) \,.
\label{Erelab2}
 \end{align} 
 
By assuming 
\[
 \tilde{B}_{0}^{r}\left(\mathbf{b},t\right)=0\,, 
\qquad \tilde{B}_{0}^{z}\left(\mathbf{b},t\right)=0\,, 
\]
and
\[
 \tilde{B}_{0}^{\phi}\left(\mathbf{b},t\right)=\mathfrak{J}B_{0}^{\phi}\left(\mathfrak{A}\left(\mathbf{b},t\right)\right)=B_{0}b^{r}\,,
\]
relabeled  equilibria are obtained by setting $\partial_{t}Q^{i}=0$, $\partial_{t}\Pi_{i} =0$, 
and $Q^{i}=b^{i}$ in  Eqs.~(\ref{Erelab1}) and \eqref{Erelab2}, which
yields
\begin{equation}
\Pi_{r}=\tilde{\rho}_{0}V^{r}, 
\quad  \Pi_{\phi}=(b^{r})^2\tilde{\rho}_{0}V^{\phi}, 
\quad   \Pi_{z}=\tilde{\rho}_{0}V^{z}\,,
\label{eq:Qdot-1-1}
\end{equation}
and 
\begin{align}
0 & =\frac{\Pi^{\phi} \Pi_{\phi}}{b^{r}\tilde{\rho}_{0}}-\frac{\partial}{\partial b^{r}}\left(\tilde{\rho}_{0}^{2}U_{\rho}\right)-\frac{\tilde{B}_{0}^{\phi}\tilde{B}_{0 {\phi}}}{4\pi b^{r}}-\frac{\partial}{\partial b^{r}}\left(\frac{\tilde{B}_{0}^{\phi}\tilde{B}_{0 {\phi}}}{8\pi}\right) \nonumber \\
& -\frac{\partial}{\partial b^{k}}\left(V^{k}\Pi_{r}\right), \label{eq:PIdot-bis-1}\\
0 & =-\frac{\partial}{\partial b^{\phi}}\left(\tilde{\rho}_{0}^{2}U_{\rho}\right)- \frac{\partial}{\partial b^{k}}\left(V^{k}\Pi_{\phi}\right),\label{eq:PIdot-bis-2}\\
0 & =-\frac{\partial}{\partial b^{z}}\left(\tilde{\rho}_{0}^{2}U_{\rho}\right)-\frac{\partial}{\partial b^{k}}\left(V^{k}\Pi_{z}\right)\,,
\label{eq:PIdot-bis-3}
\end{align}
where we used the fact that $\tilde{J}=1$.

If we consider only equilibria with both axial and translational symmetries, i.e. $\partial/\partial b^{\phi}=0$
and $\partial/\partial b^{z}$, then by substituting (\ref{eq:Qdot-1-1})
into   (\ref{eq:PIdot-bis-2}) and (\ref{eq:PIdot-bis-3}),  we obtain
\begin{equation}
 \label{relab1}
\frac{\partial}{\partial b^{r}}\left(\frac{\Pi_{\phi}\Pi_{r}}{\tilde{\rho}_{0}}\right)=0 \quad \text{and}\quad 
\frac{\partial}{\partial b^{r}}\left( \frac{\Pi_{z}\Pi_{r}}{\tilde{\rho}_{0}}\right)=0\,,
\end{equation}
which have the trivial solution $\Pi_{r}=0$. If we assume a uniform
temperature $\tilde{T}_{0}$ and an initial density field $\tilde{\rho}_{0}=\tilde{\rho}_{0}\left(b^{r}\right)$, 
such that the pressure
$
p\left(\tilde{\rho}_{0}\right)=\tilde{\rho}_{0}^{2}U_{\rho}\left(\tilde{T}_{0},\tilde{\rho}_{0}\right)
$
is the one given in  (\ref{isoteq}),   (\ref{eq:PIdot-bis-1}) results. 
Equation  (\ref{eq:PIdot-bis-1})  can be solved for $\Pi_{\phi}$ and, consequently, written in
terms of the relabeling velocity
$
V^{\phi}=\Pi_{\phi}/\left((b^{r})^2\tilde{\rho}_{0}\right)
$
in agreement with Sec.~\ref{sec:pinch}.

\section{Pinch Details}
\label{app:pinch}

Here we record some formulas needed  for the stability development of Sec.~\ref{sssec:LpinchStab}.   We use   $^*$  to denote  the complex conjugate and  $c.c.$\  to denote the complex conjugate of the  preceding term.  
From Eqs.~\eqref{dw1}--\eqref{dw6}  we obtain for the $m^{th}$ component of these equations,  the following  five terms: 
\bqy
&&\rho\left [\left({\bf v}_{\phi} \cdot\nabla {\bf v}_{\phi}  \right)  \cdot\left(  \boldsymbol{\eta}\cdot \nabla
\boldsymbol{\eta}\right) -\, |\left({\bf v}_{\phi} \cdot\nabla \boldsymbol{\eta}\right)|^2\right]_{|m|} \rightarrow 
\label{dw1c} \\
&&\hspace{2cm}  - w^2 {\hat p} \, [ \, (\eta^*_r \, \partial _r  \eta_r  + c.c)/2
\nonumber\\
&&\hspace{3cm} - i\, 3m/\, (\eta_\phi ^* \eta _r - c.c.)/2
  \nonumber \\ 
&& \hspace{4cm} + m^2 (|\eta_\phi|^2+ |\eta_r|^2 +|\eta_z|^2)]\,, 
\nonumber \\  
&& p |\left(  \mathbf{\nabla}\cdot\boldsymbol{\eta}\right)^{2} |_{|m|} \rightarrow   
 \\
 &&\hspace{2cm} ({\hat p} /r^2) [ \, |\partial _r (r \eta_r)|^2   + \, m^2 |\eta_\phi|^2 
 \nonumber\\
 && \hspace{3cm} - i \, m [\eta_\phi^*\partial _r (r \eta_r) - c.c.] \,] \,, 
 \nonumber\\
  && [(\eta_r \, \partial_r p) \left(  \mathbf{\nabla}\cdot \boldsymbol{\eta }\right)]_{|m|} \rightarrow   
\\ 
&&\hspace{2cm} [(\eta^*_r/r) \,  (\partial_r {\hat p})   [\partial _r (r \eta_r)  + i \,m \, \eta_\phi] +  c.c. \, ]/2 \,,
 \nonumber\\ 
&& |\mathbf{\nabla}\times\left(\boldsymbol{\eta}\times
\mathbf{B}\right)|^2_{|m|} \rightarrow   
\\
&&\hspace{2cm} 
  |(\partial _r (r \eta_r)|^2  \,  + m^2( |\eta_r| ^2 +  |\eta_z ^2 )]
  \nonumber\\
&&\hspace{3cm}  + (\hat{B}^2/r^2)\,  [ \,  |(\partial _r (r \eta_r)|^2  \, + m^2 |\eta_\phi| ^2
\nonumber\\
&&\hspace{3.5cm}  - i\, m (\eta_\phi \partial _r (r \eta^*_r)  -c.c. ) \, ]  \nonumber \\ 
&&\hspace{4cm} + (\hat{B}/r)\,  [ \,   i\, m  (\eta_z^* \partial _r (r \eta_r)  -c.c. ) 
 \nonumber\\
 &&\hspace{4.5cm}  - m^2\, (\eta_z^* \eta_\phi + c.c.)]\,, 
  \nonumber \\
&&\mathbf{J}\times\boldsymbol{\eta}\cdot\delta {\bf B}  \rightarrow   
 \label{dw6c} 
\\
 && \hspace{1cm}
- [ \eta_r^* \partial _r (r \eta_r)  + c.c.  + i\, m ( \eta^*_\phi \eta_r  - c.c.) ]  \,.
\nonumber
\eqy

 
\section*{Acknowledgment}

\noindent PJM was supported by U.S. Dept.\ of Energy  under contract \#DE-FG02-04ER-54742.  He would also like to acknowledge support from the Humboldt Foundation and the hospitality of the Numerical Plasma Physics Division of the IPP, Max Planck, Garching.

\bibliography{amp2b}

\begin{thebibliography}{52}%
\makeatletter
\providecommand \@ifxundefined [1]{%
 \@ifx{#1\undefined}
}%
\providecommand \@ifnum [1]{%
 \ifnum #1\expandafter \@firstoftwo
 \else \expandafter \@secondoftwo
 \fi
}%
\providecommand \@ifx [1]{%
 \ifx #1\expandafter \@firstoftwo
 \else \expandafter \@secondoftwo
 \fi
}%
\providecommand \natexlab [1]{#1}%
\providecommand \enquote  [1]{``#1''}%
\providecommand \bibnamefont  [1]{#1}%
\providecommand \bibfnamefont [1]{#1}%
\providecommand \citenamefont [1]{#1}%
\providecommand \href@noop [0]{\@secondoftwo}%
\providecommand \href [0]{\begingroup \@sanitize@url \@href}%
\providecommand \@href[1]{\@@startlink{#1}\@@href}%
\providecommand \@@href[1]{\endgroup#1\@@endlink}%
\providecommand \@sanitize@url [0]{\catcode `\\12\catcode `\$12\catcode
  `\&12\catcode `\#12\catcode `\^12\catcode `\_12\catcode `\%12\relax}%
\providecommand \@@startlink[1]{}%
\providecommand \@@endlink[0]{}%
\providecommand \url  [0]{\begingroup\@sanitize@url \@url }%
\providecommand \@url [1]{\endgroup\@href {#1}{\urlprefix }}%
\providecommand \urlprefix  [0]{URL }%
\providecommand \Eprint [0]{\href }%
\providecommand \doibase [0]{http://dx.doi.org/}%
\providecommand \selectlanguage [0]{\@gobble}%
\providecommand \bibinfo  [0]{\@secondoftwo}%
\providecommand \bibfield  [0]{\@secondoftwo}%
\providecommand \translation [1]{[#1]}%
\providecommand \BibitemOpen [0]{}%
\providecommand \bibitemStop [0]{}%
\providecommand \bibitemNoStop [0]{.\EOS\space}%
\providecommand \EOS [0]{\spacefactor3000\relax}%
\providecommand \BibitemShut  [1]{\csname bibitem#1\endcsname}%
\let\auto@bib@innerbib\@empty
\bibitem [{\citenamefont {Newcomb}(1962)}]{newcomb1962}%
  \BibitemOpen
  \bibfield  {author} {\bibinfo {author} {\bibfnamefont {W.}~\bibnamefont
  {Newcomb}},\ }\href@noop {} {\bibfield  {journal} {\bibinfo  {journal}
  {Nuclear Fusion Supp.}\ }\textbf {\bibinfo {volume} {2}},\ \bibinfo {pages}
  {451} (\bibinfo {year} {1962})}\BibitemShut {NoStop}%
\bibitem [{\citenamefont {Morrison}\ and\ \citenamefont
  {Greene}(1980)}]{morrison80}%
  \BibitemOpen
  \bibfield  {author} {\bibinfo {author} {\bibfnamefont {P.~J.}\ \bibnamefont
  {Morrison}}\ and\ \bibinfo {author} {\bibfnamefont {J.~M.}\ \bibnamefont
  {Greene}},\ }\href@noop {} {\bibfield  {journal} {\bibinfo  {journal} {Phys.
  Rev. Lett.}\ }\textbf {\bibinfo {volume} {45}},\ \bibinfo {pages} {790}
  (\bibinfo {year} {1980})}\BibitemShut {NoStop}%
\bibitem [{\citenamefont {Andreussi}, \citenamefont {Morrison},\ and\
  \citenamefont {Pegoraro}(2010)}]{amp0}%
  \BibitemOpen
  \bibfield  {author} {\bibinfo {author} {\bibfnamefont {T.}~\bibnamefont
  {Andreussi}}, \bibinfo {author} {\bibfnamefont {P.~J.}\ \bibnamefont
  {Morrison}}, \ and\ \bibinfo {author} {\bibfnamefont {F.}~\bibnamefont
  {Pegoraro}},\ }\href@noop {} {\bibfield  {journal} {\bibinfo  {journal}
  {Plasma Phys. Contr. Fusion}\ }\textbf {\bibinfo {volume} {52}},\ \bibinfo
  {pages} {5001} (\bibinfo {year} {2010})}\BibitemShut {NoStop}%
\bibitem [{\citenamefont {Andreussi}, \citenamefont {Morrison},\ and\
  \citenamefont {Pegoraro}(2012)}]{amp1}%
  \BibitemOpen
  \bibfield  {author} {\bibinfo {author} {\bibfnamefont {T.}~\bibnamefont
  {Andreussi}}, \bibinfo {author} {\bibfnamefont {P.~J.}\ \bibnamefont
  {Morrison}}, \ and\ \bibinfo {author} {\bibfnamefont {F.}~\bibnamefont
  {Pegoraro}},\ }\href@noop {} {\bibfield  {journal} {\bibinfo  {journal}
  {Phys. Plasmas}\ }\textbf {\bibinfo {volume} {19}},\ \bibinfo {pages} {2102}
  (\bibinfo {year} {2012})}\BibitemShut {NoStop}%
\bibitem [{\citenamefont {Andreussi}, \citenamefont {Morrison},\ and\
  \citenamefont {Pegoraro}(2013)}]{amp2a}%
  \BibitemOpen
  \bibfield  {author} {\bibinfo {author} {\bibfnamefont {T.}~\bibnamefont
  {Andreussi}}, \bibinfo {author} {\bibfnamefont {P.~J.}\ \bibnamefont
  {Morrison}}, \ and\ \bibinfo {author} {\bibfnamefont {F.}~\bibnamefont
  {Pegoraro}},\ }\href@noop {} {\bibfield  {journal} {\bibinfo  {journal}
  {Phys. Plasmas}\ }\textbf {\bibinfo {volume} {20}},\ \bibinfo {pages}
  {092104} (\bibinfo {year} {2013})}\BibitemShut {NoStop}%
\bibitem [{\citenamefont {Andreussi}, \citenamefont {Morrison},\ and\
  \citenamefont {Pegoraro}(2015)}]{ampE}%
  \BibitemOpen
  \bibfield  {author} {\bibinfo {author} {\bibfnamefont {T.}~\bibnamefont
  {Andreussi}}, \bibinfo {author} {\bibfnamefont {P.~J.}\ \bibnamefont
  {Morrison}}, \ and\ \bibinfo {author} {\bibfnamefont {F.}~\bibnamefont
  {Pegoraro}},\ }\href@noop {} {\bibfield  {journal} {\bibinfo  {journal}
  {Phys. Plasmas}\ }\textbf {\bibinfo {volume} {22}},\ \bibinfo {pages}
  {039903} (\bibinfo {year} {2015})}\BibitemShut {NoStop}%
\bibitem [{\citenamefont {Morrison}(2005)}]{morrison2005}%
  \BibitemOpen
  \bibfield  {author} {\bibinfo {author} {\bibfnamefont {P.~J.}\ \bibnamefont
  {Morrison}},\ }\href@noop {} {\bibfield  {journal} {\bibinfo  {journal}
  {Phys. Plasmas}\ }\textbf {\bibinfo {volume} {12}},\ \bibinfo {pages} {8102}
  (\bibinfo {year} {2005})}\BibitemShut {NoStop}%
\bibitem [{\citenamefont {Kimura}\ and\ \citenamefont
  {Morrison}(2014)}]{kimura}%
  \BibitemOpen
  \bibfield  {author} {\bibinfo {author} {\bibfnamefont {K.}~\bibnamefont
  {Kimura}}\ and\ \bibinfo {author} {\bibfnamefont {P.~J.}\ \bibnamefont
  {Morrison}},\ }\href@noop {} {\bibfield  {journal} {\bibinfo  {journal}
  {Phys. Plasmas}\ }\textbf {\bibinfo {volume} {21}} (\bibinfo {year}
  {2014})}\BibitemShut {NoStop}%
\bibitem [{\citenamefont {Charidakos}\ \emph {et~al.}(2014)\citenamefont
  {Charidakos}, \citenamefont {Lingam}, \citenamefont {Morrison}, \citenamefont
  {White},\ and\ \citenamefont {Wurm}}]{keramidas14}%
  \BibitemOpen
  \bibfield  {author} {\bibinfo {author} {\bibfnamefont {I.~K.}\ \bibnamefont
  {Charidakos}}, \bibinfo {author} {\bibfnamefont {M.}~\bibnamefont {Lingam}},
  \bibinfo {author} {\bibfnamefont {P.~J.}\ \bibnamefont {Morrison}}, \bibinfo
  {author} {\bibfnamefont {R.~L.}\ \bibnamefont {White}}, \ and\ \bibinfo
  {author} {\bibfnamefont {A.}~\bibnamefont {Wurm}},\ }\href@noop {} {\bibfield
   {journal} {\bibinfo  {journal} {Phys. Plasmas}\ }\textbf {\bibinfo {volume}
  {21}},\ \bibinfo {pages} {092118} (\bibinfo {year} {2014})}\BibitemShut
  {NoStop}%
\bibitem [{\citenamefont {Lingam}, \citenamefont {Morrison},\ and\
  \citenamefont {Tassi}(2015)}]{LMT15}%
  \BibitemOpen
  \bibfield  {author} {\bibinfo {author} {\bibfnamefont {M.}~\bibnamefont
  {Lingam}}, \bibinfo {author} {\bibfnamefont {P.~J.}\ \bibnamefont
  {Morrison}}, \ and\ \bibinfo {author} {\bibfnamefont {E.}~\bibnamefont
  {Tassi}},\ }\href@noop {} {\bibfield  {journal} {\bibinfo  {journal} {Phys.
  Lett. A}\ }\textbf {\bibinfo {volume} {379}},\ \bibinfo {pages} {570}
  (\bibinfo {year} {2015})}\BibitemShut {NoStop}%
\bibitem [{\citenamefont {Abdelhamid}, \citenamefont {Kawazura},\ and\
  \citenamefont {Yoshida}(2015)}]{HKY15}%
  \BibitemOpen
  \bibfield  {author} {\bibinfo {author} {\bibfnamefont {H.~M.}\ \bibnamefont
  {Abdelhamid}}, \bibinfo {author} {\bibfnamefont {Y.}~\bibnamefont
  {Kawazura}}, \ and\ \bibinfo {author} {\bibfnamefont {Z.}~\bibnamefont
  {Yoshida}},\ }\href@noop {} {\bibfield  {journal} {\bibinfo  {journal} {J.
  Phys. A}\ }\textbf {\bibinfo {volume} {48}} (\bibinfo {year}
  {2015})}\BibitemShut {NoStop}%
\bibitem [{\citenamefont {Lingam}, \citenamefont {Morrison},\ and\
  \citenamefont {Miloshevich}(2015)}]{LMM15}%
  \BibitemOpen
  \bibfield  {author} {\bibinfo {author} {\bibfnamefont {M.}~\bibnamefont
  {Lingam}}, \bibinfo {author} {\bibfnamefont {P.~J.}\ \bibnamefont
  {Morrison}}, \ and\ \bibinfo {author} {\bibfnamefont {G.}~\bibnamefont
  {Miloshevich}},\ }\href@noop {} {\bibfield  {journal} {\bibinfo  {journal}
  {Phys. Plasmas}\ }\textbf {\bibinfo {volume} {22}},\ \bibinfo {pages}
  {072111} (\bibinfo {year} {2015})}\BibitemShut {NoStop}%
\bibitem [{\citenamefont {Lingam}, \citenamefont {Miloshevich},\ and\
  \citenamefont {Morrison}(2016)}]{LMM16}%
  \BibitemOpen
  \bibfield  {author} {\bibinfo {author} {\bibfnamefont {M.}~\bibnamefont
  {Lingam}}, \bibinfo {author} {\bibfnamefont {G.}~\bibnamefont {Miloshevich}},
  \ and\ \bibinfo {author} {\bibfnamefont {P.~J.}\ \bibnamefont {Morrison}},\
  }\href@noop {} {\bibfield  {journal} {\bibinfo  {journal} {Phys. Lett. A}\
  }\textbf {\bibinfo {volume} {380}},\ \bibinfo {pages} {2400} (\bibinfo {year}
  {2016})}\BibitemShut {NoStop}%
\bibitem [{\citenamefont {Morrison}, \citenamefont {Lingam},\ and\
  \citenamefont {Acevedo}(2014)}]{MLA14}%
  \BibitemOpen
  \bibfield  {author} {\bibinfo {author} {\bibfnamefont {P.~J.}\ \bibnamefont
  {Morrison}}, \bibinfo {author} {\bibfnamefont {M.}~\bibnamefont {Lingam}}, \
  and\ \bibinfo {author} {\bibfnamefont {R.}~\bibnamefont {Acevedo}},\
  }\href@noop {} {\bibfield  {journal} {\bibinfo  {journal} {Phys. Plasmas}\
  }\textbf {\bibinfo {volume} {21}} (\bibinfo {year} {2014})}\BibitemShut
  {NoStop}%
\bibitem [{\citenamefont {D'Avignon}, \citenamefont {Morrison},\ and\
  \citenamefont {Pegoraro}(2015)}]{DMP15}%
  \BibitemOpen
  \bibfield  {author} {\bibinfo {author} {\bibfnamefont {E.~C.}\ \bibnamefont
  {D'Avignon}}, \bibinfo {author} {\bibfnamefont {P.~J.}\ \bibnamefont
  {Morrison}}, \ and\ \bibinfo {author} {\bibfnamefont {F.}~\bibnamefont
  {Pegoraro}},\ }\href@noop {} {\bibfield  {journal} {\bibinfo  {journal}
  {Phys. Rev. D}\ }\textbf {\bibinfo {volume} {91}},\ \bibinfo {pages} {084050}
  (\bibinfo {year} {2015})}\BibitemShut {NoStop}%
\bibitem [{\citenamefont {Kawazura}\ \emph {et~al.}(2016)\citenamefont
  {Kawazura}, \citenamefont {Morrison}, \citenamefont {Miloshevich},\ and\
  \citenamefont {DÕAvignon}}]{KMMA15}%
  \BibitemOpen
  \bibfield  {author} {\bibinfo {author} {\bibfnamefont {Y.}~\bibnamefont
  {Kawazura}}, \bibinfo {author} {\bibfnamefont {P.~J.}\ \bibnamefont
  {Morrison}}, \bibinfo {author} {\bibfnamefont {G.}~\bibnamefont
  {Miloshevich}}, \ and\ \bibinfo {author} {\bibfnamefont {E.~C.}\ \bibnamefont
  {DÕAvignon}},\ }\href@noop {} {\bibfield  {journal} {\bibinfo  {journal}
  {draft}\ } (\bibinfo {year} {2016})}\BibitemShut {NoStop}%
\bibitem [{\citenamefont {Tronci}(2010)}]{tronci10}%
  \BibitemOpen
  \bibfield  {author} {\bibinfo {author} {\bibfnamefont {C.}~\bibnamefont
  {Tronci}},\ }\href@noop {} {\bibfield  {journal} {\bibinfo  {journal} {J.
  Phys. A: Math. Theor.}\ }\textbf {\bibinfo {volume} {43}},\ \bibinfo {pages}
  {375501} (\bibinfo {year} {2010})}\BibitemShut {NoStop}%
\bibitem [{\citenamefont {Tronci}\ \emph {et~al.}(2014)\citenamefont {Tronci},
  \citenamefont {Tassi}, \citenamefont {Camporeale},\ and\ \citenamefont
  {Morrison}}]{pjmTTC14}%
  \BibitemOpen
  \bibfield  {author} {\bibinfo {author} {\bibfnamefont {C.}~\bibnamefont
  {Tronci}}, \bibinfo {author} {\bibfnamefont {E.}~\bibnamefont {Tassi}},
  \bibinfo {author} {\bibfnamefont {E.}~\bibnamefont {Camporeale}}, \ and\
  \bibinfo {author} {\bibfnamefont {P.~J.}\ \bibnamefont {Morrison}},\
  }\href@noop {} {\bibfield  {journal} {\bibinfo  {journal} {Plasma Phys. Cont.
  Fusion}\ }\textbf {\bibinfo {volume} {56}},\ \bibinfo {pages} {095008}
  (\bibinfo {year} {2014})}\BibitemShut {NoStop}%
\bibitem [{\citenamefont {Burby}\ \emph {et~al.}(2015)\citenamefont {Burby},
  \citenamefont {Brizard}, \citenamefont {Morrison},\ and\ \citenamefont
  {Qin}}]{BBMQ15}%
  \BibitemOpen
  \bibfield  {author} {\bibinfo {author} {\bibfnamefont {J.}~\bibnamefont
  {Burby}}, \bibinfo {author} {\bibfnamefont {A.}~\bibnamefont {Brizard}},
  \bibinfo {author} {\bibfnamefont {P.~J.}\ \bibnamefont {Morrison}}, \ and\
  \bibinfo {author} {\bibfnamefont {H.}~\bibnamefont {Qin}},\ }\href@noop {}
  {\bibfield  {journal} {\bibinfo  {journal} {Phys. Lett. A}\ }\textbf
  {\bibinfo {volume} {379}},\ \bibinfo {pages} {2073} (\bibinfo {year}
  {2015})}\BibitemShut {NoStop}%
\bibitem [{\citenamefont {Brizard}\ \emph {et~al.}(2016)\citenamefont
  {Brizard}, \citenamefont {Morrison}, \citenamefont {Burby}, \citenamefont
  {de~Guillebon},\ and\ \citenamefont {Vitto}}]{BMBGV65}%
  \BibitemOpen
  \bibfield  {author} {\bibinfo {author} {\bibfnamefont {A.~J.}\ \bibnamefont
  {Brizard}}, \bibinfo {author} {\bibfnamefont {P.~J.}\ \bibnamefont
  {Morrison}}, \bibinfo {author} {\bibfnamefont {J.~W.}\ \bibnamefont {Burby}},
  \bibinfo {author} {\bibfnamefont {L.}~\bibnamefont {de~Guillebon}}, \ and\
  \bibinfo {author} {\bibfnamefont {M.}~\bibnamefont {Vitto}},\ }\href@noop {}
  {\bibfield  {journal} {\bibinfo  {journal} {arXiv:1606.06652
  [physics.plasm-ph]}\ } (\bibinfo {year} {2016})}\BibitemShut {NoStop}%
\bibitem [{\citenamefont {Morrison}(1998)}]{morrison98}%
  \BibitemOpen
  \bibfield  {author} {\bibinfo {author} {\bibfnamefont {P.~J.}\ \bibnamefont
  {Morrison}},\ }\href@noop {} {\bibfield  {journal} {\bibinfo  {journal} {Rev.
  Mod. Physics}\ }\textbf {\bibinfo {volume} {70}},\ \bibinfo {pages} {467}
  (\bibinfo {year} {1998})}\BibitemShut {NoStop}%
\bibitem [{\citenamefont {Rein}(1994)}]{rein}%
  \BibitemOpen
  \bibfield  {author} {\bibinfo {author} {\bibfnamefont {G.}~\bibnamefont
  {Rein}},\ }\href@noop {} {\bibfield  {journal} {\bibinfo  {journal} {Math.
  Methods Appl. Sci.}\ }\textbf {\bibinfo {volume} {17}},\ \bibinfo {pages}
  {1129} (\bibinfo {year} {1994})}\BibitemShut {NoStop}%
\bibitem [{\citenamefont {Chandresekhar}(1961)}]{chandresek}%
  \BibitemOpen
  \bibfield  {author} {\bibinfo {author} {\bibfnamefont {S.}~\bibnamefont
  {Chandresekhar}},\ }\href@noop {} {\emph {\bibinfo {title} {Hydrodynamic and
  Hydromagnetic Stability}}}\ (\bibinfo  {publisher} {Oxford University Press,
  Oxford, U.K.},\ \bibinfo {year} {1961})\BibitemShut {NoStop}%
\bibitem [{\citenamefont {Morrison}\ and\ \citenamefont
  {Pfirsch}(1990)}]{MP90}%
  \BibitemOpen
  \bibfield  {author} {\bibinfo {author} {\bibfnamefont {P.~J.}\ \bibnamefont
  {Morrison}}\ and\ \bibinfo {author} {\bibfnamefont {D.}~\bibnamefont
  {Pfirsch}},\ }\href@noop {} {\bibfield  {journal} {\bibinfo  {journal} {Phys.
  Fluids B}\ }\textbf {\bibinfo {volume} {2}},\ \bibinfo {pages} {1105}
  (\bibinfo {year} {1990})}\BibitemShut {NoStop}%
\bibitem [{\citenamefont {Frieman}\ and\ \citenamefont
  {Rotenberg}(1960)}]{Frieman1960}%
  \BibitemOpen
  \bibfield  {author} {\bibinfo {author} {\bibfnamefont {E.}~\bibnamefont
  {Frieman}}\ and\ \bibinfo {author} {\bibfnamefont {M.}~\bibnamefont
  {Rotenberg}},\ }\href@noop {} {\bibfield  {journal} {\bibinfo  {journal}
  {Rev. Mod. Phys.}\ }\textbf {\bibinfo {volume} {32}},\ \bibinfo {pages} {898}
  (\bibinfo {year} {1960})}\BibitemShut {NoStop}%
\bibitem [{\citenamefont {Bernstein}\ \emph {et~al.}(1958)\citenamefont
  {Bernstein}, \citenamefont {Frieman}, \citenamefont {Kruskal},\ and\
  \citenamefont {Kulsrud}}]{Bernstein1958b}%
  \BibitemOpen
  \bibfield  {author} {\bibinfo {author} {\bibfnamefont {I.~B.}\ \bibnamefont
  {Bernstein}}, \bibinfo {author} {\bibfnamefont {E.~A.}\ \bibnamefont
  {Frieman}}, \bibinfo {author} {\bibfnamefont {M.~D.}\ \bibnamefont
  {Kruskal}}, \ and\ \bibinfo {author} {\bibfnamefont {R.~M.}\ \bibnamefont
  {Kulsrud}},\ }\href@noop {} {\bibfield  {journal} {\bibinfo  {journal} {Proc.
  Roy. Soc. Lond. A}\ }\textbf {\bibinfo {volume} {244}},\ \bibinfo {pages}
  {17} (\bibinfo {year} {1958})}\BibitemShut {NoStop}%
\bibitem [{\citenamefont {Woltjer}(1958)}]{W1}%
  \BibitemOpen
  \bibfield  {author} {\bibinfo {author} {\bibfnamefont {L.}~\bibnamefont
  {Woltjer}},\ }\href@noop {} {\bibfield  {journal} {\bibinfo  {journal} {Proc.
  Natl. Acad. Sci.}\ }\textbf {\bibinfo {volume} {44}},\ \bibinfo {pages} {833}
  (\bibinfo {year} {1958})}\BibitemShut {NoStop}%
\bibitem [{\citenamefont {Woltjer}(1959{\natexlab{a}})}]{W2}%
  \BibitemOpen
  \bibfield  {author} {\bibinfo {author} {\bibfnamefont {L.}~\bibnamefont
  {Woltjer}},\ }\href@noop {} {\bibfield  {journal} {\bibinfo  {journal} {Proc.
  Natl. Acad. Sci.}\ }\textbf {\bibinfo {volume} {45}},\ \bibinfo {pages} {769}
  (\bibinfo {year} {1959}{\natexlab{a}})}\BibitemShut {NoStop}%
\bibitem [{\citenamefont {Woltjer}(1959{\natexlab{b}})}]{W3}%
  \BibitemOpen
  \bibfield  {author} {\bibinfo {author} {\bibfnamefont {L.}~\bibnamefont
  {Woltjer}},\ }\href@noop {} {\bibfield  {journal} {\bibinfo  {journal}
  {Astrophys J.}\ }\textbf {\bibinfo {volume} {130}},\ \bibinfo {pages} {400}
  (\bibinfo {year} {1959}{\natexlab{b}})}\BibitemShut {NoStop}%
\bibitem [{\citenamefont {Woltjer}(1959{\natexlab{c}})}]{W4}%
  \BibitemOpen
  \bibfield  {author} {\bibinfo {author} {\bibfnamefont {L.}~\bibnamefont
  {Woltjer}},\ }\href@noop {} {\bibfield  {journal} {\bibinfo  {journal}
  {Astrophys J.}\ }\textbf {\bibinfo {volume} {130}},\ \bibinfo {pages} {404}
  (\bibinfo {year} {1959}{\natexlab{c}})}\BibitemShut {NoStop}%
\bibitem [{\citenamefont {Taylor}(1974)}]{T1}%
  \BibitemOpen
  \bibfield  {author} {\bibinfo {author} {\bibfnamefont {J.~B.}\ \bibnamefont
  {Taylor}},\ }\href@noop {} {\bibfield  {journal} {\bibinfo  {journal} {Phys.
  Rev. Lett.}\ }\textbf {\bibinfo {volume} {33}},\ \bibinfo {pages} {1139}
  (\bibinfo {year} {1974})}\BibitemShut {NoStop}%
\bibitem [{\citenamefont {Taylor}(1986)}]{T2}%
  \BibitemOpen
  \bibfield  {author} {\bibinfo {author} {\bibfnamefont {J.~B.}\ \bibnamefont
  {Taylor}},\ }\href@noop {} {\bibfield  {journal} {\bibinfo  {journal} {Rev.
  Mod. Phys.}\ }\textbf {\bibinfo {volume} {58}},\ \bibinfo {pages} {741}
  (\bibinfo {year} {1986})}\BibitemShut {NoStop}%
\bibitem [{\citenamefont {Morrison}(1982)}]{morrison82}%
  \BibitemOpen
  \bibfield  {author} {\bibinfo {author} {\bibfnamefont {P.~J.}\ \bibnamefont
  {Morrison}},\ }\href@noop {} {\bibfield  {journal} {\bibinfo  {journal} {AIP
  Conf. Series}\ }\textbf {\bibinfo {volume} {88}},\ \bibinfo {pages} {13}
  (\bibinfo {year} {1982})}\BibitemShut {NoStop}%
\bibitem [{\citenamefont {Padhye}\ and\ \citenamefont
  {Morrison}(1996{\natexlab{a}})}]{padhye96}%
  \BibitemOpen
  \bibfield  {author} {\bibinfo {author} {\bibfnamefont {N.}~\bibnamefont
  {Padhye}}\ and\ \bibinfo {author} {\bibfnamefont {P.}~\bibnamefont
  {Morrison}},\ }\href@noop {} {\bibfield  {journal} {\bibinfo  {journal}
  {Plasma Phys. Repts.}\ }\textbf {\bibinfo {volume} {22}},\ \bibinfo {pages}
  {869} (\bibinfo {year} {1996}{\natexlab{a}})}\BibitemShut {NoStop}%
\bibitem [{\citenamefont {Padhye}\ and\ \citenamefont
  {Morrison}(1996{\natexlab{b}})}]{padhye96fluid}%
  \BibitemOpen
  \bibfield  {author} {\bibinfo {author} {\bibfnamefont {N.}~\bibnamefont
  {Padhye}}\ and\ \bibinfo {author} {\bibfnamefont {P.}~\bibnamefont
  {Morrison}},\ }\href@noop {} {\bibfield  {journal} {\bibinfo  {journal}
  {Phys. Lett. A}\ }\textbf {\bibinfo {volume} {219}},\ \bibinfo {pages} {287}
  (\bibinfo {year} {1996}{\natexlab{b}})}\BibitemShut {NoStop}%
\bibitem [{\citenamefont {Yoshida}, \citenamefont {Morrison},\ and\
  \citenamefont {Dobarro}(2014)}]{YM14a}%
  \BibitemOpen
  \bibfield  {author} {\bibinfo {author} {\bibfnamefont {Z.}~\bibnamefont
  {Yoshida}}, \bibinfo {author} {\bibfnamefont {P.~J.}\ \bibnamefont
  {Morrison}}, \ and\ \bibinfo {author} {\bibfnamefont {F.}~\bibnamefont
  {Dobarro}},\ }\href@noop {} {\bibfield  {journal} {\bibinfo  {journal} {J.
  Math. Fluid Mech.}\ }\textbf {\bibinfo {volume} {16}},\ \bibinfo {pages} {41}
  (\bibinfo {year} {2014})}\BibitemShut {NoStop}%
\bibitem [{\citenamefont {Yoshida}\ and\ \citenamefont
  {Morrison}(2014)}]{YM14b}%
  \BibitemOpen
  \bibfield  {author} {\bibinfo {author} {\bibfnamefont {Z.}~\bibnamefont
  {Yoshida}}\ and\ \bibinfo {author} {\bibfnamefont {P.~J.}\ \bibnamefont
  {Morrison}},\ }\href@noop {} {\bibfield  {journal} {\bibinfo  {journal}
  {Fluid Dyn. Res.}\ }\textbf {\bibinfo {volume} {46}},\ \bibinfo {pages}
  {031412} (\bibinfo {year} {2014})}\BibitemShut {NoStop}%
\bibitem [{\citenamefont {Morrison}\ and\ \citenamefont
  {Pfirsch}(1989)}]{MP89}%
  \BibitemOpen
  \bibfield  {author} {\bibinfo {author} {\bibfnamefont {P.~J.}\ \bibnamefont
  {Morrison}}\ and\ \bibinfo {author} {\bibfnamefont {D.}~\bibnamefont
  {Pfirsch}},\ }\href@noop {} {\bibfield  {journal} {\bibinfo  {journal} {Phys.
  Rev. A}\ }\textbf {\bibinfo {volume} {40}},\ \bibinfo {pages} {3898}
  (\bibinfo {year} {1989})}\BibitemShut {NoStop}%
\bibitem [{\citenamefont {Landau}\ and\ \citenamefont
  {Lifshitz}(1975)}]{landau1975}%
  \BibitemOpen
  \bibfield  {author} {\bibinfo {author} {\bibfnamefont {L.}~\bibnamefont
  {Landau}}\ and\ \bibinfo {author} {\bibfnamefont {E.}~\bibnamefont
  {Lifshitz}},\ }\href@noop {} {\emph {\bibinfo {title} {Classical theory of
  fields}}}\ (\bibinfo  {publisher} {Pergamon Press, Oxford},\ \bibinfo {year}
  {1975})\BibitemShut {NoStop}%
\bibitem [{\citenamefont {Schwarzschild}(1906)}]{S06}%
  \BibitemOpen
  \bibfield  {author} {\bibinfo {author} {\bibfnamefont {K.}~\bibnamefont
  {Schwarzschild}},\ }\href@noop {} {\bibfield  {journal} {\bibinfo  {journal}
  {Nachr. K. Ges. Wiss., Gottingen.}\ ,\ \bibinfo {pages} {41}} (\bibinfo
  {year} {1906})}\BibitemShut {NoStop}%
\bibitem [{\citenamefont {Thompson}(1951)}]{thompson}%
  \BibitemOpen
  \bibfield  {author} {\bibinfo {author} {\bibfnamefont {W.~B.}\ \bibnamefont
  {Thompson}},\ }\href@noop {} {\bibfield  {journal} {\bibinfo  {journal}
  {Phil. Mag.}\ }\textbf {\bibinfo {volume} {42}},\ \bibinfo {pages} {1417}
  (\bibinfo {year} {1951})}\BibitemShut {NoStop}%
\bibitem [{\citenamefont {Kruskal}\ and\ \citenamefont
  {Schwarzschild}(1954)}]{Kruskal54}%
  \BibitemOpen
  \bibfield  {author} {\bibinfo {author} {\bibfnamefont {M.~D.}\ \bibnamefont
  {Kruskal}}\ and\ \bibinfo {author} {\bibfnamefont {M.}~\bibnamefont
  {Schwarzschild}},\ }\href@noop {} {\bibfield  {journal} {\bibinfo  {journal}
  {Proc. R. Soc. Lond. A}\ }\textbf {\bibinfo {volume} {223}},\ \bibinfo
  {pages} {348} (\bibinfo {year} {1954})}\BibitemShut {NoStop}%
\bibitem [{\citenamefont {Tserkovnikov}(1960)}]{tserkovnikov}%
  \BibitemOpen
  \bibfield  {author} {\bibinfo {author} {\bibfnamefont {Y.~A.}\ \bibnamefont
  {Tserkovnikov}},\ }\href@noop {} {\bibfield  {journal} {\bibinfo  {journal}
  {Doklady Akad. Nauk S. S. S. R.}\ }\textbf {\bibinfo {volume} {130}},\
  \bibinfo {pages} {295} (\bibinfo {year} {1960})}\BibitemShut {NoStop}%
\bibitem [{\citenamefont {Newcomb}(1961)}]{newcomb61}%
  \BibitemOpen
  \bibfield  {author} {\bibinfo {author} {\bibfnamefont {W.~A.}\ \bibnamefont
  {Newcomb}},\ }\href@noop {} {\bibfield  {journal} {\bibinfo  {journal} {Phys.
  Fluids}\ }\textbf {\bibinfo {volume} {4}},\ \bibinfo {pages} {391} (\bibinfo
  {year} {1961})}\BibitemShut {NoStop}%
\bibitem [{\citenamefont {Yu}(1966)}]{yu66}%
  \BibitemOpen
  \bibfield  {author} {\bibinfo {author} {\bibfnamefont {C.~P.}\ \bibnamefont
  {Yu}},\ }\href@noop {} {\bibfield  {journal} {\bibinfo  {journal} {Phys.
  Fluids}\ }\textbf {\bibinfo {volume} {9}},\ \bibinfo {pages} {412} (\bibinfo
  {year} {1966})}\BibitemShut {NoStop}%
\bibitem [{\citenamefont {Greene}\ and\ \citenamefont
  {Johnson}(1968)}]{greene68}%
  \BibitemOpen
  \bibfield  {author} {\bibinfo {author} {\bibfnamefont {J.~M.}\ \bibnamefont
  {Greene}}\ and\ \bibinfo {author} {\bibfnamefont {J.~L.}\ \bibnamefont
  {Johnson}},\ }\href@noop {} {\bibfield  {journal} {\bibinfo  {journal}
  {Plasma Phys.}\ }\textbf {\bibinfo {volume} {10}},\ \bibinfo {pages} {729}
  (\bibinfo {year} {1968})}\BibitemShut {NoStop}%
\bibitem [{\citenamefont {Morrison}, \citenamefont {Tassi},\ and\ \citenamefont
  {Tronko}(2013)}]{pjmTT13}%
  \BibitemOpen
  \bibfield  {author} {\bibinfo {author} {\bibfnamefont {P.~J.}\ \bibnamefont
  {Morrison}}, \bibinfo {author} {\bibfnamefont {E.}~\bibnamefont {Tassi}}, \
  and\ \bibinfo {author} {\bibfnamefont {N.}~\bibnamefont {Tronko}},\
  }\href@noop {} {\bibfield  {journal} {\bibinfo  {journal} {Phys. Plasmas}\
  }\textbf {\bibinfo {volume} {20}},\ \bibinfo {pages} {042109} (\bibinfo
  {year} {2013})}\BibitemShut {NoStop}%
\bibitem [{\citenamefont {Hameiri}(2003)}]{hameiri03}%
  \BibitemOpen
  \bibfield  {author} {\bibinfo {author} {\bibfnamefont {E.}~\bibnamefont
  {Hameiri}},\ }\href@noop {} {\bibfield  {journal} {\bibinfo  {journal} {Phys.
  Plasmas}\ }\textbf {\bibinfo {volume} {10}},\ \bibinfo {pages} {2643}
  (\bibinfo {year} {2003})}\BibitemShut {NoStop}%
\bibitem [{\citenamefont {Throumoulopoulos}\ and\ \citenamefont
  {Pfirsch}(1994)}]{throum94}%
  \BibitemOpen
  \bibfield  {author} {\bibinfo {author} {\bibfnamefont {G.~N.}\ \bibnamefont
  {Throumoulopoulos}}\ and\ \bibinfo {author} {\bibfnamefont {D.}~\bibnamefont
  {Pfirsch}},\ }\href@noop {} {\bibfield  {journal} {\bibinfo  {journal} {Phys.
  Rev. E}\ }\textbf {\bibinfo {volume} {49}},\ \bibinfo {pages} {3290}
  (\bibinfo {year} {1994})}\BibitemShut {NoStop}%
\bibitem [{\citenamefont {Throumoulopoulos}\ and\ \citenamefont
  {Pfirsch}(1996)}]{throum96}%
  \BibitemOpen
  \bibfield  {author} {\bibinfo {author} {\bibfnamefont {G.~N.}\ \bibnamefont
  {Throumoulopoulos}}\ and\ \bibinfo {author} {\bibfnamefont {D.}~\bibnamefont
  {Pfirsch}},\ }\href@noop {} {\bibfield  {journal} {\bibinfo  {journal} {Phys.
  Rev. E}\ }\textbf {\bibinfo {volume} {53}},\ \bibinfo {pages} {2767}
  (\bibinfo {year} {1996})}\BibitemShut {NoStop}%
\bibitem [{\citenamefont {Pfirsch}\ and\ \citenamefont
  {Correa-Restrepo}(2004)}]{pfirsch04a}%
  \BibitemOpen
  \bibfield  {author} {\bibinfo {author} {\bibfnamefont {D.}~\bibnamefont
  {Pfirsch}}\ and\ \bibinfo {author} {\bibfnamefont {D.}~\bibnamefont
  {Correa-Restrepo}},\ }\href@noop {} {\bibfield  {journal} {\bibinfo
  {journal} {J. Plasma Phys.}\ }\textbf {\bibinfo {volume} {70}},\ \bibinfo
  {pages} {719} (\bibinfo {year} {2004})}\BibitemShut {NoStop}%
\bibitem [{\citenamefont {Correa-Restrepo}\ and\ \citenamefont
  {Pfirsch}(2004)}]{pfirsch04b}%
  \BibitemOpen
  \bibfield  {author} {\bibinfo {author} {\bibfnamefont {D.}~\bibnamefont
  {Correa-Restrepo}}\ and\ \bibinfo {author} {\bibfnamefont {D.}~\bibnamefont
  {Pfirsch}},\ }\href@noop {} {\bibfield  {journal} {\bibinfo  {journal} {J.
  Plasma Phys.}\ }\textbf {\bibinfo {volume} {70}},\ \bibinfo {pages} {757}
  (\bibinfo {year} {2004})}\BibitemShut {NoStop}%
\end{thebibliography}%


\end{document}